\documentclass[%
reprint,
superscriptaddress,
 amsmath,amssymb,
 aps,
floatfix,
]{revtex4-1}

\usepackage{color}
\usepackage{siunitx}

\usepackage{graphicx}
\usepackage{dcolumn}
\usepackage{bm}
\usepackage{mathrsfs}
\usepackage[normalem]{ulem}

\newcommand{\cmmt}[1]{\textcolor{black}{#1}}
\renewcommand{\Re}{\mathrm{Re}}

\begin{document}

\preprint{APS/123-QED}


\title{Quantum description of surface-enhanced resonant Raman scattering within a hybrid-optomechanical model}

\author{Tom\'{a}\v{s} Neuman}
\email{tomas\_neuman001@ehu.eus}
\affiliation{Centro de F\'{\i}sica de Materiales de San Sebasti\'an, CFM - MPC (CSIC-UPV/EHU), Paseo Manuel Lardizabal 5, 20018 Donostia-San Sebasti\'an, Spain}
\affiliation{Donostia International Physics Center (DIPC), 20018 San Sebasti\'an-Donostia, Spain}

\author{Ruben Esteban}%
\affiliation{Donostia International Physics Center (DIPC), 20018 San Sebasti\'an-Donostia, Spain}
\affiliation{IKERBASQUE, Basque Foundation for Science, Maria Diaz de Haro 3, 48013 Bilbao, Spain}

\author{Geza Giedke}%
 \affiliation{Donostia International Physics Center (DIPC), 20018 San Sebasti\'an-Donostia, Spain}
 \affiliation{IKERBASQUE, Basque Foundation for Science, Maria Diaz de Haro 3, 48013 Bilbao, Spain}
 
\author{Miko\l{}aj K. Schmidt}%
\altaffiliation{Currently at Department of Physics and Astronomy, Macquarie University NSW 2109, Australia}
\affiliation{Centro de F\'{\i}sica de Materiales de San Sebasti\'an, CFM - MPC (CSIC-UPV/EHU), Paseo Manuel Lardizabal 5, 20018 Donostia-San Sebasti\'an, Spain}
\affiliation{Donostia International Physics Center (DIPC), 20018 San Sebasti\'an-Donostia, Spain}

\author{Javier Aizpurua}%
\email{aizpurua@ehu.eus}
\affiliation{Centro de F\'{\i}sica de Materiales de San Sebasti\'an, CFM - MPC (CSIC-UPV/EHU), Paseo Manuel Lardizabal 5, 20018 Donostia-San Sebasti\'an, Spain}
\affiliation{Donostia International Physics Center (DIPC), 20018 San Sebasti\'an-Donostia, Spain}

\begin{abstract}
Surface-Enhanced Raman Scattering (SERS) allows for detection and identification of molecular vibrational fingerprints in minute sample quantities. The SERS process can be also exploited for optical manipulation of molecular vibrations. We present a quantum description of Surface-Enhanced \textit{Resonant} Raman scattering (SERRS), in analogy to hybrid cavity optomechanics, and compare the resonant situation with the off-resonant SERS. Our model predicts the existence of a regime of coherent interaction between electronic and vibrational degrees of freedom of a molecule, mediated by a plasmonic nanocavity. This coherent mechanism can be achieved by parametrically tuning the frequency and intensity of the incident pumping laser and is related to the optomechanical pumping of molecular vibrations. We find that vibrational pumping is able to selectively activate a particular vibrational mode, thus providing a mechanism to control its population and drive plasmon-assisted chemistry.
\end{abstract}

\maketitle

\section{Introduction}

Surface plasmon excitations in metallic particles are able to squeeze and enhance electromagnetic fields down to the nanometric scale and thus dramatically enhance the interaction of nearby molecules with the incident light. The plasmonic near-field enhancement has been exploited in plasmon-enhanced spectroscopies, particularly in Surface-Enhanced Raman Spectroscopy (SERS)\cite{kneipp1996poppump, kneipp1997singlemoleculesers, xu1999singlehemoglobine, haslett2000cansers, xu2000elmag,brolo2004ratiosers, maher2004stokesantistokes, haes2005plasmatsers, moskovits2005sers, leru2006vibpump, stiles2008SERS, tong_xu_kall_2014, zrimsek2017singlemol}, which enables detection of minute quantities of molecular samples. The improved design of plasmonic cavities has allowed for spectroscopic investigation of even single molecules that are placed into ultranarrow plasmonic gaps \cite{xu1999singlehemoglobine, zhang2013chemical}. 
Current experimental strategies have taken advantage of the properties of plasmonic cavity modes that allowed reaching the plasmon-exciton strong-coupling regime with single molecules \cite{chikkaraddy2016single}, as well as intramolecular optical mapping of single-molecule vibrations in SERS \cite{zhang2013chemical} or in electroluminescence \cite{schull2017vibmapping, lee2019visualizing}. These results suggest the possibility to push the use of plasmonic modes to further actively control the quantum state of a single molecule and thus influence its chemistry \cite{stipe1997single, komeda2005chemical, crim1996bond, ho2002single, pascual2003selectivity, hahna2005orbital, herrera2016chemistry}. Recent theoretical and experimental studies \cite{roelli2015molecular, schmidt2015qed, schmidt2017linking, kamandar2017quantum, benz2016single} have revealed that off-resonant SERS can be understood as a quantum optomechanical process \cite{aspelmayer2014, roelli2015molecular, schmidt2015qed} where the single plasmon mode (sustained in a plasmonic cavity) of frequency $\omega_{\rm pl}$ plays the role of the macroscopic optical cavity and the molecular vibration of frequency $\Omega$ plays the role of the macroscopic oscillation of the mirror.
The description of such a process requires the development of concepts and methods beyond the standard classical description of SERS \cite{roelli2015molecular, schmidt2015qed, benz2016single, schmidt2017linking, kamandar2017quantum}.

In this work we address a quantum mechanical theory of Surface-Enhanced \textit{Resonant} Raman Scattering (SERRS), where an optical plasmonic mode supported by a metallic nanostructure mediates a coherent laser excitation of a nearby single molecule described as an electronic two-level system (TLS), coupled to a vibrational mode. In SERRS this laser excitation is assumed to be \textit{resonant} with the electronic transition in the molecule. We discuss the similarities and differences between the SERRS Hamiltonian and the off-resonant quantum optomechanical Hamiltonian, which has been described previously. To that end we adopt a range of optomechanical parameters available in typical resonant situations.
We then show that non-trivial phenomena emerge in SERRS under intense laser illumination, when the non-linearities of the molecule can trigger the coherent coupling of molecular electronic and vibrational degrees of freedom \cite{johansson2004prl, johansson2005surface, delValle09, Garcia-VidalPRL2014, gu2010resonance, trugler2008strong, ridolfo2010quantum}. We further exploit the analogy with quantum optomechanics to propose a mechanism of on-demand frequency-selective pumping of molecular vibrations \cmmt{\cite{kneipp1996poppump, haslett2000cansers, brolo2004ratiosers, maher2004stokesantistokes, leru2006vibpump}} via the coherent laser illumination. These phenomena may provide a means to drive plasmon-enhanced vibrational spectroscopy to the realm of single-molecule selective chemistry or engineering of single-molecule optomechanical systems involving molecular vibrations on demand.

\section{SE(R)RS as an optomechanical process}

Here we extend the analogy between optomechanics and SERS and describe SERRS in the framework of cavity quantum electrodynamics (QED) as a hybrid-optomechanical process \cite{ramos2013nonlinear, akram2015fanohybrid}. 
For further convenience we now briefly describe and compare the resonant and off-resonant scenarios. 

\subsection{Off-resonant SERS}
In off-resonant SERS all electronic transitions in the molecule are far detuned from the frequency of the probing laser. The optomechanical Hamiltonian describing off-resonant SERS reads \cite{roelli2015molecular, schmidt2015qed, schmidt2017linking,kamandar2017quantum}
\begin{align}
H_{\rm om}=H_{\rm pl}+H_{\rm vib}+H_{\rm pl-vib}+H_{\rm pump},
\end{align}
where
\begin{align}
H_{\rm pl}&=\hbar\omega_{\rm pl} a{^\dagger} a\nonumber\\
H_{\rm vib}&=\hbar\Omega b{^\dagger} b\nonumber\\
H_{\rm pl-vib}&=-\hbar g_{\rm om} a^\dagger a (b^\dagger+b)\nonumber\\
H_{\rm pump}&=\hbar\mathcal{E} \left[a \exp (\mathrm{i}\omega_{\rm L}t)+a^\dagger \exp (-\mathrm{i}\omega_{\rm L}t)\right].\label{eq:optomech}
\end{align}
Here operators $a$ ($a^\dagger$) and $b$ ($b^\dagger$) are the annihilation (creation) operators for plasmons and vibrations, respectively, $\mathcal E$ is the pumping amplitude (with $\lvert\mathcal E \rvert^2$ proportional to the laser intensity) that characterizes the interaction Hamiltonian between the plasmon mode and the classical laser illumination of frequency $\omega_{\rm L}$, and $g_{\rm om}$ is the optomechanical coupling constant \cmmt{that can be connected to the Raman tensor of the molecule \cite{roelli2015molecular, schmidt2017linking, schmidt2015qed} and to the plasmonic near field. The Raman tensor contains the influence of the off-resonant electronic transitions in the molecule that are not considered explicitly in the Hamiltonian. The pumping term $H_{\rm pump}$  is considered in the rotating-wave approximation (RWA) assuming that $\mathcal{E}$ does not reach more than $ \mathcal{E}\approx 0.1\omega_{\rm pl}$. }

In what follows it will be convenient to interpret the optomechanical interaction as a displacement of the vibrational mode by a dimensionless value $n_{\rm pl}d_{\rm om}$, dependent on the number of excitations (plasmons) in the cavity, $n_{\rm pl}$. 
This becomes apparent after rearranging the bare optomechanical Hamiltonian $H_{\rm bom}=H_{\rm pl}+H_{\rm vib}+H_{\rm pl-vib}$ into the form
\begin{align}
H_{\rm bom}=&\hbar (\omega_{\rm pl}-\Omega d_{\rm om}^2a^\dagger a) a^\dagger a+\nonumber\\
&\hbar\Omega (b^\dagger+d_{\rm om}a^\dagger a)(b+d_{\rm om}a^\dagger a),\label{eq:pressureoptomechOR}
\end{align}
\cmmt{with the dimensionless displacement $d_{\rm om}=-g_{\rm om}/\Omega$. The first line of Eq.\,\eqref{eq:pressureoptomechOR} is the non-linear Hamiltonian of the cavity excitations (plasmons). In the limit of a weakly populated cavity and $d_{\rm om}\ll 1$ we can neglect the small nonlinear term $-\hbar\Omega d_{\rm om}^2 a^\dagger a^\dagger a a $, and we re-define the plasmon frequency as $[\hbar\omega_{\rm pl}-\Omega d_{\rm om}^2] \equiv \hbar\omega'_{\rm pl}$ and recover the linear plasmon Hamiltonian $\hbar\omega'_{\rm pl}a^\dagger a$, so that:
\begin{align}
H_{\rm bom}\approx &\hbar \omega'_{\rm pl} a^\dagger a+\hbar\Omega (b^\dagger+d_{\rm om}a^\dagger a)(b+d_{\rm om}a^\dagger a).\label{eq:pressureoptomech}
\end{align}
The second line in Eq.\,\eqref{eq:pressureoptomechOR} has the sought form of a vibrational mode displaced by an amount that depends on the number of plasmonic excitations in the cavity $a^\dagger a$.
}

\begin{figure*}
\includegraphics[scale=0.27]{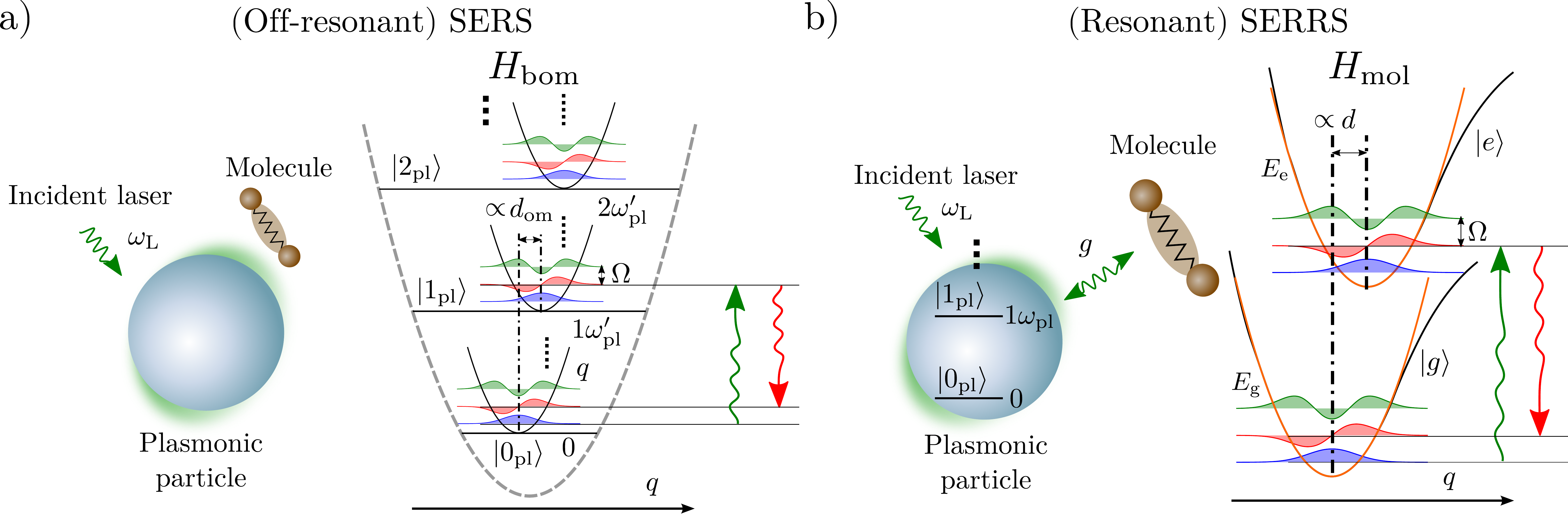}
\caption{Schematics of (a) the off-resonant SERS process in a plasmonic particle and a vibrating molecule, and (b) the SERRS process, both depicted with their corresponding level structure. (a) The plasmonic number states $|n_{\rm pl}\rangle$ have equidistant energies $\hbar n_{\rm pl}\omega'_{\rm pl}$ (vertical axis) and vibrational fine-structure for each $|n_{\rm pl}\rangle$. The vibrational parabolas are displaced by $n_{\rm pl}d_{\rm om}$ along the dimensionless normal coordinate $q$ depending on the number of plasmonic excitations. 
(b) The SERRS system consists of a plasmonic particle interacting with a molecule described by an electronic (two-level) and
vibrational (bosonic) degrees of freedom: the potential energy surfaces $\hbar E_{\rm g}(q)$ and $\hbar E_{\rm e}(q)$ for the vibrations depend on the electronic states and
are shifted with respect to the dimensionless normal coordinate $q$ by a
displacement $d$.
The plasmon mode is excited by coherent laser illumination of frequency $\omega_{\rm L}\approx \omega_{\rm pl}$.}
\label{fig:hamiltonian}
\end{figure*}

The level structure of the bare optomechanical Hamiltonian $H_{\rm bom}$ [Eq.\,\eqref{eq:pressureoptomech}] is visualized in Fig.~\ref{fig:hamiltonian}(a). 
The large grey dashed parabola illustrates an effective potential supporting the plasmonic mode. The vibrational potential is represented by the small parabolas that are displaced along the dimensionless normal coordinate $q$ by a magnitude $n_{\rm pl}d_{\rm om}$ proportional to the plasmonic number state $|n_{\rm pl}\rangle$\cite{Nunnenkamp2011spoptomech, Nunnenkamp2012sccooling}. The energies of the plasmon Hamiltonian form an equidistant ladder, schematically drawn for the three lowest plasmon number states $|n_{\rm pl}\rangle$, and contain a fine-structure of molecular vibrational sub-levels.

 
\subsection{SERRS}\label{subsec:SERRS}

We complete the previous picture of SERS by addressing the scenario where the frequency of the incident laser approaches the molecular electronic resonance. To describe SERRS we can consider the molecule as a TLS that interacts with the vibrational modes via a polaronic coupling term \cite{li1994overdampedbrownian, herrera2017absem, galperin2009raman,ramos2013nonlinear, cwik2016excitonic, zeb2018exactvibdressed} and with the plasmonic cavity via the Jaynes-Cummings coupling term [see Fig.~\ref{fig:hamiltonian}(b)]. The molecular vibrations are modeled as bosons within the Born-Oppenheimer approximation, where the effective harmonic vibrational potential is given by the ground state [$E_{\rm g}(q)$] and the excited state [$E_{\rm e}(q)$] potential energy surfaces (PESs) along a normal vibrational coordinate $q$, respectively \cite{klessinger1995excited, may2008charge, galego2015, galego2016}. We consider that the vibrational energies of the molecule,  $\hbar\Omega$, are the same for the ground and for the excited state.

The Hamiltonian of the SERRS system $H_{\rm om}^{\rm res}$, using the rotating wave approximation, can be expressed  
as \cite{betzholz2014quantum, ramos2013nonlinear, jaehne2008ground, GarciaVidal2015}:
\begin{align}\label{eq:cohelmoldr}
H_{\rm om}^{\rm res}&=H_{\rm pl}+H_{\rm mol}+H_{\rm pump}+H_{\rm pl-e},
\end{align}
with
\begin{align}
H_{\rm mol}=&\hbar [E_{\rm e}(d)-E_{\rm g}(0)]\sigma{^\dagger} \sigma\nonumber\\
&+\hbar\Omega (b{^\dagger} +\sigma_{\rm e}d)(b+\sigma_{\rm e}d),\label{eq:molham1}\\
H_{\rm pl-e}=&\hbar g a\sigma{^\dagger} +\hbar g^\ast a{^\dagger} \sigma\nonumber
\end{align}
and $H_{\rm pl}$, $H_{\rm pump}$ as defined in Eq.\,\eqref{eq:optomech} above.
Here the operator $\sigma$ ($\sigma^\dagger$) is the lowering (raising) operator of the TLS, with $\sigma_{\rm e}=\sigma^\dagger\sigma$ the TLS number operator. The parameter $d$ is the dimensionless displacement between the minima of the ground- and excited-state PESs, which is related to the Huang-Rhys factor \cite{klessinger1995excited, may2008charge, cwik2016excitonic, herrera2017absem, GarciaVidal2015}, $S$, as $S=d^2$, and is a measure of the coupling between the molecular vibration and the excitonic transition. The interaction of the localized plasmon excitation and the molecular electronic levels is mediated by the plasmon-exciton coupling constant $g$. More details about the model Hamiltonian are given in Appendix\,\ref{app:sysham}.

The level diagram describing the SERRS Hamiltonian $H_{\rm om}^{\rm res}$ [Eq.\,\eqref{eq:cohelmoldr}] is sketched in Fig. \ref{fig:hamiltonian} (b).
Strikingly, both the off-resonant Hamiltonian, $H_{\rm bom}$, and the molecular Hamiltonian in SERRS, $H_{\rm mol}$, can be represented as a series of mutually displaced harmonic vibrational PESs. The electronic states in SERRS thus play the role of the plasmon number states in the off-resonant case, an analogy which can be identified from the comparison of the Hamiltonians in Eq.\,\eqref{eq:molham1} and Eq.\,\eqref{eq:pressureoptomech}. 

In the limit of single-photon optomechanics \cite{Nunnenkamp2011spoptomech, Nunnenkamp2012sccooling, akram2010singlephoton}, where the plasmon Hilbert space is limited to the vacuum state and the singly excited state, the molecular Hamitonian $H_{\rm mol}$ and $H_{\rm bom}$ become formally identical. However, as we detail later, if the incident laser is strong, the non-linear character of the excitonic TLS Hamiltonian [Eq.\,\eqref{eq:molham1}] will lead to novel physical phenomena that cannot be achieved in the off-resonant SERS situation [Eq.\,\eqref{eq:pressureoptomech}]. This perspective makes it particularly attractive to study SERRS.
\subsection{Dissipative processes in SE(R)RS} 
In realistic systems, excitations undergo decay, pumping and dephasing processes. 
Losses and thermal pumping are considered in the dynamics of the system by solving the master equation for the density matrix, $\rho$, with incoherent damping introduced via the Lindblad-Kossakowski terms \cite{Breuer2005} for the plasmon and the vibration
\begin{align}
\mathcal{L}_{a}[\rho]&=-\frac{\gamma_{a}}{2}  \left( a^\dagger a\rho+\rho a^\dagger a -2 a\rho a^\dagger\right),\\
\mathcal{L}_{b}[\rho]&=-(n^{\rm vib}_{T}+1)\frac{\gamma_{b}}{2}\left( b^\dagger b\rho+\rho b^\dagger b -2 b\rho b^\dagger\right),\label{eq:lindvib}
\end{align}
and in case of SERRS for the electronic TLS
\begin{align}
\mathcal{L}_{\sigma}[\rho]&=- \frac{\gamma_{\sigma}}{2}\left( \sigma^\dagger\sigma\rho+\rho \sigma^\dagger\sigma -2 \sigma\rho \sigma^\dagger\right),\label{eq:elGAMMAintrinsic}
\end{align}
where $\gamma_{b}$ is the vibrational, $\gamma_{a}$ the plasmonic and $\gamma_{\sigma}$ the electronic decay rate, respectively, and $n^{\rm vib}_{T}$ is the equilibrium thermal population determined by the reservoir according to the Bose-Einstein distribution:
\begin{align}
n^{\rm vib}_{ T}=\frac{1}{\exp \left( \frac{\hbar\Omega}{k_{\rm B}T} \right)-1},
\end{align}
with the Boltzmann constant $k_{\rm B}$ and thermodynamic temperature $T$. Furthermore, for finite temperatures we consider the thermal pumping of the vibrations via:
\begin{align}
\mathcal{L}_{b^\dagger}[\rho]&=-n^{\rm vib}_{T}\frac{\gamma_{b}}{2}\left( b b^\dagger\rho+\rho b b^\dagger -2 b^\dagger\rho b\right).\label{eq:lindvibpump}
\end{align}
In most of the paper we consider the low-temperature limit ($T=0$\,K) where the thermal populations can be neglected. Finite temperatures are considered in Section\,\ref{sec:temp} where we study the effects of finite temperature on the steady-state vibrational populations. We further include into the model the pure dephasing of the molecular electronic excitations in the form of the Lindblad term \cite{Breuer2005, delValle09, esteban2014dephasing}.
\begin{align}
\mathcal{L}_{\sigma_z/2}(\rho)=-\frac{\gamma_{\phi}}{4}\left( \lbrace {\sigma_z}^\dagger {\sigma_z}, \rho\rbrace-2{\sigma_z}\rho {\sigma_z}^\dagger \right), 
\end{align}
with $\sigma_z=\sigma^\dagger\sigma-\sigma\sigma^\dagger$. \cmmt{This description of pure dephasing is valid in the plasmon-exciton weak-coupling regime considered below \cite{neuman2018origin}.}

The decay of the state-of-the-art plasmonic cavities $\gamma_a$ is ultimately limited by the material properties of the metal \cite{shen2006qualityf, Delga2014, kristensen2013modes, Sauvan2013, Koenderink10}, reaching quality factors $Q$ ($Q=\omega_{\rm pl}/\gamma_a$) of up to $Q\approx 50$ ($\hbar\gamma_a\approx 40$\,meV). However, the values of $Q$ commonly achieved in plasmonic systems are usually smaller ($Q\approx 1-20$). 
For example, the leaky gap mode formed between a tip of a scanning tunneling microscope and a metallic substrate often used for single-molecule spectroscopy \cite{zhang2013chemical, zhang2016visualizing, schull2017vibmapping, imada2017fano, zhang2017sub, lee2019visualizing} can be strongly damped, $\hbar\gamma_a\sim 10^2$\,meV, and can thus be regarded as a low-$Q$ plasmonic cavity. 

On the other hand, a typical decay rate of molecular excitations decoupled from the plasmonic cavity is much smaller than that of plasmons (as small as $\hbar\gamma_\sigma\sim 10^{-2}$\, meV $\ll \gamma_a$). The linewidth of the molecular resonance is thus mostly limited by the pure dephasing $\gamma_\phi$ which strongly scales with temperature and is highly dependent on the environment surrounding the molecule. It is thus possible to engineer conditions (low-temperature vacuum experiment) under which the pure dephasing becomes small and the linewidth of the molecular electronic excitation decreases below $< 10$\,meV and may even be limited only by the spontaneous decay. 


\subsection{Setting the SERRS regime}
Let us consider a value of $\gamma_{a}$ representing a leaky gap plasmonic mode formed between a tip of a scanning tunneling microscope and a metallic substrate as typically used for single-molecule spectroscopy \cite{schull2017vibmapping, imada2017fano, zhang2017sub, zhang2013chemical,zhang2016visualizing}, for which $\hbar\gamma_a\sim 10^2$\,meV. We consider the bad-cavity limit (weak-coupling regime) where the plasmon-exciton coupling $g$ is small compared to the plasmonic losses, but large with respect to the intrinsic decay $\gamma_{\sigma}$ ($\gamma_{\sigma}\ll g\ll \gamma_{a}$). In the bad-cavity limit, the molecular levels are only weakly perturbed by the presence of the plasmon that boosts the decay of the molecular electronic excitation via the Purcell effect, and focuses the incident light on the resonant molecule via the plasmonic near-field enhancement. The limit of the plasmon-exciton strong-coupling regime, where the coupling between the plasmon and the TLS dominates over the plasmonic losses, will be detailed elsewhere.

In the bad-cavity limit, the parameters determining the regime of the off-resonant optomechanical coupling, $d_{\rm om}$, as well as that defining the exciton-vibration coupling, $d$, in the resonant model, describe formally the same physical phenomena under weak-illumination conditions. This follows from the formal similarity between $H_{\rm bom}$ [Eq.\,\eqref{eq:pressureoptomech}] and $H_{\rm mol}$ [Eq.\,\eqref{eq:molham1}] established in Subsection\,\ref{subsec:SERRS}. In off-resonant SERS, the condition $|d_{\rm om}\Omega|>\gamma_{a}/2$ sets the so-called optomechanical strong coupling. In such situation the optomechanical non-linearity $-\Omega d_{\rm om}^2 a^\dagger a^\dagger a a$ becomes important and the system becomes interesting for quantum applications. 
It has been estimated that \cmmt{the optomechanical coupling can reach up to $d_{\rm om}\sim 10^{-1}$} for some molecular species \cite{roelli2015molecular}. 

On the other hand in SERRS, for relevant dye molecules with electronic excitations in the visible, $d$ ranges from $d\sim 0.1$ for rigid molecules (such as porphyrines \cite{huang2005porphyrins}) up to values of $d\sim 1$ for soft organic molecules \cite{herrera2017absem, SANCHEZCARRERA20101701}. SERRS might thus offer relatively high optomechanical coupling strengths even for a single organic molecule. 
Moreover, under the conditions of small molecular dephasing the width of the excitonic resonance becomes much smaller than that of the plasmon and SERRS may open the possibility to achieve large optomechanical coupling compared to the relevant linewidth \cmmt{$|d\Omega|>\Gamma_{\rm tot}/2+\gamma_\phi$ (strong optomechanical coupling), with $\Gamma_{\rm tot}=\gamma_\sigma+\Gamma_{\rm eff}$ and $\Gamma_{\rm eff}$ the width of the TLS due to the Purcell effect induced by the plasmons that we discuss later.}
In off-resonant SERS it is the cavity width $\gamma_a$ which determines the regime of optomechanical coupling and achieving the condition $|d_{\rm om}\Omega|>\gamma_a/2$ is more challenging.



In this work we consider a range of relatively large coupling strength between the plasmon and the molecular TLS ($\hbar g\approx$9 meV to $\hbar g=$50 meV), although still small enough to be in the bad-cavity limit \cite{chikkaraddy2016single, trugler2008strong}. These selected values allow us to explore different regimes of plasmon-assisted interaction between molecular excitons and vibrations, as detailed below. We also adopt $\hbar\gamma_{a}=500$ meV, $\hbar\gamma_{\sigma}=0.02$ meV, $\hbar\omega_{\rm pl}=2$ eV, \cmmt{and $\hbar\omega_{\rm eg}=2\,$eV,} as typical representative values of realistic molecules and plasmonic systems \cite{trugler2008strong, GarciaVidal2015}. We assume vibrational frequencies ranging between $\hbar\Omega=10$\,meV and $\hbar\Omega=50$\,meV, describing low-energy vibrations of typical organic molecules. In particular, when discussing phenomena emerging under \textit{weak} laser illumination (as discussed below) we adopt the value of $\hbar\Omega=50$\,meV and $\hbar\gamma_{b}=2$\,meV to describe the molecular vibration, whereas to address the regime of \textit{strong} laser illumination we choose $\hbar\Omega=10$\,meV and $\hbar\gamma_{b}=1$\,meV. This lower value of the vibrational frequency allows us to access the non-linear response of the system even for realistic laser intensities that comply with the rotating-wave approximation assumed in our model.   
With the exception of Section\,\ref{sec:temp}, throughout the manuscript we consider zero ambient temperature $T=0$\,K.

In the following we describe the inelastic emission spectra and the vibrational pumping in SERRS for (i) the linear response regime (relatively weak laser illumination in Secs.\,\ref{sec:peslr} and \ref{sec:vplr}) and (ii) strong laser illumination where the molecular levels are dressed by the intense laser field and form a qualitatively new set of light-matter states (Secs.\,\ref{sec:pessli} and \ref{sec:vpsli}).  
  
\section{Photon emission spectra in the linear regime}\label{sec:peslr}

We first discuss the spectral response and the physics of hybrid-optomechanical vibrational pumping in SERRS systems in the limit of weak incident laser intensities, for which the system can be treated within the linear-response theory (further denoted as \textit{weak illumination}). In this regime we limit the description to $T=0$ K as the thermal effects would mask the optomechanical pumping and damping processes that are weak for the low laser intensities discussed here. We relax this assumption later when we address the regime where the system is pumped by an intense laser.

For convenience, we define the detuning parameters $\Delta=\omega_{\rm pl}-\omega_{\rm L}$ and $\delta = \omega_{\rm eg}-\omega_{\rm L}$ (with the exciton frequency $\omega_{\rm eg}=E_{\rm e}(d)-E_{\rm g}(0)$) and define the coherent amplitude of the plasmon annihilation operator induced by the incident monochromatic illumination $\alpha_{\rm S}=\frac{-\mathcal{E}}{\Delta-{\rm i} \gamma_{a}/2}$.  
The solution of the dynamics of the hybrid optomechanical Hamiltonian and the respective Lindblad terms with the parameters described above, allow for calculating the steady-state emission spectrum $s_{\rm e}(\omega)$ from the plasmonic cavity using the quantum regression theorem (QRT):
\cmmt{
\begin{align}
\begin{split}
s_{\rm e}(\omega)&\propto 2\Re\int_0^\infty \langle\langle a{^\dagger} (0)a(\tau) \rangle\rangle\mathrm{d}\,\tau.
\end{split}\label{eq:espectrum}
\end{align}
In this spectrum we remove the elastic scattering contribution and use the notation $\langle\langle O_1 O_2 \rangle \rangle=\langle O_1O_2\rangle-\langle O_1\rangle\langle O_2\rangle$, where $O_1$ and $O_2$ are operators. More details about the specific implementation of this expression can be found in Appendix\,\ref{appendixA}.}

\begin{figure*}
\includegraphics[scale=0.56]{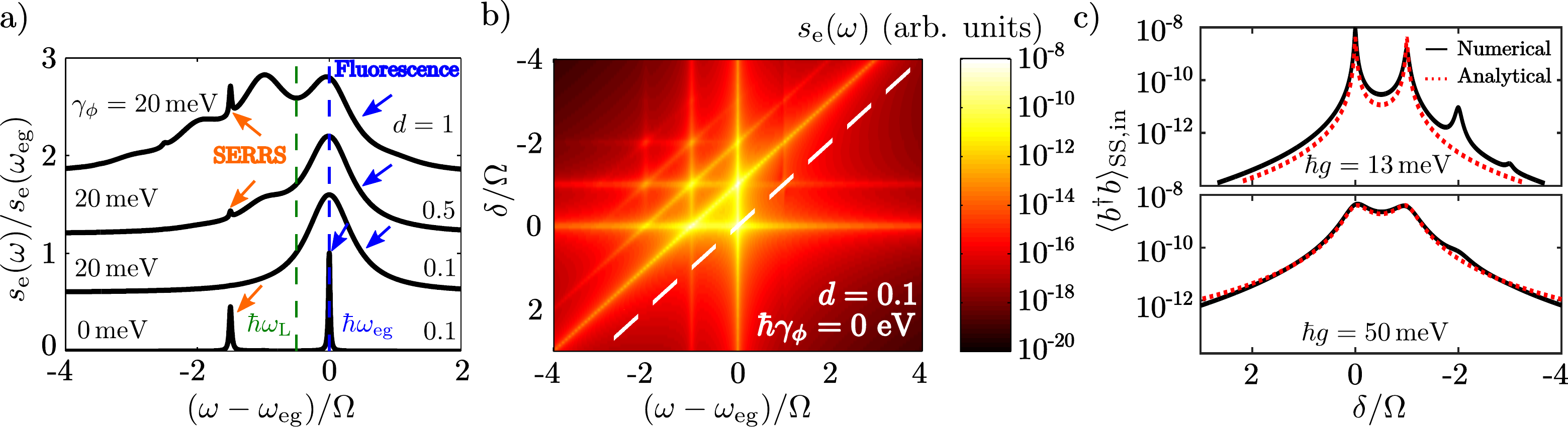}
\caption{Inelastic emission spectra and vibrational populations of a SERRS process where the molecular exciton is weakly coupled to the plasmon. (a) Normalized inelastic emission spectra $s_{\rm e}(\omega)/s_{\rm e}(\omega_{\rm eg})$ for different values of pure dephasing $\gamma_\phi$ (for the three top spectra $\hbar\gamma_\phi=20$\,meV, for the bottom spectrum $\hbar\gamma_\phi=0$\,meV) and dimensionless displacement $d$ (from the top: $d=1,\,0.5,\,0.1,\,0.1$). The blue-dashed line indicates the position of $\omega_{\rm eg}$ and the green dashed line marks the excitation frequency $\omega_{\rm L}$. The spectra are vertically shifted for clarity. The vibrational frequency is $\hbar\Omega=50$\,meV.
(b) Inelastic emission spectra as a function of detuning $\delta=\omega_{\rm eg}-\omega_{\rm L}$ of the incident laser frequency $\omega_{\rm L}$ from the exciton frequency $\omega_{\rm eg}$. The white dashed line marks the laser frequency in each emission spectrum. In (a,b) we set $\hbar g=13$\,meV. 
(c) Population of the vibrational mode as a function of laser detuning $\delta$ for illumination amplitude $\hbar\mathcal{E}=1\times10^{-2}$\,meV and different values of plasmon-exciton coupling $g$. In the upper panel $\hbar g=13$\,meV and the molecule effective broadening $\Gamma_{\rm eff}$ due to the Purcell effect is similar to the line-width of the vibrational Raman lines $\gamma_b$. In the lower panel $\hbar g=50$\,meV, a larger value of plasmon-exciton coupling that ensures $\Gamma_{\rm eff}>\gamma_b$. The red dashed line corresponds to the populations calculated analytically using Eq.\,\eqref{eq:vibpopapprox} and the black full line to the numerical results.}\label{fig:weakill}
\end{figure*}

To illustrate the emission properties of typical molecules we plot in Fig. \ref{fig:weakill} (a) the inelastic spectra in a SERRS system [normalized to $s_{\rm e}(\hbar\omega_{\rm eg})$ and vertically shifted], calculated for weak illumination, $\hbar\mathcal{E}=1\times10^{-2}$\,meV (roughly corresponding to a laser power density of $W\approx 1\times 10^{-4}$\,\si{\micro\watt/\micro\meter ^2} ), from a monochromatic laser of frequency $\hbar\omega_{\rm L}=1.975$\,eV (green dashed line) and for exciton-plasmon coupling $\hbar g=13$\,meV. The excitonic energy is $\hbar\omega_{\rm eg}=2$\,eV. We calculate the spectra for two large values of $d=1,\,0.5$ (top spectra) representing soft organic molecules and for a small value of $d=0.1$ of a rigid molecule (two bottom spectra). We choose $\hbar\gamma_\phi=20$\,meV for the three top spectra to demonstrate the effect of pure dephasing on the emission of molecules interacting with a decoherence-inducing environment. In the bottom spectrum no dephasing is considered, $\hbar\gamma_{\phi}=0$\,eV. 

The bottom spectrum in Fig. \ref{fig:weakill} (a), calculated for weak exciton-vibration coupling, $d=0.1$, and considering no pure dephasing ($\hbar\gamma_\phi=0$\,meV), features two sharp emission peaks. The fluorescence peak appears at frequency $\omega=\omega_{\rm eg}$ regardless of the incident laser frequency. The second peak, appearing at $\omega=\omega_{\rm L}-\Omega$, is the Raman-Stokes emission line. The anti-Stokes line is not visible because the vibrations are not thermally populated for $T=0$\,K. The Raman (SERRS) line always appears at a constant detuning from the laser frequency which facilitates its identification in the spectrum. When the pure dephasing is increased, the fluorescence line starts to broaden and also increases in absolute intensity (the latter is not manifested in Fig. \ref{fig:weakill} (a) due to normalization). The SERRS emission becomes hardly distinguishable on top of the strong fluorescence background for $d=0.1$. As $d$ increases, the fluorescence background becomes asymmetrical and broadens towards lower energies due to radiative transitions allowed by the simultaneous exchange of energy between electronic and vibrational states (\textit{hot luminescence}). This so-called  vibrational progression of the luminescence spectrum thus consists of a series of broad peaks, each peak positioned at frequency $\omega_{\rm eg}-n \Omega$ (with $n$ a positive integer), with its amplitude determined by the overlap of the vibrational wave functions in the electronic ground and excited states, respectively (Franck-Condon factors) \cite{Nunnenkamp2011spoptomech, Nunnenkamp2012sccooling, akram2010singlephoton, klessinger1995excited, may2008charge}. The Raman-Stokes lines appear on top of the fluorescence peaks at frequencies $\omega_{\rm L}-n\Omega$. The strength of the Raman lines is determined from a combination of the Franck-Condon overlaps, as in the case of the \textit{hot luminescence}, and from the further enhancement due to the proximity of the molecular electronic resonance that (i) enhances the interaction of the incident laser with the molecular transition and (ii) boosts the efficiency of the Raman-Stokes emission. The Raman peaks are also notably narrower than the fluorescence peaks when dephasing is large ($\hbar\gamma_\phi=20$ meV), which facilitates their identification on top of the broad and intense fluorescence background.

The difference between the physical origin of the SERRS lines and the fluorescence lines becomes clearer if we plot the emission spectra as a function of the laser detuning from the exciton frequency, $\delta=\omega_{\rm eg}-\omega_{\rm L}$. The emission spectra are shown in Fig. \ref{fig:weakill} (b) for a rigid molecule ($d=0.1$) and for no dephasing, similar to the bottom spectrum in Fig. \ref{fig:weakill} (a). The molecule is pumped by an incident laser of amplitude $\hbar \mathcal{E}=1\times 10^{-2}$\,meV and we consider an exciton-plasmon coupling, $\hbar g=13$\,meV. The incident laser frequency is marked in the spectra as a white dashed line diagonally crossing the color plot. The color plot with the spectra shows both the Raman-Stokes peaks that appear red-detuned from the incident-laser frequency $\omega_{\rm L}$ by $n\Omega$, and the fluorescence peaks emerging at the energy of the excitonic transition regardless of $\delta$. 

Furthermore, the emission shown in Fig. \ref{fig:weakill}(b) is enhanced when the exciton is resonantly pumped, $\hbar\delta=0$\,eV, or when when the frequency of the first-order Raman-Stokes line, $\omega=\omega_{\rm L}-\Omega$, corresponds to the bare excitonic resonance, $\hbar\delta=-\hbar\Omega=-50$\,meV.  
When the laser frequency is tuned to the molecular exciton ($\hbar\delta=0$\,eV), the incident laser coherently (coherent population $n_{\sigma}^{\rm coh}=|\langle\sigma\rangle|^2$) and incoherently (incoherent population $n_{\sigma}^{\rm incoh}=\langle\sigma^\dagger\sigma \rangle-|\langle\sigma\rangle|^2$) populates the electronic excited state. Thereafter, the molecule efficiently emits both the Raman-Stokes ($\propto n_{\sigma}^{\rm coh}$) and the hot luminescence ($\propto n_{\sigma}^{\rm incoh}$) photons at $\omega=\omega_{\rm eg}-\Omega$. On the other hand, when $\hbar\delta=-\hbar\Omega=-50$\,meV, the spectral position of the \cmmt{first-order} Raman-Stokes line coincides with the resonance frequency of the molecular exciton. \cmmt{In this case, the molecular fluorescence peaks, now appearing at the spectral positions of the Raman-Stokes lines, are suppressed since the off-resonance illumination does not efficiently populate the excited electronic state and the emission peaks appear mainly due to the SERRS mechanism. }
Both of these mechanisms of SERRS enhancement are closely related to the process of optomechanical vibrational pumping, described for the off-resonant case \cite{roelli2015molecular, schmidt2015qed, schmidt2017linking,kamandar2017quantum}.

Finally we remark that the spectral map in Fig. \ref{fig:weakill}(b) also features lines appearing due to higher-order Raman scattering and \textit{hot luminescence}. These lines show much lower intensity than the lines of lower orders, but they exhibit the same mechanism of emission enhancement, as is apparent from the spectral map.   




\section{Vibrational pumping in the linear regime}\label{sec:vplr}

Inasmuch as the emission of Raman-Stokes photons is accompanied by the creation of a vibrational quantum, the enhanced Raman-Stokes emission is reflected in the incoherent steady-state vibrational population $\langle b^\dagger b\rangle_{\rm SS, in}$. 
To elucidate the role of Raman-Stokes scattering in the process of vibrational pumping in SERRS, we plot in Fig. \ref{fig:weakill} (c) the vibrational population as a function of incident laser detuning for two values of plasmon-exciton coupling. The upper panel corresponds to $\hbar g=\hbar\sqrt{\gamma_{b}\gamma_{a} /6}\approx 13$\,meV, for which the broadening of the electronic resonance due to the plasmonic Purcell effect (see also Appendix\,\ref{Appendix Dressed}),
\begin{align}
\Gamma_{\rm eff}\approx\frac{g^2\gamma_{a}}{\left[\left(\frac{\gamma_{a}}{2}\right)^2+({\delta}+\Omega d^2-\Delta)^2\right]}\approx \frac{g^2\gamma_{a}}{\left[\left(\frac{\gamma_{a}}{2}\right)^2+({\delta}-\Delta)^2\right]},\label{eq:purcellgamma}
\end{align}
becomes comparable to the broadening of the vibrational line $\gamma_b$, and in the lower panel we use $\hbar g=50$\,meV, ensuring that $\Gamma_{\rm eff}>\gamma_b$ (with $\hbar\Gamma_{\rm eff}\approx 20$\,meV). In the calculations we set $\hbar\mathcal{E}=1\times 10^{-2}$\,meV to make sure that we stay in the linear regime. The numerically calculated values of $\langle b^\dagger b\rangle_{\rm SS, in}$ (black lines) are qualitatively similar in both cases. A set of peaks are clearly observed which correspond to the enhancement of the Raman-Stokes emission for detunings of $\hbar\delta=0$\,eV, $\hbar\delta=-\hbar\Omega=-50$\,meV, and higher orders ($\delta=-n\Omega$, $n>1$), respectively.
The effect of the larger plasmon-exciton coupling $g$ is to broaden the peaks and to smear off the population maxima associated with enhancement of the higher-order Raman-Stokes emission ($\delta=-n\Omega$, $n>1$). 

To shed light on the mechanism of vibrational pumping in SERRS, we derive the effective vibrational dynamics which results from the elimination of the plasmon and the TLS dynamics, following standard methods from the theory of open quantum systems \cite{Breuer2005}, in close analogy with the procedure developed in hybrid quantum optomechanics \cite{jaehne2008ground} (see description of the procedure in Appendix\,\ref{Appendix Dressed} and Appendix\,\ref{appendixC}). 
Upon elimination of the plasmon, the effective reduced TLS-vibrational Hamiltonian, $H_{\rm red}$, becomes
\begin{align}
H_{\rm red}=\hbar\delta\sigma{^\dagger} \sigma-\hbar\frac{1}{2}\mathcal{E}_{\rm pl} \sigma_x+\hbar\Omega (b{^\dagger} +\sigma_{\rm e}d)(b+\sigma_{\rm e}d),\label{eq:molhamel}
\end{align}
where $\mathcal{E}_{\rm pl}=-2g\alpha_{\rm S}$ is the coherent pumping of the molecule mediated by the plasmon, and $\sigma_x$ is the Pauli $x$ operator. Moreover, in the bad-cavity limit, the molecular excitonic TLS is effectively broadened due to the plasmon via the Purcell effect. The total TLS decay rate thus becomes $\gamma_\sigma\rightarrow\Gamma_{\rm tot}=\Gamma_{\rm eff}+\gamma_{\sigma}$. 

By further eliminating the TLS from the vibrational dynamics, assuming that the broadening of the electronic levels $\Gamma_{\rm tot}$ is larger than $|d\Omega|$ so that the Markovian approach applies, we obtain an effective vibrational Hamiltonian that includes the coherent pumping due to the TLS excited-state population
\begin{align}
H_{\rm vib}=\hbar\Omega b^\dagger b+\hbar d\Omega \langle\sigma_{\rm e}\rangle(b^\dagger+b),\label{eq:hvib}
\end{align}
which is accompanied by the effective incoherent damping $\Gamma_{-}$ and pumping  $\Gamma_{+}$ rates (see explicit expressions in Appendix\,\ref{appendixC}), which need to be added to the intrinsic vibrational dissipation rate [described by the original Lindblad term in Eq.\,\eqref{eq:lindvib}] via the following (new) Lindblad terms:
\begin{align}
\mathcal{L}_{\rm eff}^{-}[\rho]=-\frac{\Gamma_{-}}{2}\left(b{^\dagger} b\rho+\rho b{^\dagger} b-2b\rho b{^\dagger}  \right)\label{eq:effvibdec}
\end{align}
for the effective damping, and 
\begin{align}
\mathcal{L}_{\rm eff}^{+}[\rho]=-\frac{\Gamma_{+}}{2}\left(bb{^\dagger} \rho+\rho b b{^\dagger} -2b{^\dagger} \rho b \right)\label{eq:effvibpump}
\end{align}
for the effective pumping.
These rates are defined as $\Gamma_{-}=2\left(\Omega d\right)^2\Re\lbrace\tilde{S}(\Omega)\rbrace $ and $\Gamma_{+}=2\left(\Omega d\right)^2\Re\lbrace\tilde{S}(-\Omega)\rbrace$, respectively, where $\Re$ indicates the real part, and $\tilde{S}(s)=\int_0^\infty\langle\langle \sigma_{\rm e}(\tau)\sigma_{\rm e}(0) \rangle\rangle e^{{\rm i}s \tau}{\rm d}\tau$ is the spectral function corresponding to the one-sided Fourier transform of the correlation function of the TLS for a generic frequency $s$. The latter is calculated for the TLS decoupled from the vibrations\cmmt{, but coupled with the plasmon, which effectively broadens the TLS} (details about the analytical calculation of $\tilde{S}(s)$, and thus $\Gamma_{+}$ and $\Gamma_{-}$, are provided in Appendix\,\ref{appendixC}). Finally, the incoherent steady-state vibrational population induced by the effective pumping of the vibrations via the TLS in this approximation becomes 
\begin{align}
\langle b^\dagger b \rangle_{\rm SS,in}&=\frac{\Gamma_{+}}{\gamma_{b}+\Gamma_{-}-\Gamma_{+}}\nonumber\\
&\approx\frac{\Gamma_{+}}{\gamma_{b}}\propto \Re\{\tilde{S}(-\Omega)\},\label{eq:vibpopapprox}
\end{align}
where the last approximation originates from the fact that under weak pumping $\gamma_{b}\ll \Gamma_{-}-\Gamma_{+}$. We note that in the linear regime $\langle b^\dagger b \rangle\approx \langle b^\dagger b \rangle_{\rm SS,in}$ which allows for direct comparison of the numerically calculated and the analytically obtained populations. From Eq.\,\eqref{eq:vibpopapprox} it follows that the behavior of the spectral function $\tilde{S}(-\Omega;\delta)$ as a function of the incident laser frequency (i.e. $\delta$) determines the conditions for which the vibrational pumping occurs. In the linear regime we can simplify the expression for $\Re\{\tilde{S}(-\Omega)\}$, in analogy with the description of the off-resonant model \cite{roelli2015molecular, schmidt2015qed, schmidt2017linking, kamandar2017quantum} as: 
\begin{align}
\Re\{\tilde{S}(-\Omega)\}&=\Re\left\{\int_0^\infty\langle\langle \sigma_{\rm e}(\tau)\sigma_{\rm e}(0) \rangle\rangle e^{-{\rm i}\Omega \tau}{\rm d}\tau\right\}\nonumber\\
&\approx|\langle\sigma\rangle|^2\Re\left\{\int_0^\infty\langle\langle \sigma(\tau)\sigma^\dagger(0) \rangle\rangle e^{-{\rm i}(\Omega-\omega_{\rm L}) \tau}{\rm d}\tau\right\}\nonumber\\
&\approx \underbrace{\frac{|\mathcal{E}_{\rm pl}|^2}{4[\delta^2+(\Gamma_{\rm tot}/2)^2]}}_{\displaystyle \tilde S^{\rm R}_{\rm coh}}\underbrace{\frac{\Gamma_{\rm tot}/2}{(\delta+\Omega)^2+(\Gamma_{\rm tot}/2)^2}}_{\displaystyle \tilde{S}^{\rm R}_{\rm in}}.
\end{align}
The two terms, $\tilde{S}^{\rm R}_{\rm coh}\approx|\langle\sigma\rangle|^2$ and $\tilde{S}^{\rm R}_{\rm in}\approx\Re\{\int_0^\infty\langle\langle \sigma(\tau)\sigma^\dagger(0) \rangle\rangle e^{-{\rm i}(\Omega-\omega_{\rm L}) \tau}{\rm d}\tau\}$, can then be interpreted as the efficiency of the coherent driving ($\tilde{S}^{\rm R}_{\rm coh}$ resonant at $\delta=0$) and the efficiency of the spontaneous Stokes-Raman emission ($\tilde{S}^{\rm R}_{\rm in}$ resonant at $\delta=-\Omega$), respectively.

The effective vibrational dynamics and steady-state values derived in this section are an accurate approximation to the exact problem only if the decay rate,  $\Gamma_{\rm tot}$, of the dressed TLS is significantly larger than the intrinsic vibrational decay rate $\gamma_{b}$, and the exciton-vibration coupling is weak (moderate values of $d$). Realistic situations in molecular spectroscopy might not satisfy these conditions and in those situations it would be necessary to adopt a numerical treatment to obtain accurate results. Nevertheless, the properties of the TLS spectral function reveal the origin of the vibrational pumping even beyond the limits of validity of the analytical-model.

We plot the analytical result for the evolution of the vibrational populations as a function of detuning $\delta$ with a red dotted line in Fig. \ref{fig:weakill} (c). The analytical vibrational populations share with their numerically calculated counterparts (black lines) the same dominant peaks appearing for zero laser detuning from the exciton frequency, $\delta=0$, and for detuning $\delta=-\Omega$, when the frequency of the first-order Raman-Stokes line coincides with the excitonic frequency. These two values of the laser detuning also lead to an enhancement of the Raman-Stokes emission [Fig. \ref{fig:weakill} (b)], which is here the driving mechanism of the optomechanical vibrational pumping.

Although the analytical model nicely describes the main features of the fully numerically calculated vibrational populations, it cannot explain the presence of the weaker higher-order peaks.
This is due to the Markov approximation leading to Eq.\,\eqref{eq:hvib}, Eq.\,\eqref{eq:effvibdec} and Eq.\,\eqref{eq:effvibpump} which treats the exciton-vibration interaction perturbatively. In the full model, the vibrational pumping mechanism is also present for the vibrational transitions responsible for higher-order Raman scattering and hot luminescence. Moreover, in the case that $\hbar g=13$\,meV, the analytical model overestimates the vibrational populations induced by the optomechanical amplification for $\hbar \delta=-\hbar\Omega=-50$\,meV, since the effective broadening of the TLS, $\Gamma_{\rm tot}=\Gamma_{\rm eff}+\gamma_{\sigma}$, is similar to the vibrational broadening $\gamma_b$, and the Markov approximation becomes less accurate. For $\hbar g=50$\,meV the effective broadening $\Gamma_{\rm tot}>\gamma_b$ and the analytical model describes the low-order features of the vibrational populations accurately.

\section{Photon emission spectra for strong laser intensities}\label{sec:pessli}


\cmmt{Let us explore in the following the regime where the system is illuminated by a strong-power incident laser, which induces the non-linear response of the molecule, and thus requires a treatment beyond the standard optomechanical description applicable to off-resonant SERS. To observe such non-linear effects already for realistic values of intensity of the resonant laser illumination (and retaining the validity of the RWA) in this section we further consider values of vibrational frequencies $\hbar\Omega=10$\,meV at the lower end of a typical vibrational spectrum of an organic molecule.  }
\begin{figure}[t!]
	\includegraphics[scale=0.36]{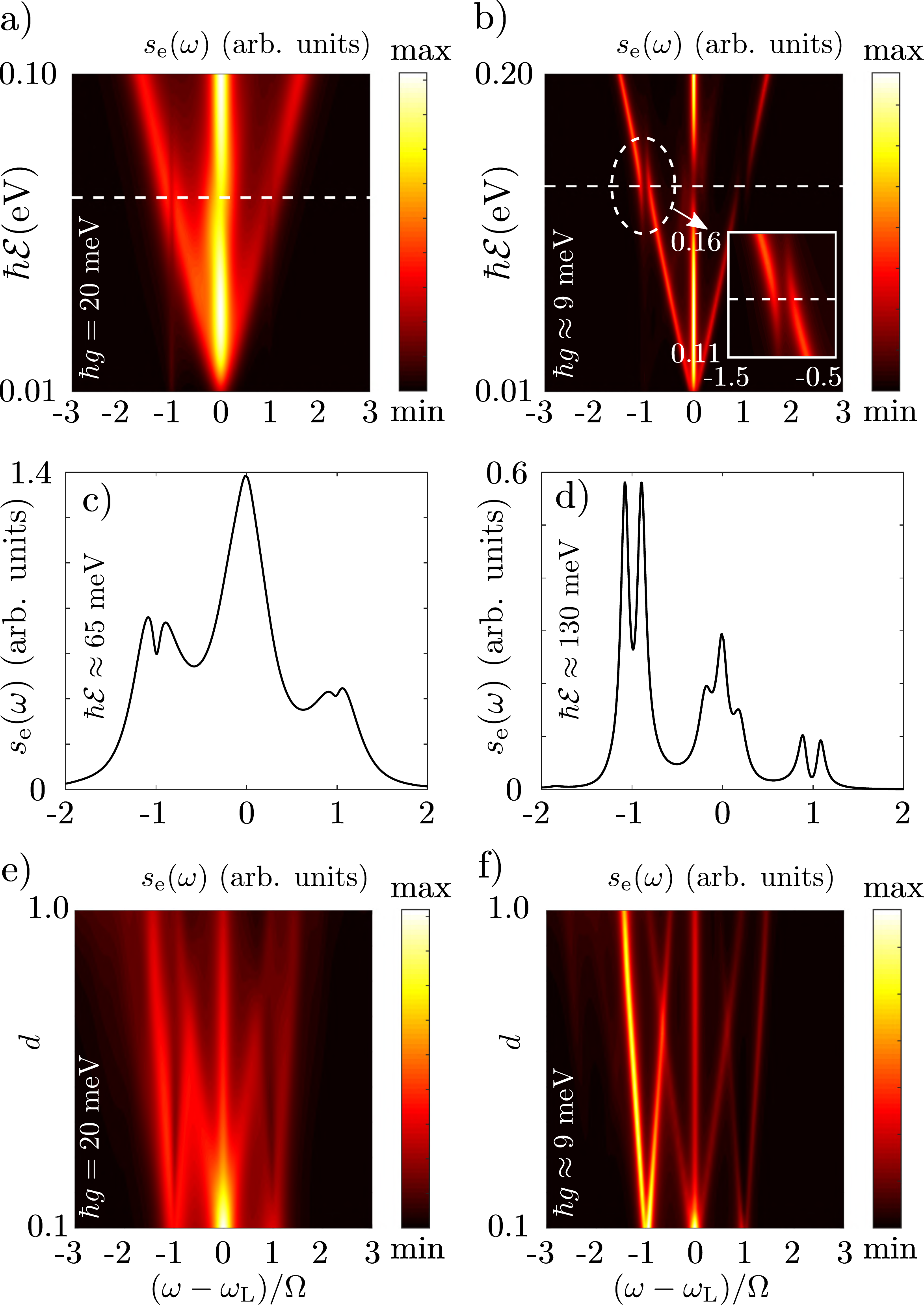}
	\caption{(a,b) Emission spectra of the coupled plasmon as a function of incident laser amplitude $\mathcal{E}$, for $d=0.2$. The molecule is coupled to the plasmonic cavity with (a) $\hbar g=20$\,meV and (b) $\hbar g=\hbar \sqrt{\gamma_{b}\gamma_{a} /6}\approx 9$\,meV. The inset in (b) shows a detail of the peak splitting due to the hybridization of the Mollow triplet side peak and the Raman line. (c,d) Cuts of the spectral map shown in (a,b) along the white dashed lines, corresponding to (c) $\hbar \mathcal{E}=65$\,meV and (d) $\hbar\mathcal{E}\approx 130$\,meV. (e,f) Emission spectra of the molecule for increasing value of coupling, $d$, setting \cmmt{$\hbar{\delta }=0 $ eV, and for (e) $\hbar\mathcal{E}\approx$65\,meV, $\hbar g=$20 meV, and (f) $\hbar\mathcal{E}\approx$130\,meV, $\hbar g\approx$9 meV.}}
	\label{fig:spectra_strong}
\end{figure}

\subsection{Influence of laser intensity}
The influence of the incident laser amplitude $\mathcal{E}$ is shown in Fig.~\ref{fig:spectra_strong}\,(a,b), where color maps of the emission spectra are displayed as a function of $\mathcal{E}$.
The incident laser frequency is tuned to the TLS electronic transition, and we consider the results for $d = 0.2$, which show the well-known Mollow triplet, a spectral structure resulting from the resonance fluorescence (RF) of the dressed TLS \cite{mollow1969power, gu2010resonance, ridolfo2010quantum, Ge2013mollow, wrigge2008efficient, SanchezMunoz2018}. The Mollow triplet consists of a strong emission line centered at the incident laser frequency and two side spectral peaks of similar spectral width that shift away as the laser intensity is increased [Figs.\,\ref{fig:spectra_strong}\,(a) and (b)]. At a specific pumping amplitude $\mathcal{E}$ [white dashed lines in Fig.\,\ref{fig:spectra_strong}\,(a) and (b)], the detuning of the Mollow triplet side peaks matches the vibrational frequency $\pm \Omega$ of the molecule and thus coherent effects emerge due to the interaction between the electronic RF and the vibrational Raman scattering. The visibility and nature of these effects depends on the width of the lines, which is dominated by the Purcell effect and hence on the coupling $g$. We thus consider again the two representative situations of interaction analysed in this work: (i) $\hbar g=20$ meV, where the electronic peak is spectrally broader than the vibrational line, as the Purcell effect strongly broadens the former [Figs.\,\ref{fig:spectra_strong}\,(a), (c)], and (ii) $\hbar g=9$ meV, a situation where the broadening of the electronic peaks is approximately equal to the vibrational broadening [Figs.~\ref{fig:spectra_strong}\,(b), (d)].

For clarity, the emission spectra for the selected values of  $\mathcal{E}$ that provide the matching ($\hbar\mathcal{E}\approx 65$ meV for $\hbar g=20$ meV, and $\hbar \mathcal{E}\approx 130$ meV for $\hbar g\approx 9$\,meV; roughly corresponding to a pumping power density of the order of $W\approx 1$\,\si{\milli\watt/\micro\meter^2} and 10 \si{\milli\watt/\micro\meter^2}, respectively) are shown in Fig.\,\ref{fig:spectra_strong}\,(c) and (d), respectively. When the RF peak is much broader than the width of the Raman line [Fig.\,\ref{fig:spectra_strong}\,(c)], the interference results in small but sharp features that might be detectable in experimental spectra and remind of Fano-resonances\cite{Miroshnichenko2010}. On the other hand, when the linewidth of the RF is similar to the linewidth of the Raman lines, the two spectral lines exhibit a clear anticrossing [inset in Fig.\,\ref{fig:spectra_strong}\,(b)] that results in a splitting of the spectral features of each branch of the Mollow triplet [Fig.\,\ref{fig:spectra_strong}\,(d)]. This splitting occurs as a result of the strong coupling between the molecule's electronic (TLS) and its vibrational degrees of freedom \cite{ramos2013nonlinear}, as predicted in the context of light emission from semiconductor quantum dots \cite{Kabuss2011}. 

The onset of strong coupling between the electronic and vibrational degrees of freedom, and thus the clear line splitting, can be understood with the help of a simplified Hamiltonian of the system (see Appendix\,\ref{Appendix Dressed}). This Hamiltonian is a result of an elimination of the plasmon cavity from the original Hamiltonian. Disregarding for the moment the vibrational part in Eq.\,\eqref{eq:molhamel}, the simplified Hamiltonian ${H}_{\rm TLS}$ becomes
\begin{align}
{H}_{\rm TLS}=\hbar\frac{1}{2}{\delta}{\sigma}_z-\hbar\frac{1}{2}\mathcal{E}_{\rm pl} {\sigma}_x+\hbar\frac{1}{2}{\delta}.\label{eq:tlsnondiag}
\end{align}
Here $\mathcal{E}_{\rm pl}=-2g\alpha_{\rm S}\propto \mathcal{E}$ again corresponds to the amplitude of the plasmon-enhanced electric field. The Hamiltonian in Eq.\,\eqref{eq:tlsnondiag} can be diagonalized by a unitary transformation that rotates the TLS Pauli matrices in the $x-z$ plane, (${\sigma}_x,\,{\sigma}_z\rightarrow{\sigma}'_x,\,{\sigma}'_z$):
\begin{align*}
\begin{split}
{\sigma}_z=\frac{{\delta}}{\lambda_{\rm TLS}}{\sigma}'_z+\frac{\mathcal E_{\rm pl}}{\lambda_{\rm TLS}}{\sigma}'_x,\\
{\sigma}_x=-\frac{\mathcal E_{\rm pl}}{\lambda_{\rm TLS}}{\sigma}'_z+\frac{{\delta}}{\lambda_{\rm TLS}}{\sigma}'_x.
\end{split}
\end{align*} 
Under those operations, the simplified Hamiltonian, $H_{\rm TLS}=\hbar\lambda_{\rm TLS}\sigma'_z+\hbar\frac{1}{2}{\delta}$, describes the dynamics of an effective electronic TLS dressed by the incident coherent illumination with an effective frequency $\lambda_{\rm TLS}=(\mathcal{E}_{\rm pl}^2+{\delta}^2)^{1/2}$.
According to this simplified Hamiltonian, the effective electronic frequency $\lambda_{\rm TLS}$ can be tuned by either changing the intensity of the incident laser (i.e. $\mathcal{E}_{\rm pl}\propto \mathcal{E}$) or by detuning the incident laser frequency ${\delta}$. This dressed TLS interacts with the molecular vibrations via the resonant Rabi interaction term \cite{ramos2013nonlinear}
\begin{align}
&\hbar\frac{1}{2}\Omega d{\sigma}_z ({b}^\dagger+{b}) \to \underbrace{\hbar\frac{1}{2}\Omega d  \frac{\mathcal{E}_{\rm pl}}{\lambda_{\rm TLS}}{\sigma}'_x({b}{^\dagger} +{b})}_{\text{Rabi term}}\nonumber\\
&+\text{residual polaronic coupling},
\end{align}
which becomes resonant if $\lambda_{\rm TLS}\approx \Omega$, the condition for the Mollow side peaks to coincide with the spectral position of the SERRS lines. 

On top of the effect of dressing the molecular levels, the plasmonic cavity increases the effective damping rate $\Gamma_{\rm eff}$ of the TLS by means of the Purcell effect (causing the broad peaks of the Mollow triplet). The condition to reach strong coupling in this situation can be derived by relating the decay rate of the molecular vibration and that of the dressed electronic transition with the exciton-vibration coupling strength:
\begin{align}
\Omega d  \frac{\mathcal{E}_{\rm pl}}{\lambda_{\rm TLS}}\gtrapprox|3\Gamma_{\rm tot}/4+\gamma_{b}/2|.
\end{align}
This condition is at the origin of the strong coupling observed in the peaks of 
Fig.\,\ref{fig:spectra_strong}\,(d) ($\hbar g =9$\,meV), but it is not reached in the case presented in Fig.\,\ref{fig:spectra_strong}\,(c) ($\hbar g =20$\,meV) where Fano-type features appear as a sign of weak coupling. When the strong coupling between the vibrational Raman scattering and the RF pathways is reached, the peak-splitting in the emission spectra in Fig.\,\ref{fig:spectra_strong}\,(d) can be also interpreted using the dressed-atom picture originally introduced by Cohen-Tannoudji \cite{ cohen1977modification, agarwal1979theory}. The dressed-atom picture allows for interpreting the splitting of the Raman and resonance-fluorescence peaks in terms of the coherent interaction among molecular vibronic states, induced by the incident coherent laser illumination. This approach shows that the final emission peaks emerge from a coherent combination of both the RF-type transitions and the Raman-type transitions, making the two mechanisms inseparably connected. We elaborate on the dressed-molecule picture assuming small $d$ in Appendix\,\ref{app:svd_dmp}, and after that provide a more general result allowing large $d$ in the next section.

\subsection{Influence of large vibrational displacement $d$}

We have so far used a moderate value of the displacement, $d=0.1$, however, in realistic molecules significant exciton-vibration coupling can lead to larger values of $d$. In Fig.~\ref{fig:spectra_strong} (e,f) we show the evolution of the spectrum with $d$ for the same two values of the plasmon-TLS coupling, i.e. $\hbar g=20$ meV [Fig.~\ref{fig:spectra_strong}(e)] and $\hbar g\approx 9$ meV [Fig.~\ref{fig:spectra_strong}(f)]. For all the values of $d$ considered, the laser intensity is chosen such that the resonance-fluorescence (RF) lines match the position of the Raman lines. For small values of $d\approx 0.1$, the RF profile follows the behaviour described in Figs.~\ref{fig:spectra_strong}(c,d). As $d$ gradually increases, the spectra start to exhibit additional features due to the increasing importance of higher-order vibronic transitions. 
When the RF line is significantly broader than the Raman lines [Fig.~\ref{fig:spectra_strong}(e)], an increase of the coupling $d$ gradually increases the spectral dip located at the frequency of the Stokes and anti-Stokes emission. Additional weak spectral features appear as $d$ is increased at larger laser detuning. When the linewidth of the Mollow side peaks is similar to the width of the Raman lines [Fig.~\ref{fig:spectra_strong}(f)], the splitting of the strongly coupled hybrid lines becomes larger as $d$ increases. 
For large values of $d$, all of the spectra in Figs.~\ref{fig:spectra_strong}(e,f) acquire a more complex structure due to the generally complicated coherent interaction between the molecular vibrational and the electronic degrees of freedom, with the emergence of the additional (weak) peaks originating from higher-order Raman and resonance-fluorescence transitions.

\section{Vibrational pumping for strong laser intensities}\label{sec:vpsli}

\begin{figure*}[ht!]
	\begin{center}
		\includegraphics[scale=0.9]{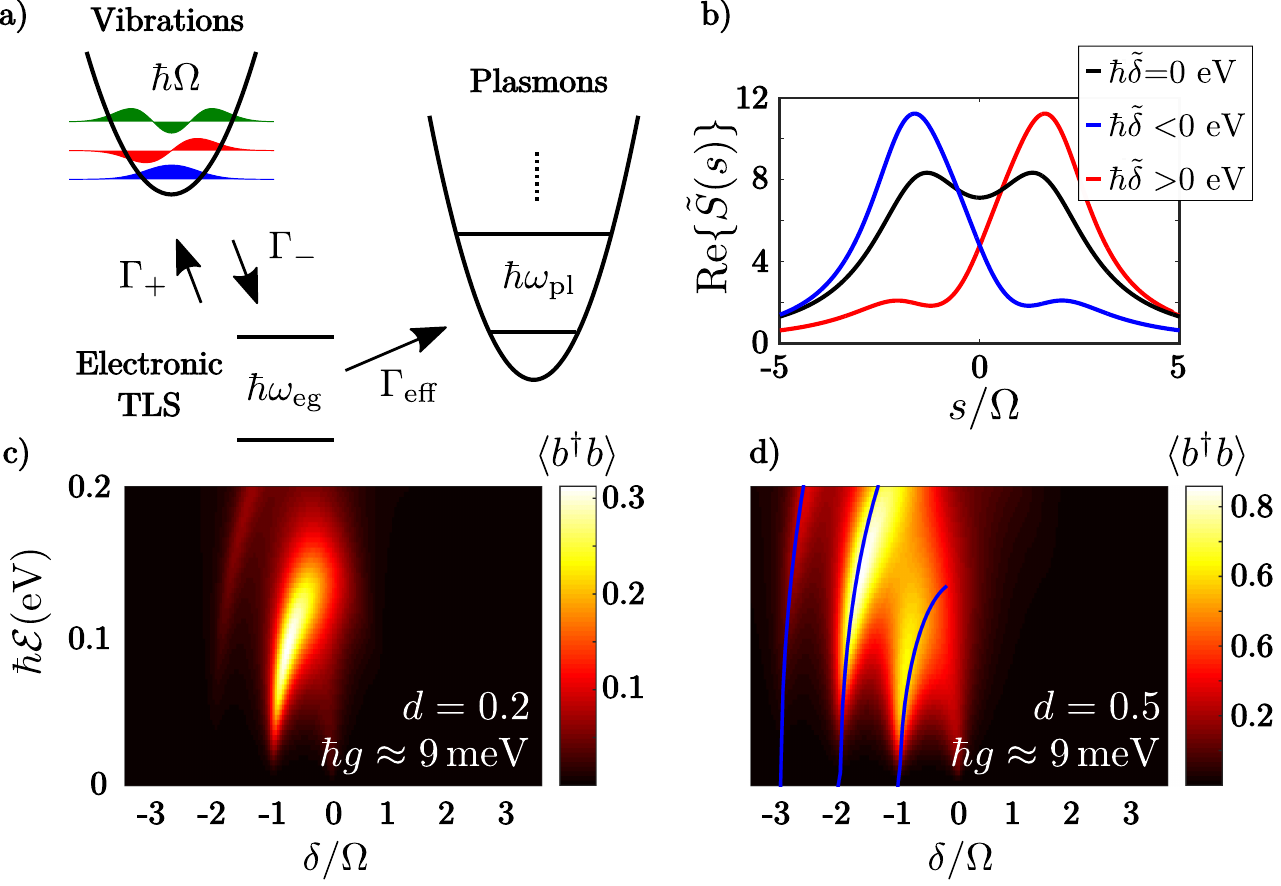}
	\end{center}
	\caption{(a) \cmmt{Schematic depiction of the hierarchy considered in the theoretical model. The plasmons serve as an effective reservoir and broaden the TLS via the Purcell effect ($\Gamma_{\rm eff}$ - effective decay of the TLS into the plasmonic reservoir, as marked by the arrow). The broadened TLS then effectively influences the incoherent dynamics of the vibrations via the effective vibrational pumping and damping ($\Gamma_{+}$ and $\Gamma_{-}$, respectively, as indicated by the arrows). (b) Real part of the spectral function (calculated from the reduced Hamiltonian where the plasmonic cavity is eliminated), $\Re\lbrace \tilde{S}(s) \rbrace$, of the operator ${\sigma}_{\rm e}$, for three different values of detuning ($\tilde{\delta}=\delta+d^2\Omega$) $\hbar \tilde{\delta}=0$ eV (black line), 10 meV (red line) and -10 meV (blue line), $\hbar \mathcal E =100$ meV, and $\hbar g=20$ meV. (c,d) Maps of vibrational population of a molecular vibration ($\hbar\Omega=10$ meV) as a function of detuning from the effective TLS energy, ${\delta}$, and of the incident laser amplitude, $\mathcal{E}$, for $\hbar g\approx 9$ meV, with  $d=0.2$ (c), $d=0.5$ (d). The blue lines in (d) indicate the condition $\lambda_{\rm TLS}=n\Omega$ with $n$ integer (only for ${\delta}<0$).}}
	\label{fig:populations1}
\end{figure*}
In this section, we extend the treatment of linear-response SERRS introduced previously towards the case of strong incident illumination, where non-linear effects become important. 

To that end we invoke the effective vibrational Hamiltonian introduced in Eq.\,\eqref{eq:hvib} together with the incoherent damping $\Gamma_{-}=2\left(\Omega d\right)^2\Re\lbrace\tilde{S}(\Omega)\rbrace $ and pumping $\Gamma_{+}=2\left(\Omega d\right)^2\Re\lbrace\tilde{S}(-\Omega)\rbrace$ rates in Eq.\,\eqref{eq:effvibdec} and Eq.\,\eqref{eq:effvibpump}, respectively. As above, the spectral function $\tilde{S}(s)$ is obtained from the effective dynamics of the TLS, which is effectively broadened by the plasmon via the Purcell effect. The hierarchy of approximations considered in this section is schematically depicted in Fig.\,\ref{fig:populations1} (a).
These effective rates are dependent on the spectral function $\tilde{S}(s)$ of the reservoir evaluated at frequencies $\Omega$ and $-\Omega$, respectively. Note that the analytical model is limited to cases where the electron-vibration coupling $\Omega d$ is smaller than the effective broadening $\Gamma_{\rm eff}$ of the electronic resonance. We thus perform full numerical calculations to obtain the results (i.e., populations) spanning the full range of model parameters, however we use the analytical model for qualitative discussion.

The value of the spectral function $\tilde{S}(s)$ at frequencies $\pm \Omega$ determines the strength of the effective vibrational pumping ($\Gamma_{+}$) or damping ($\Gamma_{-}$), which modify the vibrational populations as can be seen from the first expression in Eq.\,\eqref{eq:vibpopapprox}. It is therefore possible to achieve different regimes of interaction with the vibrations which range from pumping to damping by simply modifying the illumination conditions (laser intensity and frequency detuning) that provoke a variation of the shape of the spectral function.  When the laser intensity is large, the reservoir function $\tilde S(s)$ reflects the structure of the TLS dressed by the incident laser, and therefore it becomes qualitatively different from the weak-illumination case.

In Fig.~\ref{fig:populations1}(b) the spectral function, $\Re\lbrace\tilde{S}(s)\rbrace$, for $d=0.2$, $\hbar \mathcal E =100$ meV, and $\hbar g=20$ meV, is shown to peak around the effective frequencies of the dressed TLS ($s =\pm\lambda_{\rm TLS}$).
When the incident laser is detuned from the TLS transition ($\tilde{\delta}=\delta+d^2\Omega\neq 0$) the spectral function changes symmetry. 
For $\tilde{\delta}>0$ (red detuning marked with a red line) a regime of vibrational damping can be reached ($\Re\lbrace \tilde{S}(\Omega) \rbrace > \Re\lbrace \tilde{S}(-\Omega) \rbrace$), whereas for $\tilde{\delta}<0$ (blue detuning marked with a blue line) a regime of vibrational pumping ($\Re\lbrace \tilde{S}(\Omega) \rbrace < \Re\lbrace \tilde{S}(-\Omega) \rbrace$) is achieved. This effect is more pronounced for a situation where $\tilde{S}(\pm\Omega)$ corresponds to the maxima of $\tilde{S}(s)$.

To illustrate the possibility to achieve a controlled excitation of molecular vibrations on demand, we numerically solve the full Hamiltonian of the system [Eq.\,\eqref{eq:cohelmoldr}], and show in Figs.~\ref{fig:populations1}(c,d) the steady-state vibrational population $\langle b^\dagger b\rangle$, for an electron-plasmon coupling of $\hbar g\approx 9$ meV and two different values of the dimensionless displacement ($d=0.2$ and $d=0.5$).
The vibrational pumping is non-trivially influenced by both the detuning ${\delta}$ of the incident laser frequency from the TLS transition frequency and by the incident laser intensity ($\propto\mathcal E^2$), so that the optimal laser intensity depends on the laser detuning ${\delta}$. When the electron-vibration coupling is large ($d=0.5$), the population reaches multiple intense local maxima. In this case, by adequately tuning the laser frequency one can efficiently excite Franck-Condon transitions involving a change of more than one vibrational transition (higher-order processes). As expected, the population maxima are found when the spectral position of the side peaks of the electronic spectral function matches the frequency of the higher-order vibrational transitions ($\lambda_{\rm TLS}\approx n\Omega$, with $n$ an integer), a condition traced by the blue lines in Fig. \ref{fig:populations1} (d), and displayed only for negative detuning ${\delta}$.  

\section{Selective vibrational pumping}

\begin{figure*}[t!]
	\begin{center}
		\includegraphics[scale=0.9]{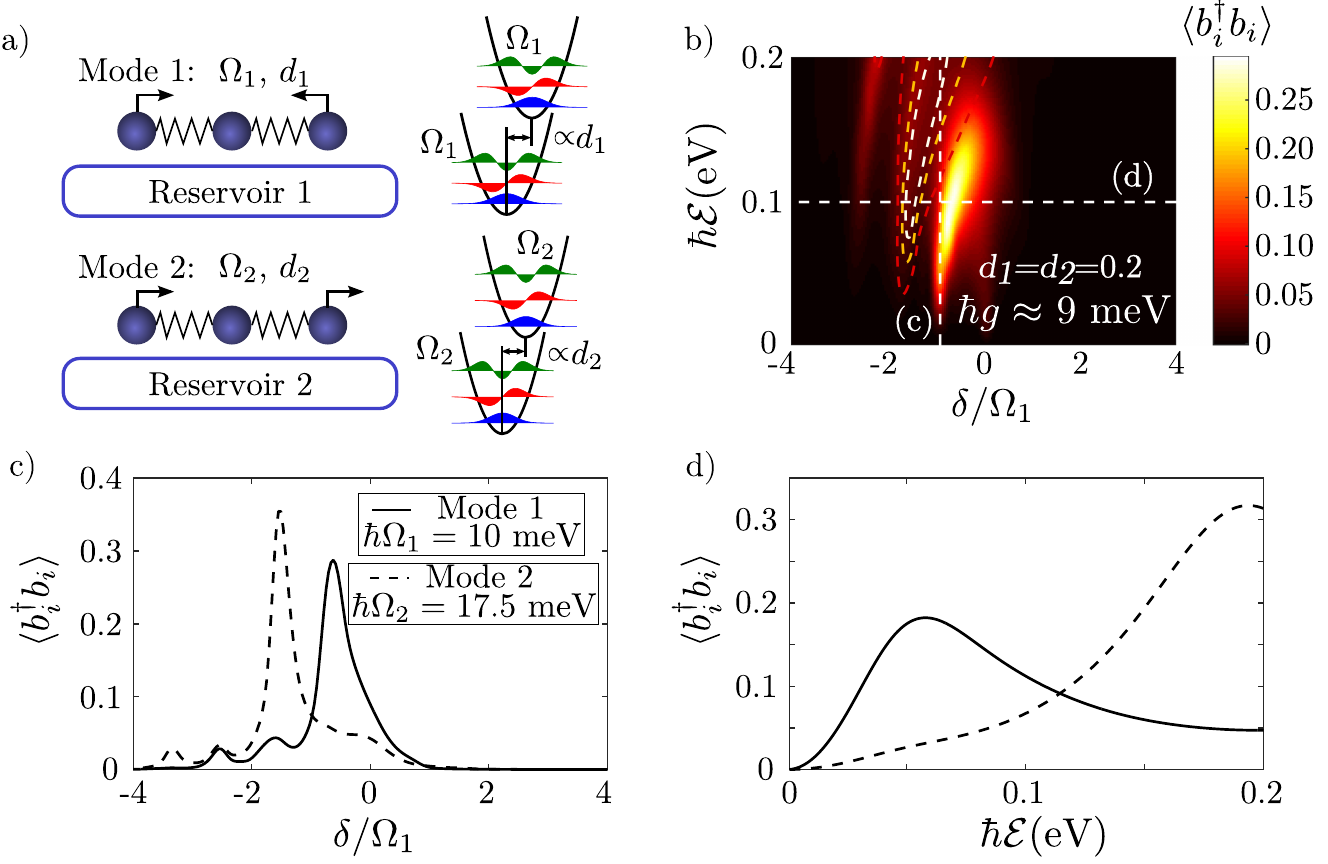}
	\end{center}
	\caption{\cmmt{Selective vibrational pumping. (a) Schematic representation of an example of two independent vibrational modes of frequencies $\Omega_1$ and $\Omega_2$, respectively, coupled with the electronic degrees of freedom via the displacement $d_1$ and $d_2$ of their respective PESs. The vibrational modes are assumed to interact independently with their corresponding reservoirs, 1 and 2. (b) Color map of the vibrational populations of two different vibrational modes present in the same molecule, with frequency $\hbar\Omega _1=10$ meV (solid colors) and $\hbar\Omega _2=17.5$ meV (values expressed by dashed contour lines).  (c,d) Populations of the modes $\Omega _1$ (solid line) and  $\Omega _2$ (dashed line) extracted along the white dashed lines in (b). In (c) $\hbar\mathcal{E}=100$ meV and $\delta$ is varied, whereas in (d) $\hbar{\delta}=-9$ meV and $\mathcal{E}$ is varied.}}
	\label{fig:populations2}
\end{figure*}

The potential to control the activation of molecular vibrations can be exploited in the selective excitation of different vibrational modes. Let us consider the coupling of a  plasmonic system with a molecule supporting two vibrations at frequencies $\hbar\Omega_1=10$\,meV  and $\hbar\Omega_2= 17.5$\,meV, both coupled to independent reservoir modes (baths) with $\hbar\gamma_{\rm vib,\, 1}=\hbar\gamma_{\rm vib,\, 2}=1$ meV, \cmmt{as schematically depicted in Fig.\,\ref{fig:populations2} (a).} We simplify the description of the system and use the effective Hamiltonian where the plasmonic degrees of freedom are eliminated (see Appendix\,\ref{app:dstvm}). We assume that the vibrational modes are coupled to the TLS via a polaronic coupling term ($d_1=d_2=0.2$), and do not consider the direct coupling between the two vibrational modes. However, this model Hamiltonian naturally couples the two vibrational modes indirectly via the electronic TLS of the molecule. Our model thus partially accounts for thermalization effects, without considering the effect of the surrounding environment that may further incoherently couple the vibrational modes.

The resulting  vibrational populations, $\langle b_i^\dagger b_i\rangle$, are shown in Fig.~\ref{fig:populations2} (b) as a function of the intensity and detuning of the incoming laser. The color map depicting the population of the vibrational mode at frequency $\Omega_1$ is displayed together with a dashed contour plot that shows the corresponding results for the mode at $\Omega_2$. Each mode presents a clear maximum for suitable illumination conditions. Noticeably, the maxima are shifted with respect to each other both in frequency and amplitude, so that changing the illumination conditions serves to pump more efficiently one mode or another. To highlight the selectivity of the vibrational pumping mechanism, we extract line cuts of Fig.~\ref{fig:populations2} (b) for constant laser pumping, $\hbar\mathcal E=100$ meV, [Fig.~\ref{fig:populations2} (c)], and for constant laser detuning $\hbar\delta=-9$ meV [Fig.~\ref{fig:populations2} (d)]. As observed in Fig.\,\ref{fig:populations2} (c,d) the conditions of intensity and detuning for maximum population of one mode give a much weaker population of the other mode (solid versus dashed lines). This scheme of interactions makes it possible to achieve selective vibrational pumping by either tuning the laser frequency for a given illumination intensity or modifying the laser intensity for a fixed illumination frequency. 

\section{Effects of temperature on vibrational pumping}\label{sec:temp}

\begin{figure*}
\includegraphics[scale=1]{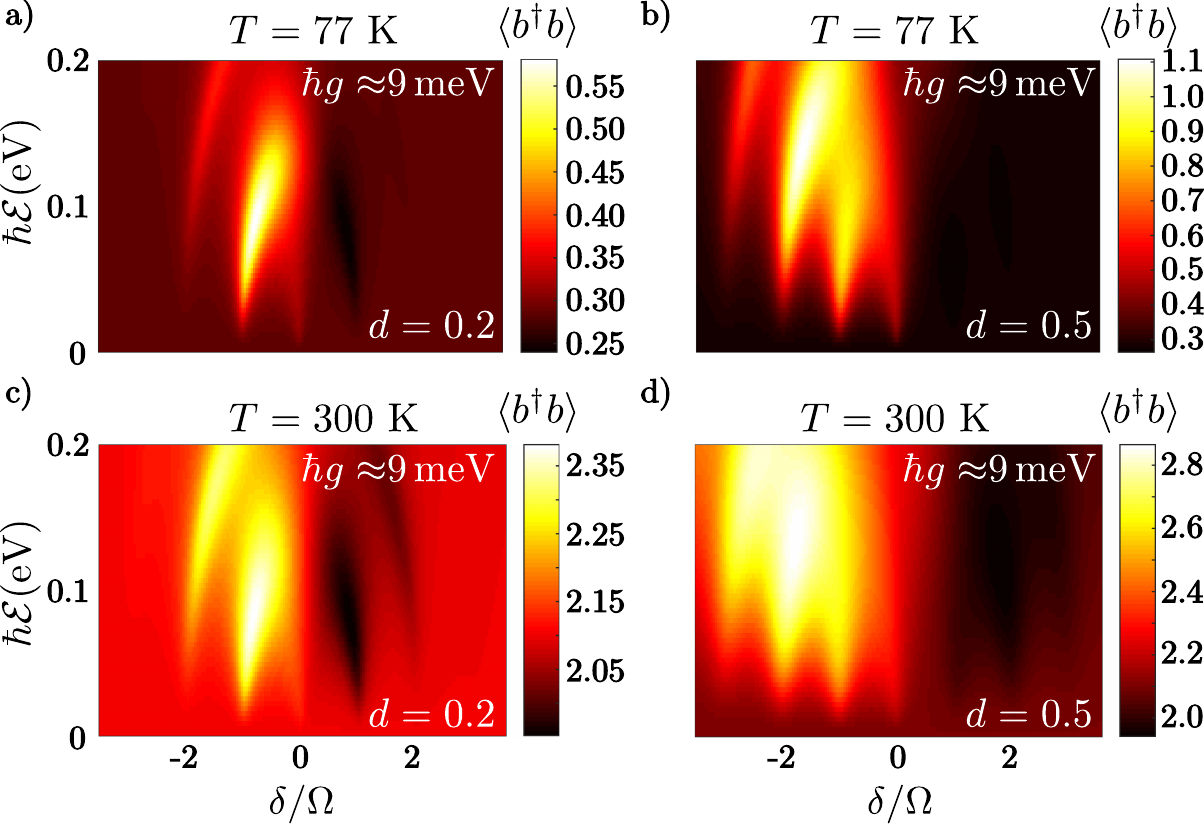}
\caption{\cmmt{Colormaps of steady-state vibrational population for finite temperature: (a,b) $T=77$\,K, (c,d) $T=300$\,K. The maps are calculated assuming weak electron-vibration coupling $d=0.2$ in (a,c) and assuming medium coupling strength of $d=0.5$ in (b,d). Other parameters are identical to the ones used in Fig.\,\ref{fig:populations1} (c,d).}}\label{fig:pumpingtemperature}
\end{figure*}

Practical SE(R)RS experiments are usually performed at a finite temperature, a situation where the thermal populations of the molecular vibrations can become considerable. We thus study this effect in this section. We calculate the steady-state vibrational populations for a temperature of $T=77$ K (liquid nitrogen temperature) and of $T=300$ K (room temperature) for different detuning and intensity of the incident laser for the same range of parameters as in Fig.\,\ref{fig:populations1} (c,d). We assume that the electronic states of the molecule interact with a single vibrational mode. The results of the vibrational population as a function of laser detuning and intensity are shown in Fig.\,\ref{fig:pumpingtemperature} as two-dimensional color maps.  

For the liquid-nitrogen temperature ($T=77$ K) the maps of the vibrational population shown in Fig.\,\ref{fig:pumpingtemperature} (a,b) for $d=0.2$ and $d=0.5$, respectively, are qualitatively similar to the results obtained when considering $T=0$\,K. However, we observe that the total vibrational populations reach larger maximal values due to the thermal population, which for a vibrational excitation of energy $10$ meV reaches at $T=77$\,K a value of $n_{77 {\rm\, K}}\approx 0.3$. Interestingly, for positive values of detuning $\delta$ we observe vibrational populations smaller than $n_{77 {\rm\, K}}$ which is a result of the optomechanical cooling process that is not observable for $T=0$\,K. 
For room temperature ($T=300$\,K) the thermal population of the vibrational mode of energy $10$\,meV reaches $n_{300 {\rm \,K}}\approx 2$, which is larger than the optomechanically induced steady-state population reached when thermal effects are disregarded. The map of vibrational populations shown in Fig.\,\ref{fig:pumpingtemperature} (c) for small $d=0.2$ and $T=300$\,K shows, as in the case of $T=77$\,K, that the optomechanical process can lead either to an increase or decrease of the steady-state vibrational population with respect to the thermal equilibrium state. 
When the detuning fulfils the condition of optomechanical vibrational pumping (i.e. $\delta\approx-\lambda_{\rm TLS}$) the population can reach values larger than the thermal population, as in this case the optomechanical pumping process enhances the effect of thermal population pumping. Conversely, when the optomechanical damping process is enhanced ($\delta\approx\lambda_{\rm TLS}$) the steady-state vibrational population features cooling. 
These effects are more pronounced in Fig.\,\ref{fig:pumpingtemperature} (d) where we consider $T=300$\,K and a stronger electron-vibration coupling of $d=0.5$ [notice the different scale of the color map in Figs.\,\ref{fig:pumpingtemperature} (a-d)]. As pointed out, the steady-state vibrational population is pumped above the value determined by the thermal equilibrium for $\delta<0$. This effect is resonantly enhanced whenever $\delta\approx -n\lambda_{\rm TLS}$ ($n$ integer). On the other hand, the system is cooled for the opposite detuning ($\delta\approx n\lambda_{\rm TLS}$). 

We thus conclude that thermal effects become important if the thermal vibrational population becomes comparable to the steady-state population induced by the optomechanical pumping or damping processes. Furthermore, when a finite temperature is considered, it is possible to observe optomechanical vibrational cooling.

\section{Conclusion}
In conclusion, we have used the formalism of quantum cavity electrodynamics to discuss SERRS as a quantum optomechanical process. First, we have shown the similarities and fundamental differences between the resonant and off-resonant SERS processes. In the linear response regime, the electronic transition appearing in the SERRS plays a role analogous to that of the plasmonic cavity in the off-resonant SERS, however, the former offers generally larger values of optomechanical coupling and smaller values of the molecular-exciton damping $\gamma_\sigma+\Gamma_{\rm eff}$, as compared to the plasmon broadening $\gamma_a$. 
We have further described vibrational pumping in SERRS and identified its two principal mechanisms: the efficient pumping of the molecule by laser resonant with the molecular exciton (optimized when $\hbar\delta\approx 0$\,eV) and the resonant enhancement of the Raman-Stokes emission ($\delta\approx-\Omega$).  

For strong laser intensities, the non-linear character of the molecular electronic TLS introduces non-trivial effects associated with the dressing of the molecular levels by the intense laser.  
We have shown that a strong exciton-vibration interaction appears when a side peak of the RF (Mollow triplet) of a single molecule \cite{wrigge2008efficient} is tuned to match the frequency of vibrational Raman lines. The fingerprint of this interaction is accessible in optical emission spectra through the presence of interference features or peak splitting. 
Finally, the regime of vibrational pumping achieved in SERRS can exploit the spectral features of the molecular electronic resonance, which are often narrower than the plasmonic resonances used in off-resonant SERS. This opens the possibility for selective pumping of different vibrational modes of the molecule.
Selective vibrational pumping can offer new ways to control chemical reactivity of molecules \cite{crim1996bond} or to engineer vibrational quantum states of experimental interest.

\section*{Acknowledgements}


The authors acknowledge projects FIS2016-80174-P from Spanish MINECO, ELKARTEK KK-2018/00001, H2020-FETOPEN project "THOR" Nr. 829067 of the European Commission, PI-2017-30 and PI-2016-41 of the Departamento de Educaci\'{o}n, Pol\'{i}tica Ling\"{u}\'{i}stica y Cultura of the Basque government.


\appendix

\section{System Hamiltonian}\label{app:sysham}

In this appendix we describe the Hamiltonian used in the main text to describe the resonant surface-enhanced Raman scattering (SERS) process [Eq.\,\eqref{eq:molham1}], where we consider the coupling between a plasmonic mode and the electronic and vibrational levels of a molecule. When a monochromatic laser excites the plasmonic mode coupled to the molecule, the system Hamiltonian can be written in the form:
\begin{align}
H_{\rm om}^{\rm res}&=H_{\rm pl}+H_{\rm mol}+H_{\rm pump}+H_{\rm pl-e},\label{eq:genHam}
\end{align}
with
\begin{align}
H_{\rm pl}&=\hbar\omega_{\rm pl} a{^\dagger} a\nonumber\\
H_{\rm mol}&=\hbar [E_{\rm e}-E_{\rm g}]\sigma{^\dagger} \sigma+\hbar\Omega (b{^\dagger} +\sigma_{\rm e}d)(b+\sigma_{\rm e}d)\nonumber\\
H_{\rm pump}&=\hbar\mathcal{E} \left[a \exp (\mathrm{i}\omega_{\rm L}t)+a^\dagger \exp (-\mathrm{i}\omega_{\rm L}t)\right]\nonumber \\  
H_{\rm pl-e}&=\hbar g a\sigma{^\dagger} +\hbar g^\ast a{^\dagger} \sigma.\nonumber
\end{align}
Here $\sigma$ ($\sigma^{\dagger}$) are the lowering (raising) operators of the two level system (TLS) representing the electronic structure of the molecule (with $\sigma^\dagger \sigma=\sigma_{\rm e}$), $b$ ($b^\dagger$) are the annihilation (creation) operators of the vibrational mode and $a$ ($a^\dagger$) are the annihilation (creation) operators of the single bosonic mode representing the plasmonic cavity. The constants appearing in the Hamiltonian have the following meaning: $g$ represents the coupling between the plasmon and the electronic TLS, $\omega_{\rm pl}$ is the plasmon frequency,  $E_{\rm g}$ and $E_{\rm e}$ are the energies of the ground and excited electronic state of the molecule, respectively, $\Omega$ is the vibrational frequency and $d$ introduces the electron-phonon coupling. The monochromatic coherent laser illumination of frequency $\omega_{\rm L}$ is coupled to the plasmonic mode via the constant $\mathcal{E}$ (that we take as real), which is proportional to the electric field amplitude of the incident light. 

The system Hamiltonian is formally split into the Hamiltonian describing the plasmonic mode as a bosonic oscillator $H_{\rm pl}$, the term describing the level structure of the bare molecule $H_{\rm mol}$, the coupling between the plasmon and the molecule $H_{\rm pl-e}$ and the pumping of the plasmon by the classical laser field $H_{\rm pump}$ in the rotating wave approximation (RWA), \cmmt{assuming that the pumping amplitude $\mathcal{E}$ is sufficiently small compared to the plasmon frequency ($\mathcal{E}\lesssim 0.1\omega_{\rm pl}$). }

The molecular Hamiltonian $H_{\rm mol}$ contains the energy splitting of the molecule's electronic levels, $\hbar [E_{\rm e}-E_{\rm g}]\sigma{^\dagger} \sigma$,  and the vibrational \cmmt{term which depends on the electronic state}, $\hbar\Omega (b+\sigma_{\rm e}d)(b^\dagger+\sigma_{\rm e}d)$. This coupling is obtained from the Born-Oppenheimer approximation, that takes into account that the vibrational states of the ground states have different equilibrium position than the vibrations defined on the excited state potential energy surface. Due to this displacement, $d$, the vibrational eigenstates in the ground state are not orthogonal to the ones in the excited state. Therefore, when the molecule is excited from the ground electronic state to the excited electronic state, it also simultaneously changes the vibrational state (according to the Franck-Condon principle).

The coupling between the molecule and the plasmon is expressed in the RWA as $H_{\rm pl-e}=\hbar g a\sigma{^\dagger} +\hbar g^\ast a{^\dagger} \sigma $. The RWA is justified in situations where $g<<\omega_{\rm pl}\approx E_{\rm e}-E_{\rm g}$. Importantly, the RWA allows for further simplifying transformations of the Hamiltonian.

To introduce incoherent effects we employ the approach based on the solution of the quantum master equation for the density matrix $\rho$: 
\begin{align}
\frac{\partial \rho}{\partial t}=\frac{1}{\mathrm{i}\hbar}\left[ H_{\rm om}^{\rm res}, \rho\right]+\mathcal{L}_{a}[\rho]+\mathcal{L}_{\sigma}[\rho]+\mathcal{L}_{b}[\rho],\label{eq:Liouvil}
\end{align}
where the bracket symbolizes the commutator and the Lindblad terms $\mathcal{L}_{c}[\rho]$ introduce the incoherent damping. In particular we use the following Lindblad superoperators:
\begin{align}
\mathcal{L}_{\sigma}[\rho]&=- \frac{\gamma_{\sigma}}{2}\left( \sigma^\dagger\sigma\rho+\rho \sigma^\dagger\sigma -2 \sigma\rho \sigma^\dagger\right),\label{eq:elGAMMAintrinsicApp}\\
\mathcal{L}_{a}[\rho]&=-\frac{\gamma_{a}}{2}  \left( a^\dagger a\rho+\rho a^\dagger a -2 a\rho a^\dagger\right),\\
\mathcal{L}_{b}[\rho]&=-(n^{\rm vib}_{T}+1)\frac{\gamma_{b}}{2}\left( b^\dagger b\rho+\rho b^\dagger b -2 b\rho b^\dagger\right),\\
\mathcal{L}_{b^\dagger}[\rho]&=-n^{\rm vib}_{T}\frac{\gamma_{b}}{2}\left( b b^\dagger\rho+\rho b b^\dagger -2 b^\dagger\rho b \right),
\end{align}
where $\gamma_{\sigma}$ is the electronic decay rate, $\gamma_{b}$ the vibrational decay rate, and $\gamma_{a}$ the plasmonic decay rate.

To facilitate the numerical calculation, we apply a unitary transformation $\tilde{H}_{\rm om}^{\rm res}=U_{\omega_{\rm L}}H_{\rm om}^{\rm res}U_{\omega_{\rm L}}^\dagger-{\rm i}\hbar U_{\omega_{\rm L}}\dot{U}_{\omega_{\rm L}}^\dagger$ with $U_{\omega_{\rm L}}=\exp ({\rm i}\sigma_{\rm e}\omega_{\rm L}t+{\rm i}a^\dagger a\omega_{\rm L}t)$ that transforms the Hamiltonian in Eq.\,\eqref{eq:genHam} into the rotating frame (interaction picture). As a consequence, the Lindblad terms remain unchanged, but the Hamiltonian in Eq.\,\eqref{eq:genHam} is modified by simply replacing (for brevity we keep the same notation for the transformed operators as for the original operators throughout the text):
\begin{align}
\omega_{\rm pl}\rightarrow \Delta=&\omega_{\rm pl}-\omega_{\rm L},\\
E_{\rm e}-E_{\rm g} \rightarrow \delta=&E_{\rm e}-E_{\rm g}-\omega_{\rm L},\\
\mathcal{E} \left[a \exp (\mathrm{i}\omega_{\rm L}t)+a^\dagger \exp (-\mathrm{i}\omega_{\rm L}t)\right]  &\rightarrow  \mathcal{E}\left[a +a^\dagger \right].
\end{align}

The resulting Hamiltonian is time independent which facilitates the numerical solution. However, it includes the direct pumping of the plasmon mode. In the numerical implementation, where we represent the plasmonic states by the number states of the plasmon Hamiltonian $H_{\rm pl}$, we would need a large number of plasmon states to correctly describe the excitation by a strong incident laser. We avoid this problem by redefining the plasmon creation and annihilation operators:
\begin{align*}
a &\rightarrow a+\alpha_{S},\\
a^\dagger &\rightarrow a^\dagger+\alpha_{S}^\ast,
\end{align*}
where $\alpha_{\rm S}=\frac{-\mathcal{E}}{\Delta-{\rm i}\frac{\gamma_{a}}{2}}$. This particular choice of $\alpha_{\rm S}$ allows us to define a new Hamiltonian $H_{\rm om}^{{\rm res},{\alpha}}$ [see also Eq. (1) of the main text]:
\begin{align}
H_{\rm om}^{{\rm res},{\alpha}}&=H_{\rm pl}^{\alpha}+H_{\rm mol}^{\alpha}+H_{\rm pump}^{\alpha}+H_{\rm pl-e}^{\alpha} \label{eq:hamsupp}
\end{align}
with
\begin{align}
H_{\rm pl}^{\alpha}&=\hbar\Delta a{^\dagger} a\nonumber\\
H_{\rm mol}^{\alpha}&=\hbar\delta\sigma{^\dagger} \sigma+\hbar\Omega (b{^\dagger} +\sigma_{\rm e}d)(b+\sigma_{\rm e}d)\nonumber\\
H_{\rm pump}^{\alpha}&=\hbar g \alpha_{\rm S}\sigma{^\dagger} +\hbar g^\ast\alpha_{\rm S}^\ast\sigma\nonumber \\  
H_{\rm pl-e}^{\alpha}&=\hbar g a\sigma{^\dagger} +\hbar g^\ast a{^\dagger} \sigma.\nonumber
\end{align}
The Lindblad terms appear unchanged provided that the plasmon operators are expressed in the shifted basis. The final form of the Master equation is thus:
\begin{align}
\frac{\partial \rho}{\partial t}=\frac{1}{\mathrm{i}\hbar}\left[ H_{\rm om}^{{\rm res},{\alpha}}, \rho\right]+\mathcal{L}_{a}[\rho]+\mathcal{L}_{\sigma}[\rho]+\mathcal{L}_{b}[\rho],\label{eq:LiouvilPOL}
\end{align}
where we have to keep in mind that we are working in the interaction picture and in the displaced basis when we evaluate the correlation functions and the operator mean values.

\section{Calculation of emission spectra}\label{appendixA}

We calculate the emission spectra numerically from the quantum regression theorem (QRT) \cite{Breuer2005} using the expression
\begin{align}
s_{\rm e}(\omega)&=2\Re\int_0^\infty \langle a{^\dagger} (0)a(\tau) \rangle\,e^{\mathrm{i}\omega \tau}\mathrm{d}\,\tau.\label{eq:spectrumgeneral}
\end{align}

In order to solve the dynamics of the damped system we need to solve the quantum master equation [Eq.\,\eqref{eq:LiouvilPOL}] for the density matrix  $\rho$ given by the Hamiltonian and by the Lindblad terms. To that end, we rewrite the quantum master equation in a form where the density matrix $\rho$ appears as a column vector (see e.g. Ref. \cite{am2015three}):
\begin{align}
\rho=\left[\begin{matrix}
\rho_{11}&\rho_{12}&\cdots \\
\rho_{21}&\rho_{22}&\cdots \\
\vdots&\vdots&\ddots 
\end{matrix}\right]\rightarrow\vec{\rho}=
\left[\begin{matrix}
\rho_{11} \\
\rho_{21} \\
\vdots \\
\rho_{12} \\
\rho_{22}\\
\vdots
\end{matrix}\right].\label{eq:origrep}
\end{align}
{
In the equations, there often appear expressions where the operators act on the density matrix from the right or from the left (e.g. the term $a\rho a^\dagger$). In the technical implementation, the expressions are transformed as:
\begin{align}
O_1\rho O_2 \rightarrow \left(O_2^{\rm T}\otimes O_1\right) \vec{\rho}.\label{eq:rightlefttransform}
\end{align}
Here $\otimes$ represents the Kronecker product, $^{\rm T}$ denotes transposition and $O_1$, $O_2$ are generic operators. In practise, if the dimension of the truncated Hilbert space is set to $N=N_{\rm vib}\times N_{\rm pl}\times 2$ (with $N_{\rm vib}$ [$N_{\rm pl}$] the maximal number of vibrational [plasmon] number states considered in the calculation), the vectorised density matrix will have a length of $N^2$ and the matrix $\left(O_2^{\rm T}\otimes O_1\right)$ will be of dimension $N^2\times N^2$.}

Equation (\ref{eq:LiouvilPOL}) becomes in this representation:
\begin{align}
\dot{\vec{\rho}}=\mathscr{L}(t)\vec{\rho},\label{eq:superop1}
\end{align}
where the $N^2\times N^2$ square matrix $\mathscr{L}(t)$ represents the {Liouville superoperator.} In general $\mathscr{L}(t)$ can depend on time, but in our model it is time independent. 
The steady-state density matrix $\rho_{\rm ss}$ is then obtained from the quantum master equation [Eq. \ref{eq:superop1}] as the eigenvector belonging to the zero eigenvalue of the matrix $\mathscr{L}$. 

\begin{widetext}
The two time correlation function [Eq.\,\eqref{eq:spectrumgeneral}] that defines the spectrum is calculated using the quantum regression theorem. Utilizing Laplace transform techniques and with the substitution of the Laplace parameter $s\rightarrow -{\rm i}\omega$:
\begin{align}
s_{\rm e}(\omega)&=2\Re\int_0^\infty \langle a{^\dagger} (0)a(\tau) \rangle\,e^{\mathrm{i}\omega \tau}\mathrm{d}\,\tau
=2\Re \left[ {\rm Tr} \left\lbrace (I \otimes a)\frac{1}{-\mathrm{i}(\omega-\omega_{\rm L})-\mathscr{L}} \left[\lbrace(a{^\dagger})^{\rm T}\otimes I\rbrace\vec\rho_{\rm ss} \right]\right\rbrace \right].\label{eq:spectrumpractical}
\end{align}

The direct implementation of Eq.\,\eqref{eq:spectrumpractical} requires to invert the Liouvillian matrix $\mathscr{L}$ [time independent in the frame rotating with $\exp(-{\rm i}\omega_{\rm L}t)$] for each frequency of interest. However, this procedure becomes inefficient for calculations of emission spectra. In such a case, and assuming that the Liouvillian is represented by a diagonalizable matrix (which we verify numerically), it is often more convenient to expand the time-dependent solution into an exponential series with the exponents being the eigenvalues of the Liouvillian superoperator \cite{tan1999quantum}. 

The time-dependent solution of the density matrix [or of the vector $\lbrace(a{^\dagger})^{\rm T}\otimes I\rbrace\vec\rho$ since according to the QRT they obey the same differential equation, Eq.\,\eqref{eq:LiouvilPOL}] is formally given by the exponential of the Liouvillian as:
\begin{subequations}
\begin{align}
\vec\rho(t)=\exp(\mathscr{L}t)\vec\rho(0).
\end{align}
\begin{align}
[\lbrace(a{^\dagger})^{\rm T}\otimes I\rbrace\vec\rho](t)=\exp(\mathscr{L}t)[\lbrace(a{^\dagger})^{\rm T}\otimes I\rbrace\vec\rho](0).
\end{align}
\end{subequations}
Equivalently, the exponential can be expressed using the eigenvalue decomposition of the Liouvillian $\mathscr{L}={S}{D}{S}^{-1}$ as follows:
\begin{align}
\exp(\mathscr{L}t)={S}\exp({D}t){S}^{-1},
\end{align}
where the operator $D$ is represented by a diagonal matrix so that its exponentiation simply indicates exponentiation of the matrix diagonal elements one by one. We can then write the correlation function $\langle a{^\dagger} (0)a(\tau) \rangle$ as
\begin{align}
g(\tau)&=\langle a{^\dagger} (0)a(\tau) \rangle\nonumber \\
&={\rm Tr} \left\lbrace (I \otimes a) \exp(\mathscr{L}\tau)\left[\lbrace(a{^\dagger})^{\rm T}\otimes I\rbrace\vec\rho \right] \right\rbrace\, e^{-{\rm i}\omega_{\rm L}\tau} ={\rm Tr}& \left\lbrace \underbrace{\left( [I \otimes a] {S}\right)}_{A}\underbrace{\exp({D}\tau)}_{B}\underbrace{\left({S}^{-1}\left[\lbrace(a{^\dagger})^{\rm T}\otimes I\rbrace\vec\rho \right]\right)}_{v} \right\rbrace\, e^{-{\rm i}\omega_{\rm L}\tau},\label{eq:corfuntrace}
\end{align} 
\end{widetext}
{where, for convenience, we have defined a full matrix $A$, a diagonal matrix $B$ containing the exponentiated eigenvalues, and a vector $v$. Note that the explicit dependence on the laser frequency, $\omega_{\rm L}$, appears because we express the operators in the interaction picture. The trace operator is defined in the original representation where the density operator has the form of a square matrix (the left hand side in Eq.\,\eqref{eq:origrep}). 
The matrix $B$ can be seen element-wise as $B_{ii}=\delta_{ij}\exp(d_it)$, where $d_i$ are the diagonal elements of the matrix $D$. The vector $c=ABv$ is therefore equal to the following exponential series:}
\begin{align}
c_l=\sum_{m=1}^{N^2} A_{lm}\exp(d_mt)v_m. 
\end{align}  
The trace can be represented as a scalar product of the vectorised operator with
vector ${\nu}_{tr}$ where $(\nu_{tr})_j = 1$ for $j=1,\, n+2,\, 2n+3,\,...,\, n^2$ ,and 0 otherwise.
The trace in Eq.\,\eqref{eq:corfuntrace} above becomes:
\begin{align}
{\rm Tr}\left\lbrace c \right \rbrace &=\sum_{l=1}^{N}c_{(l-1)N+l}\nonumber\\
&=\sum_{m=1}^{N^2}\sum_{l=1}^{N}A_{(l-1)N+l,m}v_m\exp(d_mt)\nonumber\\
&=\sum_{m=1}^{N^2} l_m\exp(d_mt),
\end{align}
with
\begin{align}
&\;l_m =\sum_{l=1}^{N}A_{(l-1)N+l,m}v_m.
\end{align}
Last, inserting this expression into Eq.\,\eqref{eq:spectrumgeneral} we get:
\begin{align}
s_{\rm e}(\omega )&=2\,\Re\int_0^\infty g(t)\exp({\rm i}\omega t)\, {\rm d}t\nonumber\\
&=2\Re\sum_{m=1}^{N^2} \frac{-l_m}{(d_m-{\rm i}\omega_{\rm L})+{\rm i}\omega}.
\end{align}

\section{Effective two-level-system Hamiltonian under coherent laser illumination}\label{Appendix Dressed}
We discuss in the following how to eliminate the plasmon cavity to obtain a new effective Hamiltonian of the molecule. We assume that the plasmonic cavity (after transforming out the coherent displacement $\alpha_{\rm S}$) acts as a fluctuating reservoir that effectively damps the molecule via the Purcell effect. To describe the effects of the reservoir, we use the standard methods of the quantum noise approach \cite{Breuer2005} and eliminate the plasmon under the assumption that the broadening of the plasmon excitation is considerably larger than the coupling strength between the plasmon and the electronic TLS. 
We obtain the electronic decay of the molecule into the plasmonic mode (the Purcell effect), assuming that the plasmon cavity is unpopulated (after removing the coherent contributions as described in Appendix\,\ref{app:sysham}), expressed by the following Lindblad term:
\begin{align*}
\mathcal{L}_{\rm eff}[\rho]=-\frac{\Gamma_{\rm eff}}{2}\left( \sigma^\dagger\sigma \rho+\rho \sigma^\dagger\sigma -2 \sigma \rho \sigma^\dagger\right),
\end{align*}
with
\begin{align*}
\Gamma_{\rm eff}= 2 g^2\Re\lbrace {S}_{a}(\delta)\rbrace,
\end{align*}
and where
\begin{align*}
{S}_{a}(s)=\int_0^\infty\langle a(\tau) a^\dagger (0) \rangle e^{{\rm i}s \tau}{\rm d}\tau,
\end{align*}
which, assuming that the plasmon obeys a dynamics unperturbed by the presence of the molecule, yields 
\begin{align}
\Gamma_{\rm eff}=\frac{g^2\gamma_{a}}{\left(\frac{\gamma_{a}}{2}\right)^2+({\delta}-\Delta)^2}.\label{eq:effectiveGamma}
\end{align}

We neglect the slight frequency shift of the TLS due to the action of the cavity which is formally given by the imaginary part of $S_{a}(s)$. The Hamiltonian of the reduced system thus has the form [Eq.\,\eqref{eq:molhamel}]:
\begin{align}
H_{\rm red}=\hbar\delta\sigma{^\dagger} \sigma-\hbar\frac{1}{2}\mathcal{E}_{\rm pl} \sigma_x+\hbar\Omega (b^\dagger+d\sigma_{\rm e}) (b+d\sigma_{\rm e}),\label{eq:molham}
\end{align}
where we defined $\mathcal{E}_{\rm pl}=-2g\alpha_{\rm S}$ and $\sigma_x$ is the Pauli $x$ operator.

The Hamiltonian in Eq.\,\eqref{eq:molham} [Eq.\,\eqref{eq:molhamel}], from which the plasmon has been eliminated, can be recast into a form where an effective electronic TLS dressed by the incident coherent illumination interacts with the molecular vibrations via the Rabi interaction term \cite{ramos2013nonlinear}. We first start by grouping the terms that correspond to the TLS Hamiltonian under strong laser illumination. In particular, we consider as the TLS Hamiltonian the terms that are free of the vibrational operators (do not contain the electron-vibration coupling, i.e. $d=0$):
\begin{align}
H_{\rm TLS}=\hbar\frac{1}{2}{\delta}\sigma_z-\hbar\frac{1}{2}\mathcal{E}_{\rm pl} \sigma_x+\hbar\frac{1}{2}{\delta},\label{eq:tlsnondiagApp}
\end{align}
where we used the fact that the operator $\sigma_{\rm e}=\sigma^\dagger\sigma$ can be rewritten with the help of the standard Pauli $z$ operator as $\sigma_{\rm e}=\frac{1}{2}(\sigma_z+I)$.

{We first apply the following rotation in the space of the standard Pauli operators $\sigma_x$, $\sigma_y$ and $\sigma_z$ (with the primed operators being the new Pauli operators, and $\lambda_{\rm TLS}= \sqrt{{\delta}^2+{\cal E}_{\rm pl}^2}$) that diagonalizes Eq.\,\eqref{eq:tlsnondiagApp}:
\begin{align}
\begin{split}
\sigma_z=\frac{{\delta}}{\lambda_{\rm TLS}}\sigma'_z+\frac{\mathcal E_{\rm pl}}{\lambda_{\rm TLS}}\sigma'_x,\\
\sigma_x=-\frac{\mathcal E_{\rm pl}}{\lambda_{\rm TLS}}\sigma'_z+\frac{{\delta}}{\lambda_{\rm TLS}}\sigma'_x,\end{split}\label{eq:rotpauli}
\end{align}
The resulting Hamiltonian for the molecule after this transformation is:
\begin{align}
\begin{split}
H_{\rm red}&=\hbar\frac{1}{2}\lambda_{\rm TLS}\sigma'_z+\hbar\Omega b{^\dagger} b\nonumber\\
& +\hbar\frac{1}{2}\Omega d \left( \frac{{\delta}}{\lambda_{\rm TLS}}\sigma'_z+\frac{\mathcal{E}_{\rm pl}}{\lambda_{\rm TLS}}\sigma'_x \right)(b{^\dagger} +b)\nonumber\\ &+\hbar\frac{1}{2}\Omega d(b{^\dagger} +b),
\end{split}\label{eq:effectiveTLSvibAppendix}
\end{align} 
\label{eq:hameffweak}
where the TLS Hamiltonian from Eq.\,\eqref{eq:tlsnondiagApp} corresponds to the diagonal term $\hbar\frac{1}{2}\lambda_{\rm TLS}\sigma'_z$.}

For completeness we note that $\sigma_y$ has not been changed by the transformation and from Eq.\,\eqref{eq:rotpauli} it follows that the original {lowering operator [$\sigma=\frac{1}{2}(\sigma_x-{\rm i}\sigma_y)$] transforms into the following form:}
\begin{align}
\sigma=\frac{1}{2}\frac{{\delta}}{\lambda_{\rm TLS}}\sigma'_x-\frac{1}{2}\frac{\mathcal{E}_{\rm pl}}{\lambda_{\rm TLS}}\sigma'_z - {\rm i}\frac{1}{2}\sigma_y.
\end{align}

\section{Vibrational pumping in the quantum-noise approximation}\label{appendixC}
{The Hamiltonian in Eq.\,\eqref{eq:molham} describes the dynamics of the molecule after the plasmon is eliminated. We describe here how it is also possible to eliminate the TLS degrees of freedom to focus on the dynamics of the vibrations, under the assumption that the plasmon-enhanced decay rate of the TLS ($\Gamma_{\rm eff}+\gamma_{\sigma}$) is much larger than the decay rate of the vibrations $\gamma_{b}$ and that the electron-phonon coupling is weak (for example, $d=0.1$ and $\hbar g=50$ meV).
We therefore divide the Hamiltonian in Eq.\,\eqref{eq:tlsnondiag} into the part representing the system, the reservoir and the system-reservoir interaction as follows:
\begin{align}
H_{\rm red}&=\underbrace{\hbar\tilde\delta\sigma{^\dagger} \sigma-\hbar\frac{1}{2}\mathcal{E}_{\rm pl} \sigma_x}_{\text{Reservoir}}+\underbrace{\hbar d\Omega (\sigma_{\rm e}-\langle\sigma_e\rangle)(b{^\dagger} +b)}_{\text{System-reservoir}}\nonumber\\
&+\underbrace{\hbar d\Omega \langle\sigma_e\rangle(b{^\dagger} +b)+\hbar\Omega b^\dagger b}_{\text{System}},\label{eq:}
\end{align}
where $\langle\sigma_e\rangle$ is the steady-state average of the TLS excited-state population calculated for the TLS decoupled from the vibrations, and $\tilde{\delta}=\delta+d^2\Omega$.
The elimination of the reservoir can be performed using the quantum noise approach \cite{Breuer2005, jaehne2008ground, rabl2010cooling} to Eq.\,\eqref{eq:molham}, giving the Hamiltonian:}
\begin{align*}
H_{\rm vib}=\hbar\Omega b^\dagger b+\hbar d\Omega \langle\sigma_{\rm e}\rangle(b^\dagger+b).
\end{align*}

The effective damping of the vibrations (appearing aside of the intrinsic damping $\gamma_{b}$) can be expressed via the Lindblad superoperator [Eqs.\,\eqref{eq:effvibdec} and \eqref{eq:effvibpump}]
\begin{align*}
\mathcal{L}_{\rm eff}^{-}[\rho]=-\frac{\Gamma_{-}}{2}\left(b{^\dagger} b\rho+\rho b{^\dagger} b-2b\rho b{^\dagger}  \right)
\end{align*}
and the effective vibrational pumping via the superoperator 
\begin{align*}
\mathcal{L}_{\rm eff}^{+}[\rho]=-\frac{\Gamma_{+}}{2}\left(bb{^\dagger} \rho+\rho b b{^\dagger} -2b{^\dagger} \rho b \right).
\end{align*}
The damping and pumping rates that appear in the Lindblad superoperators are given by 
\begin{align}
\Gamma_{-}=2\left(\Omega d\right)^2\Re\lbrace\tilde{S}(\Omega)\rbrace\label{eq:vibdecap}
\end{align}
and
\begin{align}
\Gamma_{+}=2\left(\Omega d\right)^2\Re\lbrace\tilde{S}(-\Omega)\rbrace,\label{eq:vibpumpap}
\end{align}
respectively. Here $\tilde{S}(s)=\int_0^\infty\langle\langle \sigma_{\rm e}(\tau)\sigma_{\rm e}(0) \rangle\rangle e^{{\rm i}s \tau}{\rm d}\tau$ (with $\sigma^\dagger\sigma=\sigma_{\rm e}$) is the spectral function that comprises the properties of the electronic TLS "bath" uncoupled from the vibrations and broadened by the plasmon.

We show below that the analytical expression for the spectral function $\Re \lbrace \tilde{S}(s) \rbrace$ can be expressed in the form:
\begin{widetext}
\begin{align}
&\Re \lbrace \tilde{S}(s)\rbrace=\Re\left\lbrace\int_0^\infty\langle\langle \sigma_{\rm e}(\tau)\sigma_{\rm e}(0) \rangle\rangle e^{{\rm i}s \tau}{\rm d}\tau\right\rbrace=\nonumber\\
&{ \textstyle
\frac{\mathcal{E}_{\rm pl}^2 {\frac{\Gamma_{\rm tot}}{2}} \left({\tilde{\delta}}^2+2 {\tilde{\delta}} {s}+{\left(\frac{\Gamma_{\rm tot}}{2}\right)}^2+{s}^2\right) \left(\mathcal{E}_{\rm pl}^2+8 {\left(\frac{\Gamma_{\rm tot}}{2}\right)}^2+2{s}^2\right)}{4\left(\mathcal{E}_{\rm pl}^2+2{\tilde{\delta}}^2+2{\left(\frac{\Gamma_{\rm tot}}{2}\right)}^2\right) \left[2 {\tilde{\delta}}^2 \left\lbrace{\left(\frac{\Gamma_{\rm tot}}{2}\right)}^2 \left(2\mathcal{E}_{\rm pl}^2-3 {s}^2\right)+\mathcal{E}_{\rm pl}^2 {s}^2 +4 {\left(\frac{\Gamma_{\rm tot}}{2}\right)}^4-{s}^4\right\rbrace+\left({\left(\frac{\Gamma_{\rm tot}}{2}\right)}^2+{s}^2\right) \left\lbrace{\left(\frac{\Gamma_{\rm tot}}{2}\right)}^2 \left(4 \mathcal{E}_{\rm pl}^2+5 {s}^2\right)+\left({s}^2-\mathcal{E}_{\rm pl}^2\right)^2+4 {\left(\frac{\Gamma_{\rm tot}}{2}\right)}^4\right\rbrace+{\tilde{\delta}}^4 \left( {\Gamma_{\rm tot}}^2+{s}^2\right)\right]}
}.
\label{eq:corfunTLS}
\end{align}
\end{widetext}

Here $\Gamma_{\rm tot}=\Gamma_{\rm eff}+\gamma_{\sigma}$ [see Eq.\,\eqref{eq:effectiveGamma} for definition of $\Gamma_{\rm eff}$] and $\langle\langle\sigma_{\rm e}(\tau)\sigma_{\rm e}(0)\rangle\rangle=\langle \sigma_{\rm e}(\tau)\sigma_{\rm e}(0)\rangle-\langle\sigma_{\rm e}\rangle\langle\sigma_{\rm e}\rangle$ is the part of the correlation function that corresponds to the fluctuations of the operators around the steady-state value. {We generally define the fluctuating correlation function as $\langle\langle O_1(\tau)O_2(0)\rangle\rangle=\langle O_1(\tau)O_2(0)\rangle-\langle O_1\rangle\langle O_2\rangle$.}

We obtain the spectral function $\Re \lbrace \tilde{S}(s) \rbrace$ in Eq.\,\eqref{eq:corfunTLS} following the procedure described in references \cite{jaehne2008ground, rabl2010cooling}. We start with the Liouville equation describing the TLS dynamics after effective elimination of the plasmonic cavity.  In the case of no electron-phonon coupling, it contains the Hamiltonian:
\begin{align*}
H_{\rm TLS}=\hbar\tilde{\delta}\sigma_{\rm e}-\hbar\frac{1}{2}\mathcal{E}_{\rm pl} \sigma_x
\end{align*}
together with the Lindblad term
\begin{align*}
\mathcal{L}_{\rm tot}[\rho]=-\frac{\Gamma_{\rm tot}}{2}\left(\sigma{^\dagger} \sigma\rho+\rho \sigma{^\dagger} \sigma-2\sigma\rho \sigma{^\dagger}  \right).
\end{align*}
\begin{widetext}
We obtain the following equations of motion for the mean values of the operators:
\begin{align}
\frac{\rm d}{{\rm d}t}\left[ \begin{array}{c}
{\langle\sigma\rangle}\\
{\langle\sigma^\dagger\rangle}\\
{\langle\sigma_{\rm e}\rangle}
\end{array}\right]=
\left[\begin{array}{c c c}
-{\rm i}\tilde{\delta}-\frac{\Gamma_{\rm tot}}{2} & 0 & -{\rm i}\mathcal{E}_{\rm pl}\\
0  & {\rm i}\tilde{\delta}-\frac{\Gamma_{\rm tot}}{2} & {\rm i}\mathcal{E}^\ast_{\rm pl} \\
-{\rm i}\mathcal{E}^\ast_{\rm pl}/2 & {\rm i}\mathcal{E}_{\rm pl}/2& -\Gamma_{\rm tot}
\end{array}\right]\cdot
\left[ \begin{array}{c}
\langle\sigma\rangle\\
\langle\sigma^\dagger\rangle\\
\langle\sigma_{\rm e}\rangle
\end{array}\right]+
\left[\begin{array}{c}
{\rm i}\mathcal{E}_{\rm pl}/2\\
-{\rm i}\mathcal{E}^\ast_{\rm pl}/2\\
0
\end{array}\right].\label{eq:TLSsige}
\end{align}
\end{widetext}
According to the quantum regression theorem \cite{Breuer2005}, the correlation functions $\langle\langle\sigma(\tau)\sigma_{\rm e}(0)\rangle\rangle$, $\langle\langle\sigma^\dagger(\tau)\sigma_{\rm e}(0)\rangle\rangle$ and $\langle\langle\sigma_{\rm e}(\tau)\sigma_{\rm e}(0)\rangle\rangle$ obey the same time evolution as $\langle\sigma (\tau)\rangle$, $\langle\sigma^\dagger(\tau)\rangle$ and $\langle\sigma_{\rm e}(\tau)\rangle$, respectively:
\begin{widetext}
\begin{align}
\frac{\rm d}{{\rm d}t}\left[ \begin{array}{c}
{\langle\langle\sigma(\tau)\sigma_{\rm e}(0)\rangle\rangle}\\
{\langle\langle\sigma^\dagger(\tau)\sigma_{\rm e}(0)\rangle\rangle}\\
{\langle\langle\sigma_{\rm e}(\tau)\sigma_{\rm e}(0)\rangle\rangle}
\end{array}\right]=
\left[\begin{array}{c c c}
-{\rm i}\tilde{\delta}-\frac{\Gamma_{\rm tot}}{2} & 0 & -{\rm i}\mathcal{E}_{\rm pl}\\
0  & {\rm i}\tilde{\delta}-\frac{\Gamma_{\rm tot}}{2} & {\rm i}\mathcal{E}^\ast_{\rm pl} \\
-{\rm i}\mathcal{E}^\ast_{\rm pl}/2 & {\rm i}\mathcal{E}_{\rm pl}/2& -\Gamma_{\rm tot}
\end{array}\right]\cdot
\left[ \begin{array}{c}
{\langle\langle\sigma(\tau)\sigma_{\rm e}(0)\rangle\rangle}\\
{\langle\langle\sigma^\dagger(\tau)\sigma_{\rm e}(0)\rangle\rangle}\\
{\langle\langle\sigma_{\rm e}(\tau)\sigma_{\rm e}(0)\rangle\rangle}
\end{array}\right] \label{eq:TLScorfundifeq}
\end{align}
\end{widetext}
with the initial values:
\begin{align*}
\langle\langle\sigma(0)\sigma_{\rm e}(0)\rangle\rangle &=\langle\sigma\rangle(1-\langle \sigma_{\rm e}\rangle),\\
\langle\langle\sigma^\dagger(0)\sigma_{\rm e}(0)\rangle\rangle &=-\langle\sigma^\dagger\rangle\langle \sigma_{\rm e}\rangle,\\
\langle\langle\sigma_{\rm e}(0)\sigma_{\rm e}(0)\rangle\rangle &=\langle\sigma_{\rm e}\rangle(1-\langle \sigma_{\rm e}\rangle).\\
\end{align*}
Here all the mean values are evaluated in the steady state. The inhomogeneous part in Eq.\,\eqref{eq:TLScorfundifeq} disappears because of the conveniently chosen value of the correlation functions when $\tau \rightarrow \infty$. 

The direct solution of Eq.\,\eqref{eq:TLScorfundifeq} yields the result in Eq.\,\eqref{eq:corfunTLS} for $\Re \lbrace \tilde{S}(s)\rbrace$, which in turn provides the analytical expressions for the effective vibrational pumping $\Gamma_{+}$ and damping $\Gamma_{-}$. We can then solve the effective vibrational dynamics:
\begin{align}
\frac{\rm d}{{\rm d}t}{\langle b\rangle}&=-{\rm i}\Omega\langle b\rangle-{\rm i}\Omega d \langle\sigma_{\rm e}\rangle\nonumber\\
&-\left(\frac{\gamma_{b}}{2}+\frac{\Gamma_{-}}{2}-\frac{\Gamma_{+}}{2}\right)\langle b\rangle,\\
\frac{\rm d}{{\rm d}t}{\langle b{^\dagger} b\rangle}&=-{\rm i}\Omega d \langle\sigma_{\rm e}\rangle\left(\langle b{^\dagger} \rangle -\langle b\rangle\right) \nonumber\\
&-(\gamma_{b}+\Gamma_{-}-\Gamma_{+})\langle b{^\dagger} b\rangle + \Gamma_{+},
\end{align} 
{from which we obtain the steady-state values:
\begin{align}
\langle b \rangle&=-\frac{{\rm i}\Omega d\langle\sigma_{\rm e}\rangle}{\frac{\gamma_{b}}{2}+{\rm i}\Omega},\\
\langle b{^\dagger} b\rangle&=|\langle b \rangle|^2+\frac{\Gamma_{+}}{\gamma_{b}+\Gamma_{-}-\Gamma_{+}}.\label{eq:cohplusincoh}
\end{align}}
Eq.\,\eqref{eq:cohplusincoh} shows two different sources of vibrational pumping that are involved in the process. The first one, represented by the term $|\langle b \rangle|^2$ is the coherent pumping of the vibrations due to the TLS. {The second term, $\langle b^\dagger b \rangle_{\rm SS,in}=\frac{\Gamma_{+}}{\gamma_{b}+\Gamma_{-}-\Gamma_{+}}$, is due to the "incoherent" vibrational pumping induced by the fluctuating part of the TLS quantum dynamics. 
 
{Under the considered conditions, the incoherent pumping term dominates over the weak coherent pumping term so that the shape of the correlation function $\Re \lbrace \tilde{S}(s) \rbrace$ governs the population of the vibrations as indicated by Eqs.\,\eqref{eq:vibdecap} and \eqref{eq:vibpumpap}. As we have demonstrated in Section\,\ref{sec:temp}, cooling is also possible \cite{jaehne2008ground, rabl2010cooling}. }

\section{Small vibrational displacement $d$: dressed-molecule picture}\label{app:svd_dmp}

The regime where the linewidth of the RF peaks is comparable to the width of the Raman peaks is a limiting case of Raman scattering in intense fields that has been studied in the context of atomic physics \cite{cohen1994dressed, Tanoudji77dressedmulti, agarwal1979theory, cohen1977modification}. To understand the splitting of the lines that appear when the Mollow triplet side peaks have the frequency of the Raman lines, it is useful to rewrite the Hamiltonian into a form where the coupling among vibrational states is explicitly present.

This Hamiltonian can be derived from the reduced Hamiltonian ${H}_{\rm red}$ appearing in Eq.\,\eqref{eq:molhamel} by applying the so called \textit{small polaron transformation} ${H}_{\rm red}\rightarrow {{H}_{\rm red}^{'}}={U}_{\sigma_{\rm e}} {H}_{\rm red} {U}_{\sigma_{\rm e}}^\dagger$, which is represented by the unitary matrix in the form of a displacement operator ${U}_{\sigma_{\rm e}}=\exp\left[ d{\sigma}_{\rm e}({b}^\dagger-{b}) \right]$. This transformation has two effects on the Hamiltonian. First, the vibrational term in ${H}_{\rm red}$ transforms as:
\begin{align}
\hbar\Omega ({b}{^\dagger} +{\sigma}_{\rm e}d)({b}+{\sigma}_{\rm e}d)\rightarrow \hbar\Omega {b}{^\dagger}{b},
\end{align}
and, second, the pumping term of the TLS acquires an additional factor that includes the vibrational operators (which yield the well known Franck-Condon factors):
\begin{align}
&\hbar g \alpha_{\rm S}{\sigma}{^\dagger}+{\rm H.c.}\rightarrow\hbar g \alpha_{\rm S}{\sigma}{^\dagger}\exp\left[ d({b}^\dagger-{b}) \right] +{\rm H.c.}\nonumber\\
&=-\hbar\frac{\mathcal{E}_{\rm pl}}{2}{\sigma}{^\dagger}\exp\left[ d({b}^\dagger-{b}) \right] +{\rm H.c.}
\end{align}
In this approach we assume that the influence of the plasmon (given by ${H}_{\rm pl}$ and ${H}_{\rm pl-e}$ in Eq.\,\eqref{eq:cohelmoldr}) is effectively included as an enhancement of the incident laser field $\mathcal{E}_{\rm pl}$. Furthermore we consider only the case where the splitting of the fluorescence and Raman spectral peaks [see e.g. Fig.\,\ref{fig:spectra_strong} (b,d,f)] is larger than their broadening (the so called secular limit). In such a case, the incoherent broadening does not influence the peaks positions and, therefore, in the following we consider that the system can be described only by the the simplified Hermitian Hamiltonian (we do not consider the Lindblad terms as we are mainly interested in the nature of the transitions). We further assume weak electron-phonon coupling in the molecule and expand the exponential terms containing the vibrational operators to the first order: $\exp\left[ \pm d({b}^\dagger-{b}) \right]\approx I \pm d({b}^\dagger-{b})$. Last, we reduce the system comprising the vibrations and the TLS into an effective four-level system that consists of the ground and excited electronic states considering zero or one vibrational excitation for each electronic state.
\begin{figure}[t!]
	\begin{center}
	\includegraphics[scale=0.85]{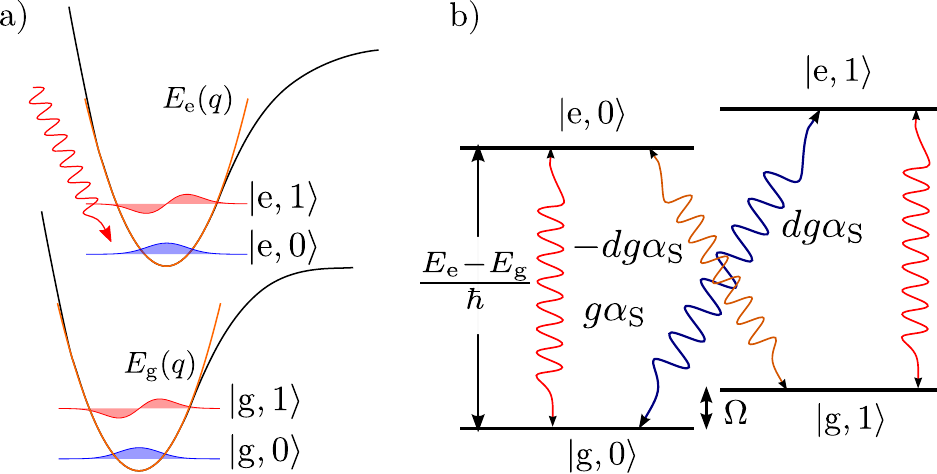}
	\end{center}
	\caption{(a,b) Schematic representation of the energies of the simplified model of a TLS molecule with one vibrational excited state. In (b) the interaction terms of the simplified Hamiltonian, ${{H}_{\rm red}^{'}}$, obtained after applying the \textit{small polaron transformation}, are graphically depicted.}
	\label{fig:strongdrivingmodel1}
\end{figure}
The diagram of the resulting effective system is drawn in Fig.\,\ref{fig:strongdrivingmodel1}\,(a) and (b). By diagonalizing this $4\times 4$ Hamiltonian we achieve a new level structure of the system that, in the dressed-molecule picture, provides the positions of the emission peaks (for detailed discussion of the dressed-molecule (atom) picture see e.g. chapter 10 of reference \cite{Scully1997}). Below we briefly describe how the emission spectra can be understood using this dressed-molecule picture.

In the dressed-molecule picture we consider the simplified Hamiltonian, which can be formally defined in the basis of states $[|{\rm g},\mathcal{N},0\rangle, |{\rm e},\mathcal{N}-1,0\rangle, |{\rm g},\mathcal{N},1\rangle, |{\rm e},\mathcal{N}-1,1\rangle ]$, with e (g) labelling the electronic excited (ground) state, $\mathcal{N}$ labelling the "photon" number state of the \textit{exciting field} and 0 (1) labelling the number of vibrational excitations, respectively. 
The exciting field is not quantized explicitly in the original Hamiltonian [Eq.\,\eqref{eq:cohelmoldr}] where it is represented by the plasmon coherent-state amplitude $\alpha_{\rm S}$. We therefore assume that the exciting field is a highly populated bosonic field which peaks sharply around a (mean) occupation number $\tilde{\mathcal{N}}$ yielding $\alpha_{\rm S} = \sqrt{\tilde{\mathcal{N}}}g_{\rm PL-L}$, with $g_{\rm PL-L}$ formally defined as a small coupling constant (such that $\tilde{\mathcal{N}}\gg 1$) between the equivalent exciting field and the molecule. Note that the formal definition of the exciting field is not important for the following discussion as by introducing the quantized exciting field we only aim at mimicking the action of the semi-classical pumping term. However, the number states $|\mathcal{N}\rangle$ of the exciting field are convenient to discuss the dressing of the molecular excited states in terms of the hybridization of the quantum-mechanical states.

\begin{figure*}[t!]
	\begin{center}
		\includegraphics[scale=1]{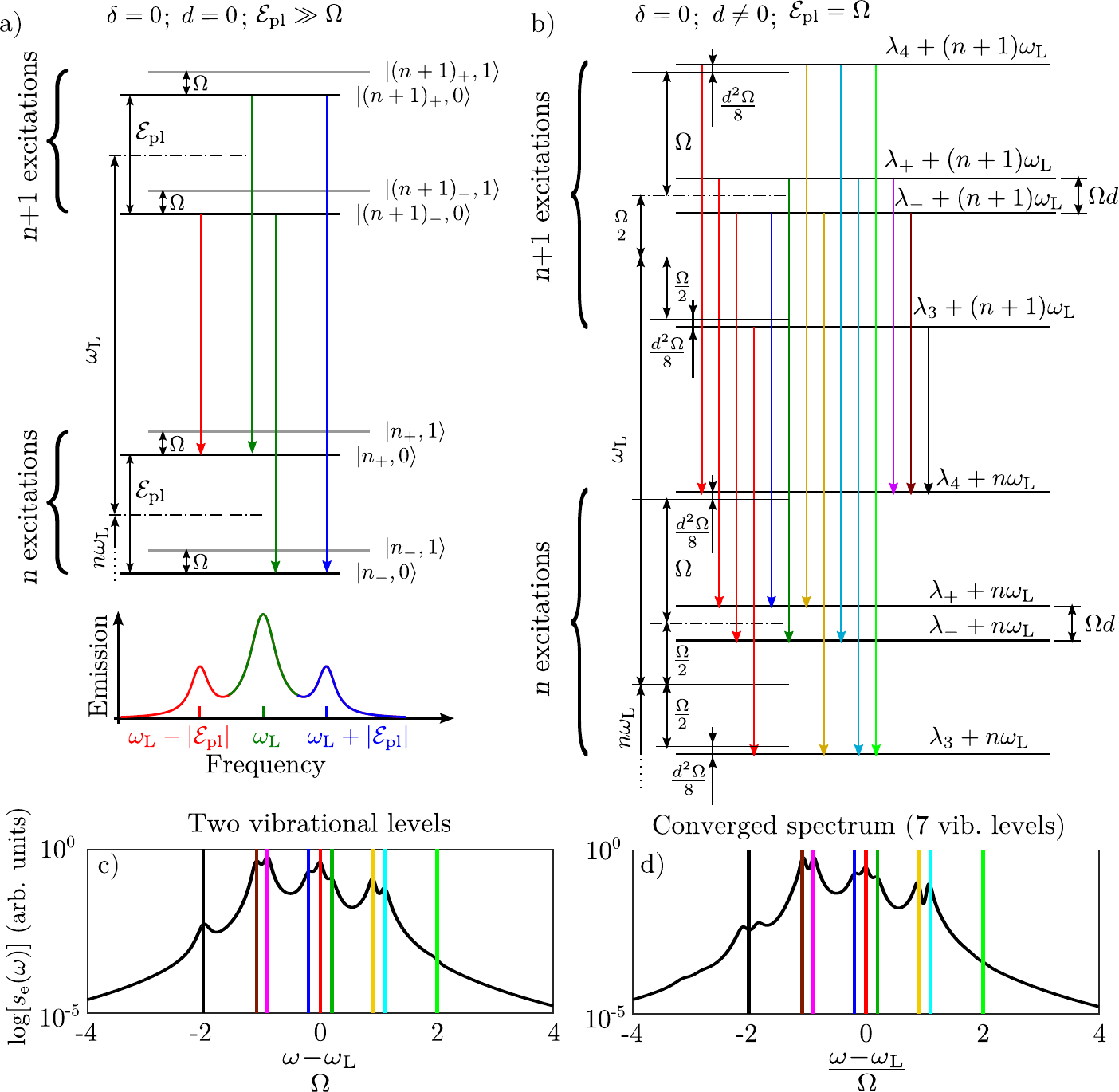}
	\end{center}
	\caption{Energy level diagram in the dressed-molecule picture where the level structure of the effective four-level system describing the molecule is repeated for each manifold containing $n$ excitation quanta: (a) a situation where no electron-phonon interaction is present ($d=0$), and (b) a situation where all the interactions are present [the energies $\lambda_i$ are defined in Eq.\,\eqref{eq:lambdas}]. In (a) we further mark the transitions that give rise to Mollow triplet by coloured arrows (connecting only $m=0$ states, for simplicity), and use the colour code to assign the transitions to the respective emission peaks in the schematically depicted spectrum below. The coloured lines in (b) represent all possible transitions that can contribute to the emission spectrum [as shown in (c,d)].       
(c,d) Particular example of an emission spectrum of a TLS in a plasmonic resonator obtained from the full model [Eq.\,\eqref{eq:cohelmoldr}] using (a) two and (b) seven vibrational levels (converged spectrum) in both the ground and the excited electronic states.  \cmmt{The spectrum is calculated for $d=0.2$, $\hbar\Omega=10$ meV, $\hbar g\approx 9$ meV, $\hbar\mathcal{E}\approx 130$ meV, $\hbar\gamma_{\sigma}=2\times 10^{-5}$ eV, $\hbar\gamma_{b}=1$ meV, $\hbar\gamma_{a}=500$ meV, $\hbar\delta=0$ eV, $\hbar\Delta=0$ eV and temperature $T=0$ K. The colored lines [calculated according to Eq.\,\eqref{eq:lambdas}] represent the different transitions graphically depicted in (b) using the same color code.}}
	\label{fig:strongdrivingmodel2}
\end{figure*}
We further define the total number of excitations as $n=\mathcal{N}+\delta_{i{\rm e}}$, with $i={\rm e,\, g}$ and $\delta_{ij}$ the Kronecker delta. We consider that the electronic levels, carrying the fine vibrational structure, are dressed by the strong laser illumination. The Hamiltonian can be expressed in the interaction picture of the incident laser field which is exactly tuned to the electronic transition, $\hbar\delta=0$ eV. In the basis $[|{\rm g},\mathcal{N},0\rangle, |{\rm e},\mathcal{N}-1,0\rangle, |{\rm g},\mathcal{N},1\rangle, |{\rm e},\mathcal{N}-1,1\rangle ]$ the Hamiltonian, ${H}_{\rm red}^{'}$, can be represented by the matrix:
\begin{align}
\mathbf{H}^{'}_{\rm red}\approx\hbar\left[
\begin{array}{cccc}
0 & -\mathcal{E}_{\rm pl}/2 & 0 & -d\mathcal{E}_{\rm pl}/2 \\
-\mathcal{E}_{\rm pl}/2 & 0 & d\mathcal{E}_{\rm pl}/2  & 0 \\
0 & d\mathcal{E}_{\rm pl}/2 & \Omega & -\mathcal{E}_{\rm pl}/2 \\
-d\mathcal{E}_{\rm pl}/2 & 0 & -\mathcal{E}_{\rm pl}/2 & \Omega \\
\end{array}
\right].\label{eq:sup_hpol_4x4}
\end{align}
For vanishing electron-phonon coupling, $d=0$, the Hamiltonian in Eq.\,\eqref{eq:sup_hpol_4x4} reduces to the form describing a pair of TLSs dressed by the incident laser illumination. The process of dressing (i.e. diagonalization of the above Hamiltonian with $d=0$) can be viewed as a mixing of the electronic states with the high number states of the exciting laser field, giving rise to the basis of hybridized states $[|n_{-},0\rangle, |n_{+},0\rangle,|n_{-},1\rangle, |n_{+},1\rangle]$ where the first quantum number, $n$, labels the total number of electronic plus laser excitations, and the second quantum number, $m$, belongs to the vibrational states [see Fig.\,\ref{fig:strongdrivingmodel2}\,(a) for schematics of the corresponding energy levels]. The hybrid states are defined as $|n_\pm,m\rangle \equiv (|{\rm g},\mathcal{N},m\rangle \pm |{\rm e},\mathcal{N}-1,m\rangle)/\sqrt{2}$ with $+$ labelling the state with higher energy. In the new basis of such dressed states, we can represent the Hamiltonian as:
	\begin{align}
	\mathbf{H}_{\rm red}^{', dr}&\approx\hbar\left[
	\begin{array}{cccc}
	\mathcal{E}_{\rm pl}/2 & 0 & 0 & 0 \\
	0 & -\mathcal{E}_{\rm pl}/2 & 0  & 0 \\
	0 & 0 & \Omega+\mathcal{E}_{\rm pl}/2 & 0 \\
	0 & 0 & 0 & \Omega-\mathcal{E}_{\rm pl}/2 \\
	\end{array}
	\right]\nonumber\\
	&-\hbar\left[
	\begin{array}{cccc}
	0 & 0 & 0 & -d\mathcal{E}_{\rm pl}/2 \\
	0 & 0 & -d\mathcal{E}_{\rm pl}/2  & 0 \\
	0 & -d\mathcal{E}_{\rm pl}/2 & 0 & 0 \\
	-d\mathcal{E}_{\rm pl}/2 & 0 & 0 & 0 \\
	\end{array}
	\right],\label{eq:sup_ham4x4trans}
	\end{align}
For $d=0$, the splitting of the states $|n_{\pm}\rangle$ for the TLS in each vibrational Fock state is $|2g\alpha_{\rm S}|=|\mathcal{E}_{\rm pl}|$. The two dressed TLS defined for each vibrational Fock state are mutually shifted by the vibrational frequency $\Omega$ along the energy axis. In the absence of electron-phonon coupling $d$, the RF emission (dominating in this case the inelastic emission) is given purely by the transitions conserving the vibrational number state and changing the total number of excitations, $n$, by one. In particular, the central Mollow peak is given by transitions between $|(n+1)_{+},0(1)\rangle\rightarrow |n_{+},0(1)\rangle$ and $|(n+1)_{-},0(1)\rangle\rightarrow |n_{-},0(1)\rangle$, while the side peaks contain transitions $|(n+1)_{-},0(1)\rangle\rightarrow |n_{+},0(1)\rangle$ (red detuned) and $|(n+1)_{+},0(1)\rangle\rightarrow |n_{-},0(1)\rangle$ (blue detuned), respectively. The respective transitions and their corresponding emission peaks (the Mollow triplet) are schematically marked in Fig.\,\ref{fig:strongdrivingmodel2}\,(a), where the colouring of the spectral emission peaks (bottom) corresponds to the colour of the respective arrows marking the transitions (top). 

If we switch on the electron-phonon interaction $d$, a mixing between the levels belonging to the two vibrational Fock states is introduced, simultaneously allowing additional transitions yielding the Raman emission (i.e. changing the vibrational Fock state). The details of the level mixing and the subsequent emission spectra depend on the particular choice of pumping strength, $\mathcal{E}_{\rm pl}$, in combination with the value of the electron-phonon coupling, $d$. 
In the following we consider a particular case where the Mollow triplet side peaks overlap with the Raman lines with the laser frequency exactly tuned to the TLS energy splitting ($\hbar\delta=0$ eV and $|\mathcal{E}_{\rm pl}|=\Omega$). Upon diagonalization, the Hamiltonian in Eq.\,\eqref{eq:sup_hpol_4x4} [Eq.\,\eqref{eq:sup_ham4x4trans}] yields the following spectrum of energy levels:
\begin{align}
\begin{split}
\lambda_{-}&=-\frac{1}{2} (d-1) \Omega,\\
\lambda_{+}&=\frac{1}{2} (d+1) \Omega, \\
\lambda_3&=-\frac{1}{2} \left(\sqrt{d^2+4}-1\right) \Omega\approx -\frac{1}{2}\left( 1+\frac{d^2}{4} \right)\Omega,\\
\lambda_4&=\frac{1}{2} \left(\sqrt{d^2+4}+1\right) \Omega\approx \frac{1}{2}\left( 3+\frac{d^2}{4} \right)\Omega,
\end{split}\label{eq:lambdas}
\end{align}
where we used the assumption that $d\ll 1$ to perform the Taylor expansion of the square root up to the first order. The states having energy $\lambda_{\pm}$ are a coherent admixture of states containing zero and one vibrational excitation $|n_{-},1\rangle$ and $|n_{+},0\rangle$, as discussed above, and the states of energy $\lambda_{3,4}$ can be identified (up to small $o(d)$ admixtures of other states) with $|n_{-},0\rangle$ and $|n_{+},1\rangle$ whose energy is renormalized due to the off-resonant electron-phonon coupling. This level structure of the molecule does not explicitly contain the quantized electromagnetic field of the incident laser. However, in the dressed-molecule picture the molecular level structure [Eq.\,\eqref{eq:lambdas}] is periodically repeated for each manifold represented by a specific number of excitations, $n$, and thus appears repeated along the energy axis displaced by integer values of the laser frequency $\omega_{\rm L}$, as schematically illustrated in Fig.\,\ref{fig:strongdrivingmodel2}\,(b). In this picture, the emission events are represented by transitions between manifolds that differ by one excitation quantum of the electronic and effective photonic states, i.e. transitions between the manifolds containing $n$ and $n+1$ excitation quanta [represented by colored lines in Fig.\,\ref{fig:strongdrivingmodel2}\,(b)]. 


The dressed-molecule picture above nicely allows us to identify the spectral peaks which appear in the complex photon emission spectra obtained from the numerical calculation of the complete Hamiltonian in Eq.\,\eqref{eq:cohelmoldr} with its corresponding Lindblad terms. Figure\,\ref{fig:strongdrivingmodel2}\,(c) shows such a situation where two vibrational levels corresponding to the ground vibrational state and the first excited vibrational state are considered. Nine main frequencies [coloured lines in Fig.\,\ref{fig:strongdrivingmodel2}\,(c)] are identified in the spectrum which nicely coincide with the nine transitions marked in the energy diagram of Fig.\,\ref{fig:strongdrivingmodel2}\,(b) [vertical lines marking $\lambda_i-\lambda_j$, where $i, j\,\in\{ +,-, 3,4 \}$, and $\lambda_i$ are defined in Eq.\,\eqref{eq:lambdas}].
For comparison, we show in Fig.\,\ref{fig:strongdrivingmodel2}\,(d) the results obtained using a sufficiently large number of vibrations to achieve results converged with respect to the size of the vibrational subspace. In this case, more spectral features appear [we observe higher order transitions and further peak splitting when compared with Fig.\,\ref{fig:strongdrivingmodel2}\,(c)]. Nonetheless, the simple model introduced in this section still explains very satisfactorily the spectral positions of the strongest peaks.

\section{Description of the system for two vibrational modes}\label{app:dstvm}
We consider next a molecule with two vibrational modes. To that end, we consider the plasmon as a bath (as described in Appendix\,\ref{Appendix Dressed}), which allows writing the following reduced Hamiltonian describing the dynamics of the molecule:
\begin{align}
\begin{split}
H_{\rm red,two}&=\hbar \delta\sigma_{\rm e}+\hbar\Omega_1 (b_1{^\dagger}+d_1\sigma_{\rm e} )(b_1+d_1\sigma_{\rm e})\\
&+\hbar\Omega_2 (b_2{^\dagger} +d_2\sigma_{\rm e})(b_2+d_2\sigma_{\rm e} ) +\hbar\frac{1}{2}\mathcal{E}_{\rm PL}\sigma_x,
\end{split}\label{eq:effectiveTLS2vib}
\end{align} 
where $b_{1(2)}[b^\dagger_{1(2)}]$ is the annihilation (creation) operator of the vibrational mode 1 (2), $\Omega_{1(2)}$ is the vibrational frequency of the respective mode and $d_{1(2)}$ is the displacement of the respective excited PESs with respect to the ground state ones. According to Eq.\,\eqref{eq:Liouvil} we define two Lindblad superoperators that effectively account for the Markovian damping of the two vibrational modes:
\begin{align*}
\mathcal{L}_{b_1}[\rho]&=-\frac{\gamma_{b_1}}{2}\left( b_1^\dagger b_1\rho+\rho b_1^\dagger b_1 -2 b_1\rho b_1^\dagger\right),\\
\mathcal{L}_{b_2}[\rho]&=-\frac{\gamma_{b_2}}{2}\left( b_2^\dagger b_2\rho+\rho b_2^\dagger b_2 -2 b_2\rho b_2^\dagger\right),
\end{align*}
with $\gamma_{b_1}$ ($\gamma_{b_2}$) being the respective decay rates of the vibrational modes.

\subsection*{}

\bibliography{assampBibNEW}

\providecommand{\noopsort}[1]{}\providecommand{\singleletter}[1]{#1}%
\begin{thebibliography}{81}%
\makeatletter
\providecommand \@ifxundefined [1]{%
 \@ifx{#1\undefined}
}%
\providecommand \@ifnum [1]{%
 \ifnum #1\expandafter \@firstoftwo
 \else \expandafter \@secondoftwo
 \fi
}%
\providecommand \@ifx [1]{%
 \ifx #1\expandafter \@firstoftwo
 \else \expandafter \@secondoftwo
 \fi
}%
\providecommand \natexlab [1]{#1}%
\providecommand \enquote  [1]{``#1''}%
\providecommand \bibnamefont  [1]{#1}%
\providecommand \bibfnamefont [1]{#1}%
\providecommand \citenamefont [1]{#1}%
\providecommand \href@noop [0]{\@secondoftwo}%
\providecommand \href [0]{\begingroup \@sanitize@url \@href}%
\providecommand \@href[1]{\@@startlink{#1}\@@href}%
\providecommand \@@href[1]{\endgroup#1\@@endlink}%
\providecommand \@sanitize@url [0]{\catcode `\\12\catcode `\$12\catcode
  `\&12\catcode `\#12\catcode `\^12\catcode `\_12\catcode `\%12\relax}%
\providecommand \@@startlink[1]{}%
\providecommand \@@endlink[0]{}%
\providecommand \url  [0]{\begingroup\@sanitize@url \@url }%
\providecommand \@url [1]{\endgroup\@href {#1}{\urlprefix }}%
\providecommand \urlprefix  [0]{URL }%
\providecommand \Eprint [0]{\href }%
\providecommand \doibase [0]{http://dx.doi.org/}%
\providecommand \selectlanguage [0]{\@gobble}%
\providecommand \bibinfo  [0]{\@secondoftwo}%
\providecommand \bibfield  [0]{\@secondoftwo}%
\providecommand \translation [1]{[#1]}%
\providecommand \BibitemOpen [0]{}%
\providecommand \bibitemStop [0]{}%
\providecommand \bibitemNoStop [0]{.\EOS\space}%
\providecommand \EOS [0]{\spacefactor3000\relax}%
\providecommand \BibitemShut  [1]{\csname bibitem#1\endcsname}%
\let\auto@bib@innerbib\@empty
\bibitem [{\citenamefont {Kneipp}\ \emph {et~al.}(1996)\citenamefont {Kneipp},
  \citenamefont {Wang}, \citenamefont {Kneipp}, \citenamefont {Itzkan},
  \citenamefont {Dasari},\ and\ \citenamefont {Feld}}]{kneipp1996poppump}%
  \BibitemOpen
  \bibfield  {author} {\bibinfo {author} {\bibfnamefont {K.}~\bibnamefont
  {Kneipp}}, \bibinfo {author} {\bibfnamefont {Y.}~\bibnamefont {Wang}},
  \bibinfo {author} {\bibfnamefont {H.}~\bibnamefont {Kneipp}}, \bibinfo
  {author} {\bibfnamefont {I.}~\bibnamefont {Itzkan}}, \bibinfo {author}
  {\bibfnamefont {R.~R.}\ \bibnamefont {Dasari}}, \ and\ \bibinfo {author}
  {\bibfnamefont {M.~S.}\ \bibnamefont {Feld}},\ }\href {\doibase
  10.1103/PhysRevLett.76.2444} {\bibfield  {journal} {\bibinfo  {journal}
  {Phys. Rev. Lett.}\ }\textbf {\bibinfo {volume} {76}},\ \bibinfo {pages}
  {2444} (\bibinfo {year} {1996})}\BibitemShut {NoStop}%
\bibitem [{\citenamefont {Kneipp}\ \emph {et~al.}(1997)\citenamefont {Kneipp},
  \citenamefont {Wang}, \citenamefont {Kneipp}, \citenamefont {Perelman},
  \citenamefont {Itzkan}, \citenamefont {Dasari},\ and\ \citenamefont
  {Feld}}]{kneipp1997singlemoleculesers}%
  \BibitemOpen
  \bibfield  {author} {\bibinfo {author} {\bibfnamefont {K.}~\bibnamefont
  {Kneipp}}, \bibinfo {author} {\bibfnamefont {Y.}~\bibnamefont {Wang}},
  \bibinfo {author} {\bibfnamefont {H.}~\bibnamefont {Kneipp}}, \bibinfo
  {author} {\bibfnamefont {L.~T.}\ \bibnamefont {Perelman}}, \bibinfo {author}
  {\bibfnamefont {I.}~\bibnamefont {Itzkan}}, \bibinfo {author} {\bibfnamefont
  {R.~R.}\ \bibnamefont {Dasari}}, \ and\ \bibinfo {author} {\bibfnamefont
  {M.~S.}\ \bibnamefont {Feld}},\ }\href {\doibase 10.1103/PhysRevLett.78.1667}
  {\bibfield  {journal} {\bibinfo  {journal} {Phys. Rev. Lett.}\ }\textbf
  {\bibinfo {volume} {78}},\ \bibinfo {pages} {1667} (\bibinfo {year}
  {1997})}\BibitemShut {NoStop}%
\bibitem [{\citenamefont {Xu}\ \emph {et~al.}(1999)\citenamefont {Xu},
  \citenamefont {Bjerneld}, \citenamefont {K\"all},\ and\ \citenamefont
  {B\"orjesson}}]{xu1999singlehemoglobine}%
  \BibitemOpen
  \bibfield  {author} {\bibinfo {author} {\bibfnamefont {H.}~\bibnamefont
  {Xu}}, \bibinfo {author} {\bibfnamefont {E.~J.}\ \bibnamefont {Bjerneld}},
  \bibinfo {author} {\bibfnamefont {M.}~\bibnamefont {K\"all}}, \ and\ \bibinfo
  {author} {\bibfnamefont {L.}~\bibnamefont {B\"orjesson}},\ }\href {\doibase
  10.1103/PhysRevLett.83.4357} {\bibfield  {journal} {\bibinfo  {journal}
  {Phys. Rev. Lett.}\ }\textbf {\bibinfo {volume} {83}},\ \bibinfo {pages}
  {4357} (\bibinfo {year} {1999})}\BibitemShut {NoStop}%
\bibitem [{\citenamefont {Haslett}\ \emph {et~al.}(2000)\citenamefont
  {Haslett}, \citenamefont {Tay},\ and\ \citenamefont
  {Moskovits}}]{haslett2000cansers}%
  \BibitemOpen
  \bibfield  {author} {\bibinfo {author} {\bibfnamefont {T.~L.}\ \bibnamefont
  {Haslett}}, \bibinfo {author} {\bibfnamefont {L.}~\bibnamefont {Tay}}, \ and\
  \bibinfo {author} {\bibfnamefont {M.}~\bibnamefont {Moskovits}},\ }\href
  {\doibase 10.1063/1.481952} {\bibfield  {journal} {\bibinfo  {journal} {J.
  Chem. Phys.}\ }\textbf {\bibinfo {volume} {113}},\ \bibinfo {pages} {1641}
  (\bibinfo {year} {2000})}\BibitemShut {NoStop}%
\bibitem [{\citenamefont {Xu}\ \emph {et~al.}(2000)\citenamefont {Xu},
  \citenamefont {Aizpurua}, \citenamefont {K\"all},\ and\ \citenamefont
  {Apell}}]{xu2000elmag}%
  \BibitemOpen
  \bibfield  {author} {\bibinfo {author} {\bibfnamefont {H.}~\bibnamefont
  {Xu}}, \bibinfo {author} {\bibfnamefont {J.}~\bibnamefont {Aizpurua}},
  \bibinfo {author} {\bibfnamefont {M.}~\bibnamefont {K\"all}}, \ and\ \bibinfo
  {author} {\bibfnamefont {P.}~\bibnamefont {Apell}},\ }\href {\doibase
  10.1103/PhysRevE.62.4318} {\bibfield  {journal} {\bibinfo  {journal} {Phys.
  Rev. E}\ }\textbf {\bibinfo {volume} {62}},\ \bibinfo {pages} {4318}
  (\bibinfo {year} {2000})}\BibitemShut {NoStop}%
\bibitem [{\citenamefont {Brolo}\ \emph {et~al.}(2004)\citenamefont {Brolo},
  \citenamefont {Sanderson},\ and\ \citenamefont {Smith}}]{brolo2004ratiosers}%
  \BibitemOpen
  \bibfield  {author} {\bibinfo {author} {\bibfnamefont {A.~G.}\ \bibnamefont
  {Brolo}}, \bibinfo {author} {\bibfnamefont {A.~C.}\ \bibnamefont
  {Sanderson}}, \ and\ \bibinfo {author} {\bibfnamefont {A.~P.}\ \bibnamefont
  {Smith}},\ }\href {\doibase 10.1103/PhysRevB.69.045424} {\bibfield  {journal}
  {\bibinfo  {journal} {Phys. Rev. B}\ }\textbf {\bibinfo {volume} {69}},\
  \bibinfo {pages} {045424} (\bibinfo {year} {2004})}\BibitemShut {NoStop}%
\bibitem [{\citenamefont {Maher}\ \emph {et~al.}(2004)\citenamefont {Maher},
  \citenamefont {Cohen}, \citenamefont {Etchegoin}, \citenamefont {Hartigan},
  \citenamefont {Brown},\ and\ \citenamefont
  {Milton}}]{maher2004stokesantistokes}%
  \BibitemOpen
  \bibfield  {author} {\bibinfo {author} {\bibfnamefont {R.~C.}\ \bibnamefont
  {Maher}}, \bibinfo {author} {\bibfnamefont {L.~F.}\ \bibnamefont {Cohen}},
  \bibinfo {author} {\bibfnamefont {P.}~\bibnamefont {Etchegoin}}, \bibinfo
  {author} {\bibfnamefont {H.~J.~N.}\ \bibnamefont {Hartigan}}, \bibinfo
  {author} {\bibfnamefont {R.~J.~C.}\ \bibnamefont {Brown}}, \ and\ \bibinfo
  {author} {\bibfnamefont {M.~J.~T.}\ \bibnamefont {Milton}},\ }\href {\doibase
  10.1063/1.1739398} {\bibfield  {journal} {\bibinfo  {journal} {J. Chem.
  Phys.}\ }\textbf {\bibinfo {volume} {120}},\ \bibinfo {pages} {11746}
  (\bibinfo {year} {2004})}\BibitemShut {NoStop}%
\bibitem [{\citenamefont {Haes}\ \emph {et~al.}(2005)\citenamefont {Haes},
  \citenamefont {Haynes}, \citenamefont {McFarland}, \citenamefont {Schatz},
  \citenamefont {Van~Duyne},\ and\ \citenamefont {Zou}}]{haes2005plasmatsers}%
  \BibitemOpen
  \bibfield  {author} {\bibinfo {author} {\bibfnamefont {A.~J.}\ \bibnamefont
  {Haes}}, \bibinfo {author} {\bibfnamefont {C.~L.}\ \bibnamefont {Haynes}},
  \bibinfo {author} {\bibfnamefont {A.~D.}\ \bibnamefont {McFarland}}, \bibinfo
  {author} {\bibfnamefont {G.~C.}\ \bibnamefont {Schatz}}, \bibinfo {author}
  {\bibfnamefont {R.~P.}\ \bibnamefont {Van~Duyne}}, \ and\ \bibinfo {author}
  {\bibfnamefont {S.}~\bibnamefont {Zou}},\ }\href {\doibase
  10.1557/mrs2005.100} {\bibfield  {journal} {\bibinfo  {journal} {MRS
  Bulletin}\ }\textbf {\bibinfo {volume} {30}},\ \bibinfo {pages} {368–375}
  (\bibinfo {year} {2005})}\BibitemShut {NoStop}%
\bibitem [{\citenamefont {Moskovits}(2005)}]{moskovits2005sers}%
  \BibitemOpen
  \bibfield  {author} {\bibinfo {author} {\bibfnamefont {M.}~\bibnamefont
  {Moskovits}},\ }\href {\doibase 10.1002/jrs.1362} {\bibfield  {journal}
  {\bibinfo  {journal} {J. Raman Spectrosc.}\ }\textbf {\bibinfo {volume}
  {36}},\ \bibinfo {pages} {485} (\bibinfo {year} {2005})}\BibitemShut
  {NoStop}%
\bibitem [{\citenamefont {Le~Ru}\ and\ \citenamefont
  {Etchegoin}(2006)}]{leru2006vibpump}%
  \BibitemOpen
  \bibfield  {author} {\bibinfo {author} {\bibfnamefont {E.~C.}\ \bibnamefont
  {Le~Ru}}\ and\ \bibinfo {author} {\bibfnamefont {P.~G.}\ \bibnamefont
  {Etchegoin}},\ }\href {\doibase 10.1039/B505343A} {\bibfield  {journal}
  {\bibinfo  {journal} {Faraday Discuss.}\ }\textbf {\bibinfo {volume} {132}},\
  \bibinfo {pages} {63} (\bibinfo {year} {2006})}\BibitemShut {NoStop}%
\bibitem [{\citenamefont {Stiles}\ \emph {et~al.}(2008)\citenamefont {Stiles},
  \citenamefont {Dieringer}, \citenamefont {Shah},\ and\ \citenamefont
  {Duyne}}]{stiles2008SERS}%
  \BibitemOpen
  \bibfield  {author} {\bibinfo {author} {\bibfnamefont {P.~L.}\ \bibnamefont
  {Stiles}}, \bibinfo {author} {\bibfnamefont {J.~A.}\ \bibnamefont
  {Dieringer}}, \bibinfo {author} {\bibfnamefont {N.~C.}\ \bibnamefont {Shah}},
  \ and\ \bibinfo {author} {\bibfnamefont {R.~P.~V.}\ \bibnamefont {Duyne}},\
  }\href {\doibase 10.1146/annurev.anchem.1.031207.112814} {\bibfield
  {journal} {\bibinfo  {journal} {Annu. Rev. Anal. Chem.}\ }\textbf {\bibinfo
  {volume} {1}},\ \bibinfo {pages} {601} (\bibinfo {year} {2008})}\BibitemShut
  {NoStop}%
\bibitem [{\citenamefont {Tong}\ \emph {et~al.}(2014)\citenamefont {Tong},
  \citenamefont {Xu},\ and\ \citenamefont {K{\"a}ll}}]{tong_xu_kall_2014}%
  \BibitemOpen
  \bibfield  {author} {\bibinfo {author} {\bibfnamefont {L.}~\bibnamefont
  {Tong}}, \bibinfo {author} {\bibfnamefont {H.}~\bibnamefont {Xu}}, \ and\
  \bibinfo {author} {\bibfnamefont {M.}~\bibnamefont {K{\"a}ll}},\ }\href
  {\doibase 10.1557/mrs.2014.2} {\bibfield  {journal} {\bibinfo  {journal} {MRS
  Bulletin}\ }\textbf {\bibinfo {volume} {39}},\ \bibinfo {pages} {163–168}
  (\bibinfo {year} {2014})}\BibitemShut {NoStop}%
\bibitem [{\citenamefont {Zrimsek}\ \emph {et~al.}(2017)\citenamefont
  {Zrimsek}, \citenamefont {Chiang}, \citenamefont {Mattei}, \citenamefont
  {Zaleski}, \citenamefont {McAnally}, \citenamefont {Chapman}, \citenamefont
  {Henry}, \citenamefont {Schatz},\ and\ \citenamefont
  {Van~Duyne}}]{zrimsek2017singlemol}%
  \BibitemOpen
  \bibfield  {author} {\bibinfo {author} {\bibfnamefont {A.~B.}\ \bibnamefont
  {Zrimsek}}, \bibinfo {author} {\bibfnamefont {N.}~\bibnamefont {Chiang}},
  \bibinfo {author} {\bibfnamefont {M.}~\bibnamefont {Mattei}}, \bibinfo
  {author} {\bibfnamefont {S.}~\bibnamefont {Zaleski}}, \bibinfo {author}
  {\bibfnamefont {M.~O.}\ \bibnamefont {McAnally}}, \bibinfo {author}
  {\bibfnamefont {C.~T.}\ \bibnamefont {Chapman}}, \bibinfo {author}
  {\bibfnamefont {A.-I.}\ \bibnamefont {Henry}}, \bibinfo {author}
  {\bibfnamefont {G.~C.}\ \bibnamefont {Schatz}}, \ and\ \bibinfo {author}
  {\bibfnamefont {R.~P.}\ \bibnamefont {Van~Duyne}},\ }\href {\doibase
  10.1021/acs.chemrev.6b00552} {\bibfield  {journal} {\bibinfo  {journal}
  {Chem. Rev.}\ }\textbf {\bibinfo {volume} {117}},\ \bibinfo {pages} {7583}
  (\bibinfo {year} {2017})}\BibitemShut {NoStop}%
\bibitem [{\citenamefont {Zhang}\ \emph {et~al.}(2013)\citenamefont {Zhang},
  \citenamefont {Zhang}, \citenamefont {Dong}, \citenamefont {Jiang},
  \citenamefont {Zhang}, \citenamefont {Chen}, \citenamefont {Zhang},
  \citenamefont {Liao}, \citenamefont {Aizpurua}, \citenamefont {Luo},
  \citenamefont {Yang},\ and\ \citenamefont {Hou}}]{zhang2013chemical}%
  \BibitemOpen
  \bibfield  {author} {\bibinfo {author} {\bibfnamefont {R.}~\bibnamefont
  {Zhang}}, \bibinfo {author} {\bibfnamefont {Y.}~\bibnamefont {Zhang}},
  \bibinfo {author} {\bibfnamefont {Z.}~\bibnamefont {Dong}}, \bibinfo {author}
  {\bibfnamefont {S.}~\bibnamefont {Jiang}}, \bibinfo {author} {\bibfnamefont
  {C.}~\bibnamefont {Zhang}}, \bibinfo {author} {\bibfnamefont
  {L.}~\bibnamefont {Chen}}, \bibinfo {author} {\bibfnamefont {L.}~\bibnamefont
  {Zhang}}, \bibinfo {author} {\bibfnamefont {Y.}~\bibnamefont {Liao}},
  \bibinfo {author} {\bibfnamefont {J.}~\bibnamefont {Aizpurua}}, \bibinfo
  {author} {\bibfnamefont {Y.}~\bibnamefont {Luo}}, \bibinfo {author}
  {\bibfnamefont {J.~L.}\ \bibnamefont {Yang}}, \ and\ \bibinfo {author}
  {\bibfnamefont {J.~G.}\ \bibnamefont {Hou}},\ }\href@noop {} {\bibfield
  {journal} {\bibinfo  {journal} {Nature}\ }\textbf {\bibinfo {volume} {498}},\
  \bibinfo {pages} {82} (\bibinfo {year} {2013})}\BibitemShut {NoStop}%
\bibitem [{\citenamefont {Chikkaraddy}\ \emph {et~al.}(2016)\citenamefont
  {Chikkaraddy}, \citenamefont {de~Nijs}, \citenamefont {Benz}, \citenamefont
  {Barrow}, \citenamefont {Scherman}, \citenamefont {Rosta}, \citenamefont
  {Demetriadou}, \citenamefont {Fox}, \citenamefont {Hess},\ and\ \citenamefont
  {Baumberg}}]{chikkaraddy2016single}%
  \BibitemOpen
  \bibfield  {author} {\bibinfo {author} {\bibfnamefont {R.}~\bibnamefont
  {Chikkaraddy}}, \bibinfo {author} {\bibfnamefont {B.}~\bibnamefont
  {de~Nijs}}, \bibinfo {author} {\bibfnamefont {F.}~\bibnamefont {Benz}},
  \bibinfo {author} {\bibfnamefont {S.~J.}\ \bibnamefont {Barrow}}, \bibinfo
  {author} {\bibfnamefont {O.~A.}\ \bibnamefont {Scherman}}, \bibinfo {author}
  {\bibfnamefont {E.}~\bibnamefont {Rosta}}, \bibinfo {author} {\bibfnamefont
  {A.}~\bibnamefont {Demetriadou}}, \bibinfo {author} {\bibfnamefont
  {P.}~\bibnamefont {Fox}}, \bibinfo {author} {\bibfnamefont {O.}~\bibnamefont
  {Hess}}, \ and\ \bibinfo {author} {\bibfnamefont {J.~J.}\ \bibnamefont
  {Baumberg}},\ }\href@noop {} {\bibfield  {journal} {\bibinfo  {journal}
  {Nature}\ }\textbf {\bibinfo {volume} {535}},\ \bibinfo {pages} {127}
  (\bibinfo {year} {2016})}\BibitemShut {NoStop}%
\bibitem [{\citenamefont {Doppagne}\ \emph {et~al.}(2017)\citenamefont
  {Doppagne}, \citenamefont {Chong}, \citenamefont {Lorchat}, \citenamefont
  {Berciaud}, \citenamefont {Romeo}, \citenamefont {Bulou}, \citenamefont
  {Boeglin}, \citenamefont {Scheurer},\ and\ \citenamefont
  {Schull}}]{schull2017vibmapping}%
  \BibitemOpen
  \bibfield  {author} {\bibinfo {author} {\bibfnamefont {B.}~\bibnamefont
  {Doppagne}}, \bibinfo {author} {\bibfnamefont {M.~C.}\ \bibnamefont {Chong}},
  \bibinfo {author} {\bibfnamefont {E.}~\bibnamefont {Lorchat}}, \bibinfo
  {author} {\bibfnamefont {S.}~\bibnamefont {Berciaud}}, \bibinfo {author}
  {\bibfnamefont {M.}~\bibnamefont {Romeo}}, \bibinfo {author} {\bibfnamefont
  {H.}~\bibnamefont {Bulou}}, \bibinfo {author} {\bibfnamefont
  {A.}~\bibnamefont {Boeglin}}, \bibinfo {author} {\bibfnamefont
  {F.}~\bibnamefont {Scheurer}}, \ and\ \bibinfo {author} {\bibfnamefont
  {G.}~\bibnamefont {Schull}},\ }\href {\doibase
  10.1103/PhysRevLett.118.127401} {\bibfield  {journal} {\bibinfo  {journal}
  {Phys. Rev. Lett.}\ }\textbf {\bibinfo {volume} {118}},\ \bibinfo {pages}
  {127401} (\bibinfo {year} {2017})}\BibitemShut {NoStop}%
\bibitem [{\citenamefont {Lee}\ \emph {et~al.}(2019)\citenamefont {Lee},
  \citenamefont {Crampton}, \citenamefont {Tallarida},\ and\ \citenamefont
  {Apkarian}}]{lee2019visualizing}%
  \BibitemOpen
  \bibfield  {author} {\bibinfo {author} {\bibfnamefont {J.}~\bibnamefont
  {Lee}}, \bibinfo {author} {\bibfnamefont {K.~T.}\ \bibnamefont {Crampton}},
  \bibinfo {author} {\bibfnamefont {N.}~\bibnamefont {Tallarida}}, \ and\
  \bibinfo {author} {\bibfnamefont {V.~A.}\ \bibnamefont {Apkarian}},\
  }\href@noop {} {\bibfield  {journal} {\bibinfo  {journal} {Nature}\ }\textbf
  {\bibinfo {volume} {568}},\ \bibinfo {pages} {78} (\bibinfo {year}
  {2019})}\BibitemShut {NoStop}%
\bibitem [{\citenamefont {Stipe}\ \emph {et~al.}(1997)\citenamefont {Stipe},
  \citenamefont {Rezaei}, \citenamefont {Ho}, \citenamefont {Gao},
  \citenamefont {Persson},\ and\ \citenamefont {Lundqvist}}]{stipe1997single}%
  \BibitemOpen
  \bibfield  {author} {\bibinfo {author} {\bibfnamefont {B.~C.}\ \bibnamefont
  {Stipe}}, \bibinfo {author} {\bibfnamefont {M.~A.}\ \bibnamefont {Rezaei}},
  \bibinfo {author} {\bibfnamefont {W.}~\bibnamefont {Ho}}, \bibinfo {author}
  {\bibfnamefont {S.}~\bibnamefont {Gao}}, \bibinfo {author} {\bibfnamefont
  {M.}~\bibnamefont {Persson}}, \ and\ \bibinfo {author} {\bibfnamefont
  {B.~I.}\ \bibnamefont {Lundqvist}},\ }\href@noop {} {\bibfield  {journal}
  {\bibinfo  {journal} {Phys. Rev. Lett.}\ }\textbf {\bibinfo {volume} {78}},\
  \bibinfo {pages} {4410} (\bibinfo {year} {1997})}\BibitemShut {NoStop}%
\bibitem [{\citenamefont {Komeda}(2005)}]{komeda2005chemical}%
  \BibitemOpen
  \bibfield  {author} {\bibinfo {author} {\bibfnamefont {T.}~\bibnamefont
  {Komeda}},\ }\href@noop {} {\bibfield  {journal} {\bibinfo  {journal} {Prog.
  Surf. Sci.}\ }\textbf {\bibinfo {volume} {78}},\ \bibinfo {pages} {41}
  (\bibinfo {year} {2005})}\BibitemShut {NoStop}%
\bibitem [{\citenamefont {Crim}(1996)}]{crim1996bond}%
  \BibitemOpen
  \bibfield  {author} {\bibinfo {author} {\bibfnamefont {F.~F.}\ \bibnamefont
  {Crim}},\ }\href@noop {} {\bibfield  {journal} {\bibinfo  {journal} {J. Phys.
  Chem.}\ }\textbf {\bibinfo {volume} {100}},\ \bibinfo {pages} {12725}
  (\bibinfo {year} {1996})}\BibitemShut {NoStop}%
\bibitem [{\citenamefont {Ho}(2002)}]{ho2002single}%
  \BibitemOpen
  \bibfield  {author} {\bibinfo {author} {\bibfnamefont {W.}~\bibnamefont
  {Ho}},\ }\href@noop {} {\bibfield  {journal} {\bibinfo  {journal} {J. Chem.
  Phys.}\ }\textbf {\bibinfo {volume} {117}},\ \bibinfo {pages} {11033}
  (\bibinfo {year} {2002})}\BibitemShut {NoStop}%
\bibitem [{\citenamefont {Pascual}\ \emph {et~al.}(2003)\citenamefont
  {Pascual}, \citenamefont {Lorente}, \citenamefont {Song}, \citenamefont
  {Conrad},\ and\ \citenamefont {Rust}}]{pascual2003selectivity}%
  \BibitemOpen
  \bibfield  {author} {\bibinfo {author} {\bibfnamefont {J.~I.}\ \bibnamefont
  {Pascual}}, \bibinfo {author} {\bibfnamefont {N.}~\bibnamefont {Lorente}},
  \bibinfo {author} {\bibfnamefont {Z.}~\bibnamefont {Song}}, \bibinfo {author}
  {\bibfnamefont {H.}~\bibnamefont {Conrad}}, \ and\ \bibinfo {author}
  {\bibfnamefont {H.-P.}\ \bibnamefont {Rust}},\ }\href@noop {} {\bibfield
  {journal} {\bibinfo  {journal} {Nature}\ }\textbf {\bibinfo {volume} {423}},\
  \bibinfo {pages} {525} (\bibinfo {year} {2003})}\BibitemShut {NoStop}%
\bibitem [{\citenamefont {Hahna}\ and\ \citenamefont
  {Ho}(2005)}]{hahna2005orbital}%
  \BibitemOpen
  \bibfield  {author} {\bibinfo {author} {\bibfnamefont {J.}~\bibnamefont
  {Hahna}}\ and\ \bibinfo {author} {\bibfnamefont {W.}~\bibnamefont {Ho}},\
  }\href@noop {} {\bibfield  {journal} {\bibinfo  {journal} {J. Chem. Phys.}\
  }\textbf {\bibinfo {volume} {11}},\ \bibinfo {pages} {16} (\bibinfo {year}
  {2005})}\BibitemShut {NoStop}%
\bibitem [{\citenamefont {Herrera}\ and\ \citenamefont
  {Spano}(2016)}]{herrera2016chemistry}%
  \BibitemOpen
  \bibfield  {author} {\bibinfo {author} {\bibfnamefont {F.}~\bibnamefont
  {Herrera}}\ and\ \bibinfo {author} {\bibfnamefont {F.~C.}\ \bibnamefont
  {Spano}},\ }\href {\doibase 10.1103/PhysRevLett.116.238301} {\bibfield
  {journal} {\bibinfo  {journal} {Phys. Rev. Lett.}\ }\textbf {\bibinfo
  {volume} {116}},\ \bibinfo {pages} {238301} (\bibinfo {year}
  {2016})}\BibitemShut {NoStop}%
\bibitem [{\citenamefont {Roelli}\ \emph {et~al.}(2015)\citenamefont {Roelli},
  \citenamefont {Galland}, \citenamefont {Piro},\ and\ \citenamefont
  {Kippenberg}}]{roelli2015molecular}%
  \BibitemOpen
  \bibfield  {author} {\bibinfo {author} {\bibfnamefont {P.}~\bibnamefont
  {Roelli}}, \bibinfo {author} {\bibfnamefont {C.}~\bibnamefont {Galland}},
  \bibinfo {author} {\bibfnamefont {N.}~\bibnamefont {Piro}}, \ and\ \bibinfo
  {author} {\bibfnamefont {T.~J.}\ \bibnamefont {Kippenberg}},\ }\href@noop {}
  {\bibfield  {journal} {\bibinfo  {journal} {Nat. Nanotechnol.}\ }\textbf
  {\bibinfo {volume} {11}},\ \bibinfo {pages} {164} (\bibinfo {year}
  {2015})}\BibitemShut {NoStop}%
\bibitem [{\citenamefont {Schmidt}\ \emph {et~al.}(2016)\citenamefont
  {Schmidt}, \citenamefont {Esteban}, \citenamefont {Gonz{\'a}lez-Tudela},
  \citenamefont {Giedke},\ and\ \citenamefont {Aizpurua}}]{schmidt2015qed}%
  \BibitemOpen
  \bibfield  {author} {\bibinfo {author} {\bibfnamefont {M.~K.}\ \bibnamefont
  {Schmidt}}, \bibinfo {author} {\bibfnamefont {R.}~\bibnamefont {Esteban}},
  \bibinfo {author} {\bibfnamefont {A.}~\bibnamefont {Gonz{\'a}lez-Tudela}},
  \bibinfo {author} {\bibfnamefont {G.}~\bibnamefont {Giedke}}, \ and\ \bibinfo
  {author} {\bibfnamefont {J.}~\bibnamefont {Aizpurua}},\ }\href@noop {}
  {\bibfield  {journal} {\bibinfo  {journal} {ACS Nano}\ }\textbf {\bibinfo
  {volume} {10}},\ \bibinfo {pages} {6291} (\bibinfo {year}
  {2016})}\BibitemShut {NoStop}%
\bibitem [{\citenamefont {Schmidt}\ \emph {et~al.}(2017)\citenamefont
  {Schmidt}, \citenamefont {Esteban}, \citenamefont {Benz}, \citenamefont
  {Baumberg},\ and\ \citenamefont {Aizpurua}}]{schmidt2017linking}%
  \BibitemOpen
  \bibfield  {author} {\bibinfo {author} {\bibfnamefont {M.~K.}\ \bibnamefont
  {Schmidt}}, \bibinfo {author} {\bibfnamefont {R.}~\bibnamefont {Esteban}},
  \bibinfo {author} {\bibfnamefont {F.}~\bibnamefont {Benz}}, \bibinfo {author}
  {\bibfnamefont {J.~J.}\ \bibnamefont {Baumberg}}, \ and\ \bibinfo {author}
  {\bibfnamefont {J.}~\bibnamefont {Aizpurua}},\ }\href@noop {} {\bibfield
  {journal} {\bibinfo  {journal} {Faraday Discuss.}\ }\textbf {\bibinfo
  {volume} {205}},\ \bibinfo {pages} {31} (\bibinfo {year} {2017})}\BibitemShut
  {NoStop}%
\bibitem [{\citenamefont {Kamandar~Dezfouli}\ and\ \citenamefont
  {Hughes}(2017)}]{kamandar2017quantum}%
  \BibitemOpen
  \bibfield  {author} {\bibinfo {author} {\bibfnamefont {M.}~\bibnamefont
  {Kamandar~Dezfouli}}\ and\ \bibinfo {author} {\bibfnamefont {S.}~\bibnamefont
  {Hughes}},\ }\href@noop {} {\bibfield  {journal} {\bibinfo  {journal} {ACS
  Photonics}\ }\textbf {\bibinfo {volume} {4}},\ \bibinfo {pages} {1245}
  (\bibinfo {year} {2017})}\BibitemShut {NoStop}%
\bibitem [{\citenamefont {Benz}\ \emph {et~al.}(2016)\citenamefont {Benz},
  \citenamefont {Schmidt}, \citenamefont {Dreismann}, \citenamefont
  {Chikkaraddy}, \citenamefont {Zhang}, \citenamefont {Demetriadou},
  \citenamefont {Carnegie}, \citenamefont {Ohadi}, \citenamefont {de~Nijs},
  \citenamefont {Esteban}, \citenamefont {Aizpurua},\ and\ \citenamefont
  {Baumberg}}]{benz2016single}%
  \BibitemOpen
  \bibfield  {author} {\bibinfo {author} {\bibfnamefont {F.}~\bibnamefont
  {Benz}}, \bibinfo {author} {\bibfnamefont {M.~K.}\ \bibnamefont {Schmidt}},
  \bibinfo {author} {\bibfnamefont {A.}~\bibnamefont {Dreismann}}, \bibinfo
  {author} {\bibfnamefont {R.}~\bibnamefont {Chikkaraddy}}, \bibinfo {author}
  {\bibfnamefont {Y.}~\bibnamefont {Zhang}}, \bibinfo {author} {\bibfnamefont
  {A.}~\bibnamefont {Demetriadou}}, \bibinfo {author} {\bibfnamefont
  {C.}~\bibnamefont {Carnegie}}, \bibinfo {author} {\bibfnamefont
  {H.}~\bibnamefont {Ohadi}}, \bibinfo {author} {\bibfnamefont
  {B.}~\bibnamefont {de~Nijs}}, \bibinfo {author} {\bibfnamefont
  {R.}~\bibnamefont {Esteban}}, \bibinfo {author} {\bibfnamefont
  {J.}~\bibnamefont {Aizpurua}}, \ and\ \bibinfo {author} {\bibfnamefont
  {J.}~\bibnamefont {Baumberg}},\ }\href@noop {} {\bibfield  {journal}
  {\bibinfo  {journal} {Science}\ }\textbf {\bibinfo {volume} {354}},\ \bibinfo
  {pages} {726} (\bibinfo {year} {2016})}\BibitemShut {NoStop}%
\bibitem [{\citenamefont {Aspelmeyer}\ \emph {et~al.}(2014)\citenamefont
  {Aspelmeyer}, \citenamefont {Kippenberg},\ and\ \citenamefont
  {Marquardt}}]{aspelmayer2014}%
  \BibitemOpen
  \bibfield  {author} {\bibinfo {author} {\bibfnamefont {M.}~\bibnamefont
  {Aspelmeyer}}, \bibinfo {author} {\bibfnamefont {T.~J.}\ \bibnamefont
  {Kippenberg}}, \ and\ \bibinfo {author} {\bibfnamefont {F.}~\bibnamefont
  {Marquardt}},\ }\href {\doibase 10.1103/RevModPhys.86.1391} {\bibfield
  {journal} {\bibinfo  {journal} {Rev. Mod. Phys.}\ }\textbf {\bibinfo {volume}
  {86}},\ \bibinfo {pages} {1391} (\bibinfo {year} {2014})}\BibitemShut
  {NoStop}%
\bibitem [{\citenamefont {Xu}\ \emph {et~al.}(2004)\citenamefont {Xu},
  \citenamefont {Wang}, \citenamefont {Persson}, \citenamefont {Xu},
  \citenamefont {K\"all},\ and\ \citenamefont {Johansson}}]{johansson2004prl}%
  \BibitemOpen
  \bibfield  {author} {\bibinfo {author} {\bibfnamefont {H.}~\bibnamefont
  {Xu}}, \bibinfo {author} {\bibfnamefont {X.-H.}\ \bibnamefont {Wang}},
  \bibinfo {author} {\bibfnamefont {M.~P.}\ \bibnamefont {Persson}}, \bibinfo
  {author} {\bibfnamefont {H.~Q.}\ \bibnamefont {Xu}}, \bibinfo {author}
  {\bibfnamefont {M.}~\bibnamefont {K\"all}}, \ and\ \bibinfo {author}
  {\bibfnamefont {P.}~\bibnamefont {Johansson}},\ }\href {\doibase
  10.1103/PhysRevLett.93.243002} {\bibfield  {journal} {\bibinfo  {journal}
  {Phys. Rev. Lett.}\ }\textbf {\bibinfo {volume} {93}},\ \bibinfo {pages}
  {243002} (\bibinfo {year} {2004})}\BibitemShut {NoStop}%
\bibitem [{\citenamefont {Johansson}\ \emph {et~al.}(2005)\citenamefont
  {Johansson}, \citenamefont {Xu},\ and\ \citenamefont
  {K{\"a}ll}}]{johansson2005surface}%
  \BibitemOpen
  \bibfield  {author} {\bibinfo {author} {\bibfnamefont {P.}~\bibnamefont
  {Johansson}}, \bibinfo {author} {\bibfnamefont {H.}~\bibnamefont {Xu}}, \
  and\ \bibinfo {author} {\bibfnamefont {M.}~\bibnamefont {K{\"a}ll}},\
  }\href@noop {} {\bibfield  {journal} {\bibinfo  {journal} {Phys. Rev. B}\
  }\textbf {\bibinfo {volume} {72}},\ \bibinfo {pages} {035427} (\bibinfo
  {year} {2005})}\BibitemShut {NoStop}%
\bibitem [{\citenamefont {del Valle}\ \emph {et~al.}(2009)\citenamefont {del
  Valle}, \citenamefont {Laussy},\ and\ \citenamefont {Tejedor}}]{delValle09}%
  \BibitemOpen
  \bibfield  {author} {\bibinfo {author} {\bibfnamefont {E.}~\bibnamefont {del
  Valle}}, \bibinfo {author} {\bibfnamefont {F.~P.}\ \bibnamefont {Laussy}}, \
  and\ \bibinfo {author} {\bibfnamefont {C.}~\bibnamefont {Tejedor}},\ }\href
  {\doibase 10.1103/PhysRevB.79.235326} {\bibfield  {journal} {\bibinfo
  {journal} {Phys. Rev. B}\ }\textbf {\bibinfo {volume} {79}},\ \bibinfo
  {pages} {235326} (\bibinfo {year} {2009})}\BibitemShut {NoStop}%
\bibitem [{\citenamefont {Delga}\ \emph
  {et~al.}(2014{\natexlab{a}})\citenamefont {Delga}, \citenamefont {Feist},
  \citenamefont {Bravo-Abad},\ and\ \citenamefont
  {Garcia-Vidal}}]{Garcia-VidalPRL2014}%
  \BibitemOpen
  \bibfield  {author} {\bibinfo {author} {\bibfnamefont {A.}~\bibnamefont
  {Delga}}, \bibinfo {author} {\bibfnamefont {J.}~\bibnamefont {Feist}},
  \bibinfo {author} {\bibfnamefont {J.}~\bibnamefont {Bravo-Abad}}, \ and\
  \bibinfo {author} {\bibfnamefont {F.~J.}\ \bibnamefont {Garcia-Vidal}},\
  }\href@noop {} {\bibfield  {journal} {\bibinfo  {journal} {Phys. Rev. Lett.}\
  }\textbf {\bibinfo {volume} {112}},\ \bibinfo {pages} {253601} (\bibinfo
  {year} {2014}{\natexlab{a}})}\BibitemShut {NoStop}%
\bibitem [{\citenamefont {Gu}\ \emph {et~al.}(2010)\citenamefont {Gu},
  \citenamefont {Huang}, \citenamefont {Martin},\ and\ \citenamefont
  {Gong}}]{gu2010resonance}%
  \BibitemOpen
  \bibfield  {author} {\bibinfo {author} {\bibfnamefont {Y.}~\bibnamefont
  {Gu}}, \bibinfo {author} {\bibfnamefont {L.}~\bibnamefont {Huang}}, \bibinfo
  {author} {\bibfnamefont {O.~J.~F.}\ \bibnamefont {Martin}}, \ and\ \bibinfo
  {author} {\bibfnamefont {Q.}~\bibnamefont {Gong}},\ }\href@noop {} {\bibfield
   {journal} {\bibinfo  {journal} {Phys. Rev. B}\ }\textbf {\bibinfo {volume}
  {81}},\ \bibinfo {pages} {193103} (\bibinfo {year} {2010})}\BibitemShut
  {NoStop}%
\bibitem [{\citenamefont {Tr{\"u}gler}\ and\ \citenamefont
  {Hohenester}(2008)}]{trugler2008strong}%
  \BibitemOpen
  \bibfield  {author} {\bibinfo {author} {\bibfnamefont {A.}~\bibnamefont
  {Tr{\"u}gler}}\ and\ \bibinfo {author} {\bibfnamefont {U.}~\bibnamefont
  {Hohenester}},\ }\href@noop {} {\bibfield  {journal} {\bibinfo  {journal}
  {Phys. Rev. B}\ }\textbf {\bibinfo {volume} {77}},\ \bibinfo {pages} {115403}
  (\bibinfo {year} {2008})}\BibitemShut {NoStop}%
\bibitem [{\citenamefont {Ridolfo}\ \emph {et~al.}(2010)\citenamefont
  {Ridolfo}, \citenamefont {Di~Stefano}, \citenamefont {Fina}, \citenamefont
  {Saija},\ and\ \citenamefont {Savasta}}]{ridolfo2010quantum}%
  \BibitemOpen
  \bibfield  {author} {\bibinfo {author} {\bibfnamefont {A.}~\bibnamefont
  {Ridolfo}}, \bibinfo {author} {\bibfnamefont {O.}~\bibnamefont {Di~Stefano}},
  \bibinfo {author} {\bibfnamefont {N.}~\bibnamefont {Fina}}, \bibinfo {author}
  {\bibfnamefont {R.}~\bibnamefont {Saija}}, \ and\ \bibinfo {author}
  {\bibfnamefont {S.}~\bibnamefont {Savasta}},\ }\href@noop {} {\bibfield
  {journal} {\bibinfo  {journal} {Phys. Rev. Lett.}\ }\textbf {\bibinfo
  {volume} {105}},\ \bibinfo {pages} {263601} (\bibinfo {year}
  {2010})}\BibitemShut {NoStop}%
\bibitem [{\citenamefont {Ramos}\ \emph {et~al.}(2013)\citenamefont {Ramos},
  \citenamefont {Sudhir}, \citenamefont {Stannigel}, \citenamefont {Zoller},\
  and\ \citenamefont {Kippenberg}}]{ramos2013nonlinear}%
  \BibitemOpen
  \bibfield  {author} {\bibinfo {author} {\bibfnamefont {T.}~\bibnamefont
  {Ramos}}, \bibinfo {author} {\bibfnamefont {V.}~\bibnamefont {Sudhir}},
  \bibinfo {author} {\bibfnamefont {K.}~\bibnamefont {Stannigel}}, \bibinfo
  {author} {\bibfnamefont {P.}~\bibnamefont {Zoller}}, \ and\ \bibinfo {author}
  {\bibfnamefont {T.~J.}\ \bibnamefont {Kippenberg}},\ }\href@noop {}
  {\bibfield  {journal} {\bibinfo  {journal} {Phys. Rev. Lett.}\ }\textbf
  {\bibinfo {volume} {110}},\ \bibinfo {pages} {193602} (\bibinfo {year}
  {2013})}\BibitemShut {NoStop}%
\bibitem [{\citenamefont {Akram}\ \emph {et~al.}(2015)\citenamefont {Akram},
  \citenamefont {Ghafoor},\ and\ \citenamefont {Saif}}]{akram2015fanohybrid}%
  \BibitemOpen
  \bibfield  {author} {\bibinfo {author} {\bibfnamefont {M.~J.}\ \bibnamefont
  {Akram}}, \bibinfo {author} {\bibfnamefont {F.}~\bibnamefont {Ghafoor}}, \
  and\ \bibinfo {author} {\bibfnamefont {F.}~\bibnamefont {Saif}},\ }\href
  {http://stacks.iop.org/0953-4075/48/i=6/a=065502} {\bibfield  {journal}
  {\bibinfo  {journal} {J. Phys. B}\ }\textbf {\bibinfo {volume} {48}},\
  \bibinfo {pages} {065502} (\bibinfo {year} {2015})}\BibitemShut {NoStop}%
\bibitem [{\citenamefont {Nunnenkamp}\ \emph {et~al.}(2011)\citenamefont
  {Nunnenkamp}, \citenamefont {B\o{}rkje},\ and\ \citenamefont
  {Girvin}}]{Nunnenkamp2011spoptomech}%
  \BibitemOpen
  \bibfield  {author} {\bibinfo {author} {\bibfnamefont {A.}~\bibnamefont
  {Nunnenkamp}}, \bibinfo {author} {\bibfnamefont {K.}~\bibnamefont
  {B\o{}rkje}}, \ and\ \bibinfo {author} {\bibfnamefont {S.~M.}\ \bibnamefont
  {Girvin}},\ }\href {\doibase 10.1103/PhysRevLett.107.063602} {\bibfield
  {journal} {\bibinfo  {journal} {Phys. Rev. Lett.}\ }\textbf {\bibinfo
  {volume} {107}},\ \bibinfo {pages} {063602} (\bibinfo {year}
  {2011})}\BibitemShut {NoStop}%
\bibitem [{\citenamefont {Nunnenkamp}\ \emph {et~al.}(2012)\citenamefont
  {Nunnenkamp}, \citenamefont {B\o{}rkje},\ and\ \citenamefont
  {Girvin}}]{Nunnenkamp2012sccooling}%
  \BibitemOpen
  \bibfield  {author} {\bibinfo {author} {\bibfnamefont {A.}~\bibnamefont
  {Nunnenkamp}}, \bibinfo {author} {\bibfnamefont {K.}~\bibnamefont
  {B\o{}rkje}}, \ and\ \bibinfo {author} {\bibfnamefont {S.~M.}\ \bibnamefont
  {Girvin}},\ }\href {\doibase 10.1103/PhysRevA.85.051803} {\bibfield
  {journal} {\bibinfo  {journal} {Phys. Rev. A}\ }\textbf {\bibinfo {volume}
  {85}},\ \bibinfo {pages} {051803(R)} (\bibinfo {year} {2012})}\BibitemShut
  {NoStop}%
\bibitem [{\citenamefont {Li}\ \emph {et~al.}(1994)\citenamefont {Li},
  \citenamefont {Johnson}, \citenamefont {Mukamel},\ and\ \citenamefont
  {Myers}}]{li1994overdampedbrownian}%
  \BibitemOpen
  \bibfield  {author} {\bibinfo {author} {\bibfnamefont {B.}~\bibnamefont
  {Li}}, \bibinfo {author} {\bibfnamefont {A.~E.}\ \bibnamefont {Johnson}},
  \bibinfo {author} {\bibfnamefont {S.}~\bibnamefont {Mukamel}}, \ and\
  \bibinfo {author} {\bibfnamefont {A.~B.}\ \bibnamefont {Myers}},\ }\href
  {\doibase 10.1021/ja00103a020} {\bibfield  {journal} {\bibinfo  {journal} {J.
  Am. Chem. Soc.}\ }\textbf {\bibinfo {volume} {116}},\ \bibinfo {pages}
  {11039} (\bibinfo {year} {1994})}\BibitemShut {NoStop}%
\bibitem [{\citenamefont {Herrera}\ and\ \citenamefont
  {Spano}(2017)}]{herrera2017absem}%
  \BibitemOpen
  \bibfield  {author} {\bibinfo {author} {\bibfnamefont {F.}~\bibnamefont
  {Herrera}}\ and\ \bibinfo {author} {\bibfnamefont {F.~C.}\ \bibnamefont
  {Spano}},\ }\href {\doibase 10.1103/PhysRevA.95.053867} {\bibfield  {journal}
  {\bibinfo  {journal} {Phys. Rev. A}\ }\textbf {\bibinfo {volume} {95}},\
  \bibinfo {pages} {053867} (\bibinfo {year} {2017})}\BibitemShut {NoStop}%
\bibitem [{\citenamefont {Galperin}\ \emph {et~al.}(2009)\citenamefont
  {Galperin}, \citenamefont {Ratner},\ and\ \citenamefont
  {Nitzan}}]{galperin2009raman}%
  \BibitemOpen
  \bibfield  {author} {\bibinfo {author} {\bibfnamefont {M.}~\bibnamefont
  {Galperin}}, \bibinfo {author} {\bibfnamefont {M.~A.}\ \bibnamefont
  {Ratner}}, \ and\ \bibinfo {author} {\bibfnamefont {A.}~\bibnamefont
  {Nitzan}},\ }\href@noop {} {\bibfield  {journal} {\bibinfo  {journal} {J.
  Chem. Phys.}\ }\textbf {\bibinfo {volume} {130}},\ \bibinfo {pages} {144109}
  (\bibinfo {year} {2009})}\BibitemShut {NoStop}%
\bibitem [{\citenamefont {{\'C}wik}\ \emph {et~al.}(2016)\citenamefont
  {{\'C}wik}, \citenamefont {Kirton}, \citenamefont {De~Liberato},\ and\
  \citenamefont {Keeling}}]{cwik2016excitonic}%
  \BibitemOpen
  \bibfield  {author} {\bibinfo {author} {\bibfnamefont {J.~A.}\ \bibnamefont
  {{\'C}wik}}, \bibinfo {author} {\bibfnamefont {P.}~\bibnamefont {Kirton}},
  \bibinfo {author} {\bibfnamefont {S.}~\bibnamefont {De~Liberato}}, \ and\
  \bibinfo {author} {\bibfnamefont {J.}~\bibnamefont {Keeling}},\ }\href@noop
  {} {\bibfield  {journal} {\bibinfo  {journal} {Phys. Rev. A}\ }\textbf
  {\bibinfo {volume} {93}},\ \bibinfo {pages} {033840} (\bibinfo {year}
  {2016})}\BibitemShut {NoStop}%
\bibitem [{\citenamefont {Zeb}\ \emph {et~al.}(2018)\citenamefont {Zeb},
  \citenamefont {Kirton},\ and\ \citenamefont
  {Keeling}}]{zeb2018exactvibdressed}%
  \BibitemOpen
  \bibfield  {author} {\bibinfo {author} {\bibfnamefont {M.~A.}\ \bibnamefont
  {Zeb}}, \bibinfo {author} {\bibfnamefont {P.~G.}\ \bibnamefont {Kirton}}, \
  and\ \bibinfo {author} {\bibfnamefont {J.}~\bibnamefont {Keeling}},\ }\href
  {\doibase 10.1021/acsphotonics.7b00916} {\bibfield  {journal} {\bibinfo
  {journal} {ACS Photonics}\ }\textbf {\bibinfo {volume} {5}},\ \bibinfo
  {pages} {249} (\bibinfo {year} {2018})}\BibitemShut {NoStop}%
\bibitem [{\citenamefont {Klessinger}\ and\ \citenamefont
  {Michl}(1995)}]{klessinger1995excited}%
  \BibitemOpen
  \bibfield  {author} {\bibinfo {author} {\bibfnamefont {M.}~\bibnamefont
  {Klessinger}}\ and\ \bibinfo {author} {\bibfnamefont {J.}~\bibnamefont
  {Michl}},\ }\href {https://books.google.fr/books?id=0FeVPQAACAAJ} {\emph
  {\bibinfo {title} {Excited States and Photo-Chemistry of Organic
  Molecules}}}\ (\bibinfo  {publisher} {Wiley},\ \bibinfo {year}
  {1995})\BibitemShut {NoStop}%
\bibitem [{\citenamefont {May}\ and\ \citenamefont
  {K{\"u}hn}(2008)}]{may2008charge}%
  \BibitemOpen
  \bibfield  {author} {\bibinfo {author} {\bibfnamefont {V.}~\bibnamefont
  {May}}\ and\ \bibinfo {author} {\bibfnamefont {O.}~\bibnamefont {K{\"u}hn}},\
  }\href@noop {} {\emph {\bibinfo {title} {Charge and energy transfer dynamics
  in molecular systems}}}\ (\bibinfo  {publisher} {John Wiley \& Sons},\
  \bibinfo {year} {2008})\ p.~\bibinfo {pages} {15}\BibitemShut {NoStop}%
\bibitem [{\citenamefont {Galego}\ \emph {et~al.}(2015)\citenamefont {Galego},
  \citenamefont {Garcia-Vidal},\ and\ \citenamefont {Feist}}]{galego2015}%
  \BibitemOpen
  \bibfield  {author} {\bibinfo {author} {\bibfnamefont {J.}~\bibnamefont
  {Galego}}, \bibinfo {author} {\bibfnamefont {F.~J.}\ \bibnamefont
  {Garcia-Vidal}}, \ and\ \bibinfo {author} {\bibfnamefont {J.}~\bibnamefont
  {Feist}},\ }\href {\doibase 10.1103/PhysRevX.5.041022} {\bibfield  {journal}
  {\bibinfo  {journal} {Phys. Rev. X}\ }\textbf {\bibinfo {volume} {5}},\
  \bibinfo {pages} {041022} (\bibinfo {year} {2015})}\BibitemShut {NoStop}%
\bibitem [{\citenamefont {Galego}\ \emph {et~al.}(2016)\citenamefont {Galego},
  \citenamefont {Garcia-Vidal},\ and\ \citenamefont {Feist}}]{galego2016}%
  \BibitemOpen
  \bibfield  {author} {\bibinfo {author} {\bibfnamefont {J.}~\bibnamefont
  {Galego}}, \bibinfo {author} {\bibfnamefont {F.~J.}\ \bibnamefont
  {Garcia-Vidal}}, \ and\ \bibinfo {author} {\bibfnamefont {J.}~\bibnamefont
  {Feist}},\ }\href@noop {} {\bibfield  {journal} {\bibinfo  {journal} {Nature
  Communications}\ }\textbf {\bibinfo {volume} {7}} (\bibinfo {year}
  {2016})}\BibitemShut {NoStop}%
\bibitem [{\citenamefont {Betzholz}\ \emph {et~al.}(2014)\citenamefont
  {Betzholz}, \citenamefont {Torres},\ and\ \citenamefont
  {Bienert}}]{betzholz2014quantum}%
  \BibitemOpen
  \bibfield  {author} {\bibinfo {author} {\bibfnamefont {R.}~\bibnamefont
  {Betzholz}}, \bibinfo {author} {\bibfnamefont {J.~M.}\ \bibnamefont
  {Torres}}, \ and\ \bibinfo {author} {\bibfnamefont {M.}~\bibnamefont
  {Bienert}},\ }\href@noop {} {\bibfield  {journal} {\bibinfo  {journal} {Phys.
  Rev. A}\ }\textbf {\bibinfo {volume} {90}},\ \bibinfo {pages} {063818}
  (\bibinfo {year} {2014})}\BibitemShut {NoStop}%
\bibitem [{\citenamefont {Jaehne}\ \emph {et~al.}(2008)\citenamefont {Jaehne},
  \citenamefont {Hammerer},\ and\ \citenamefont
  {Wallquist}}]{jaehne2008ground}%
  \BibitemOpen
  \bibfield  {author} {\bibinfo {author} {\bibfnamefont {K.}~\bibnamefont
  {Jaehne}}, \bibinfo {author} {\bibfnamefont {K.}~\bibnamefont {Hammerer}}, \
  and\ \bibinfo {author} {\bibfnamefont {M.}~\bibnamefont {Wallquist}},\
  }\href@noop {} {\bibfield  {journal} {\bibinfo  {journal} {New J. Phys.}\
  }\textbf {\bibinfo {volume} {10}},\ \bibinfo {pages} {095019} (\bibinfo
  {year} {2008})}\BibitemShut {NoStop}%
\bibitem [{\citenamefont {del Pino}\ \emph {et~al.}(2015)\citenamefont {del
  Pino}, \citenamefont {Feist},\ and\ \citenamefont
  {Garcia-Vidal}}]{GarciaVidal2015}%
  \BibitemOpen
  \bibfield  {author} {\bibinfo {author} {\bibfnamefont {J.}~\bibnamefont {del
  Pino}}, \bibinfo {author} {\bibfnamefont {J.}~\bibnamefont {Feist}}, \ and\
  \bibinfo {author} {\bibfnamefont {F.~J.}\ \bibnamefont {Garcia-Vidal}},\
  }\href@noop {} {\bibfield  {journal} {\bibinfo  {journal} {Phys. Chem. C}\
  }\textbf {\bibinfo {volume} {119}},\ \bibinfo {pages} {29132} (\bibinfo
  {year} {2015})}\BibitemShut {NoStop}%
\bibitem [{\citenamefont {Akram}\ \emph {et~al.}(2010)\citenamefont {Akram},
  \citenamefont {Kiesel}, \citenamefont {Aspelmeyer},\ and\ \citenamefont
  {Milburn}}]{akram2010singlephoton}%
  \BibitemOpen
  \bibfield  {author} {\bibinfo {author} {\bibfnamefont {U.}~\bibnamefont
  {Akram}}, \bibinfo {author} {\bibfnamefont {N.}~\bibnamefont {Kiesel}},
  \bibinfo {author} {\bibfnamefont {M.}~\bibnamefont {Aspelmeyer}}, \ and\
  \bibinfo {author} {\bibfnamefont {G.~J.}\ \bibnamefont {Milburn}},\ }\href
  {http://stacks.iop.org/1367-2630/12/i=8/a=083030} {\bibfield  {journal}
  {\bibinfo  {journal} {New J. Phys}\ }\textbf {\bibinfo {volume} {12}},\
  \bibinfo {pages} {083030} (\bibinfo {year} {2010})}\BibitemShut {NoStop}%
\bibitem [{\citenamefont {Breuer}\ and\ \citenamefont
  {Petruccione}(2003)}]{Breuer2005}%
  \BibitemOpen
  \bibfield  {author} {\bibinfo {author} {\bibfnamefont {H.-P.}\ \bibnamefont
  {Breuer}}\ and\ \bibinfo {author} {\bibfnamefont {F.}~\bibnamefont
  {Petruccione}},\ }\href@noop {} {\emph {\bibinfo {title} {The theory of open
  quantum systems}}}\ (\bibinfo  {publisher} {Oxford {U}niversity {P}ress},\
  \bibinfo {year} {2003})\BibitemShut {NoStop}%
\bibitem [{\citenamefont {Esteban}\ \emph {et~al.}(2014)\citenamefont
  {Esteban}, \citenamefont {Aizpurua},\ and\ \citenamefont
  {Bryant}}]{esteban2014dephasing}%
  \BibitemOpen
  \bibfield  {author} {\bibinfo {author} {\bibfnamefont {R.}~\bibnamefont
  {Esteban}}, \bibinfo {author} {\bibfnamefont {J.}~\bibnamefont {Aizpurua}}, \
  and\ \bibinfo {author} {\bibfnamefont {G.~W.}\ \bibnamefont {Bryant}},\
  }\href {http://stacks.iop.org/1367-2630/16/i=1/a=013052} {\bibfield
  {journal} {\bibinfo  {journal} {New Journal of Physics}\ }\textbf {\bibinfo
  {volume} {16}},\ \bibinfo {pages} {013052} (\bibinfo {year}
  {2014})}\BibitemShut {NoStop}%
\bibitem [{\citenamefont {Neuman}\ and\ \citenamefont
  {Aizpurua}(2018)}]{neuman2018origin}%
  \BibitemOpen
  \bibfield  {author} {\bibinfo {author} {\bibfnamefont {T.}~\bibnamefont
  {Neuman}}\ and\ \bibinfo {author} {\bibfnamefont {J.}~\bibnamefont
  {Aizpurua}},\ }\href {\doibase 10.1364/OPTICA.5.001247} {\bibfield  {journal}
  {\bibinfo  {journal} {Optica}\ }\textbf {\bibinfo {volume} {5}},\ \bibinfo
  {pages} {1247} (\bibinfo {year} {2018})}\BibitemShut {NoStop}%
\bibitem [{\citenamefont {Wang}\ and\ \citenamefont
  {Shen}(2006)}]{shen2006qualityf}%
  \BibitemOpen
  \bibfield  {author} {\bibinfo {author} {\bibfnamefont {F.}~\bibnamefont
  {Wang}}\ and\ \bibinfo {author} {\bibfnamefont {Y.~R.}\ \bibnamefont
  {Shen}},\ }\href {\doibase 10.1103/PhysRevLett.97.206806} {\bibfield
  {journal} {\bibinfo  {journal} {Phys. Rev. Lett.}\ }\textbf {\bibinfo
  {volume} {97}},\ \bibinfo {pages} {206806} (\bibinfo {year}
  {2006})}\BibitemShut {NoStop}%
\bibitem [{\citenamefont {Delga}\ \emph
  {et~al.}(2014{\natexlab{b}})\citenamefont {Delga}, \citenamefont {Feist},
  \citenamefont {Bravo-Abad},\ and\ \citenamefont {Garcia-Vidal}}]{Delga2014}%
  \BibitemOpen
  \bibfield  {author} {\bibinfo {author} {\bibfnamefont {A.}~\bibnamefont
  {Delga}}, \bibinfo {author} {\bibfnamefont {J.}~\bibnamefont {Feist}},
  \bibinfo {author} {\bibfnamefont {J.}~\bibnamefont {Bravo-Abad}}, \ and\
  \bibinfo {author} {\bibfnamefont {F.~J.}\ \bibnamefont {Garcia-Vidal}},\
  }\href {\doibase 10.1103/PhysRevLett.112.253601} {\bibfield  {journal}
  {\bibinfo  {journal} {Phys. Rev. Lett.}\ }\textbf {\bibinfo {volume} {112}},\
  \bibinfo {pages} {253601} (\bibinfo {year} {2014}{\natexlab{b}})}\BibitemShut
  {NoStop}%
\bibitem [{\citenamefont {Kristensen}\ and\ \citenamefont
  {Hughes}(2013)}]{kristensen2013modes}%
  \BibitemOpen
  \bibfield  {author} {\bibinfo {author} {\bibfnamefont {P.~T.}\ \bibnamefont
  {Kristensen}}\ and\ \bibinfo {author} {\bibfnamefont {S.}~\bibnamefont
  {Hughes}},\ }\href@noop {} {\bibfield  {journal} {\bibinfo  {journal} {ACS
  Photonics}\ }\textbf {\bibinfo {volume} {1}},\ \bibinfo {pages} {2} (\bibinfo
  {year} {2013})}\BibitemShut {NoStop}%
\bibitem [{\citenamefont {Sauvan}\ \emph {et~al.}(2013)\citenamefont {Sauvan},
  \citenamefont {Hugonin}, \citenamefont {Maksymov},\ and\ \citenamefont
  {Lalanne}}]{Sauvan2013}%
  \BibitemOpen
  \bibfield  {author} {\bibinfo {author} {\bibfnamefont {C.}~\bibnamefont
  {Sauvan}}, \bibinfo {author} {\bibfnamefont {J.~P.}\ \bibnamefont {Hugonin}},
  \bibinfo {author} {\bibfnamefont {I.~S.}\ \bibnamefont {Maksymov}}, \ and\
  \bibinfo {author} {\bibfnamefont {P.}~\bibnamefont {Lalanne}},\ }\href
  {\doibase 10.1103/PhysRevLett.110.237401} {\bibfield  {journal} {\bibinfo
  {journal} {Phys. Rev. Lett.}\ }\textbf {\bibinfo {volume} {110}},\ \bibinfo
  {pages} {237401} (\bibinfo {year} {2013})}\BibitemShut {NoStop}%
\bibitem [{\citenamefont {Koenderink}(2010)}]{Koenderink10}%
  \BibitemOpen
  \bibfield  {author} {\bibinfo {author} {\bibfnamefont {A.~F.}\ \bibnamefont
  {Koenderink}},\ }\href {\doibase 10.1364/OL.35.004208} {\bibfield  {journal}
  {\bibinfo  {journal} {Opt. Lett.}\ }\textbf {\bibinfo {volume} {35}},\
  \bibinfo {pages} {4208} (\bibinfo {year} {2010})}\BibitemShut {NoStop}%
\bibitem [{\citenamefont {Zhang}\ \emph {et~al.}(2016)\citenamefont {Zhang},
  \citenamefont {Luo}, \citenamefont {Zhang}, \citenamefont {Yu}, \citenamefont
  {Kuang}, \citenamefont {Zhang}, \citenamefont {Meng}, \citenamefont {Luo},
  \citenamefont {Yang}, \citenamefont {Dong} \emph
  {et~al.}}]{zhang2016visualizing}%
  \BibitemOpen
  \bibfield  {author} {\bibinfo {author} {\bibfnamefont {Y.}~\bibnamefont
  {Zhang}}, \bibinfo {author} {\bibfnamefont {Y.}~\bibnamefont {Luo}}, \bibinfo
  {author} {\bibfnamefont {Y.}~\bibnamefont {Zhang}}, \bibinfo {author}
  {\bibfnamefont {Y.-J.}\ \bibnamefont {Yu}}, \bibinfo {author} {\bibfnamefont
  {Y.-M.}\ \bibnamefont {Kuang}}, \bibinfo {author} {\bibfnamefont
  {L.}~\bibnamefont {Zhang}}, \bibinfo {author} {\bibfnamefont {Q.-S.}\
  \bibnamefont {Meng}}, \bibinfo {author} {\bibfnamefont {Y.}~\bibnamefont
  {Luo}}, \bibinfo {author} {\bibfnamefont {J.-L.}\ \bibnamefont {Yang}},
  \bibinfo {author} {\bibfnamefont {Z.-C.}\ \bibnamefont {Dong}},  \emph
  {et~al.},\ }\href@noop {} {\bibfield  {journal} {\bibinfo  {journal}
  {Nature}\ }\textbf {\bibinfo {volume} {531}},\ \bibinfo {pages} {623}
  (\bibinfo {year} {2016})}\BibitemShut {NoStop}%
\bibitem [{\citenamefont {Imada}\ \emph {et~al.}(2017)\citenamefont {Imada},
  \citenamefont {Miwa}, \citenamefont {Imai-Imada}, \citenamefont {Kawahara},
  \citenamefont {Kimura},\ and\ \citenamefont {Kim}}]{imada2017fano}%
  \BibitemOpen
  \bibfield  {author} {\bibinfo {author} {\bibfnamefont {H.}~\bibnamefont
  {Imada}}, \bibinfo {author} {\bibfnamefont {K.}~\bibnamefont {Miwa}},
  \bibinfo {author} {\bibfnamefont {M.}~\bibnamefont {Imai-Imada}}, \bibinfo
  {author} {\bibfnamefont {S.}~\bibnamefont {Kawahara}}, \bibinfo {author}
  {\bibfnamefont {K.}~\bibnamefont {Kimura}}, \ and\ \bibinfo {author}
  {\bibfnamefont {Y.}~\bibnamefont {Kim}},\ }\href {\doibase
  10.1103/PhysRevLett.119.013901} {\bibfield  {journal} {\bibinfo  {journal}
  {Phys. Rev. Lett.}\ }\textbf {\bibinfo {volume} {119}},\ \bibinfo {pages}
  {013901} (\bibinfo {year} {2017})}\BibitemShut {NoStop}%
\bibitem [{\citenamefont {Zhang}\ \emph {et~al.}(2017)\citenamefont {Zhang},
  \citenamefont {Meng}, \citenamefont {Zhang}, \citenamefont {Luo},
  \citenamefont {Yu}, \citenamefont {Yang}, \citenamefont {Zhang},
  \citenamefont {Esteban}, \citenamefont {Aizpurua}, \citenamefont {Luo},
  \citenamefont {Yang}, \citenamefont {Dong},\ and\ \citenamefont
  {Hou}}]{zhang2017sub}%
  \BibitemOpen
  \bibfield  {author} {\bibinfo {author} {\bibfnamefont {Y.}~\bibnamefont
  {Zhang}}, \bibinfo {author} {\bibfnamefont {Q.-S.}\ \bibnamefont {Meng}},
  \bibinfo {author} {\bibfnamefont {L.}~\bibnamefont {Zhang}}, \bibinfo
  {author} {\bibfnamefont {Y.}~\bibnamefont {Luo}}, \bibinfo {author}
  {\bibfnamefont {Y.-J.}\ \bibnamefont {Yu}}, \bibinfo {author} {\bibfnamefont
  {B.}~\bibnamefont {Yang}}, \bibinfo {author} {\bibfnamefont {Y.}~\bibnamefont
  {Zhang}}, \bibinfo {author} {\bibfnamefont {R.}~\bibnamefont {Esteban}},
  \bibinfo {author} {\bibfnamefont {J.}~\bibnamefont {Aizpurua}}, \bibinfo
  {author} {\bibfnamefont {Y.}~\bibnamefont {Luo}}, \bibinfo {author}
  {\bibfnamefont {J.-L.}\ \bibnamefont {Yang}}, \bibinfo {author}
  {\bibfnamefont {Z.-C.}\ \bibnamefont {Dong}}, \ and\ \bibinfo {author}
  {\bibfnamefont {J.~G.}\ \bibnamefont {Hou}},\ }\href@noop {} {\bibfield
  {journal} {\bibinfo  {journal} {Nat. Commun.}\ }\textbf {\bibinfo {volume}
  {8}} (\bibinfo {year} {2017})}\BibitemShut {NoStop}%
\bibitem [{\citenamefont {Huang}\ \emph {et~al.}(2005)\citenamefont {Huang},
  \citenamefont {Medforth},\ and\ \citenamefont
  {Schweitzer-Stenner}}]{huang2005porphyrins}%
  \BibitemOpen
  \bibfield  {author} {\bibinfo {author} {\bibfnamefont {Q.}~\bibnamefont
  {Huang}}, \bibinfo {author} {\bibfnamefont {C.~J.}\ \bibnamefont {Medforth}},
  \ and\ \bibinfo {author} {\bibfnamefont {R.}~\bibnamefont
  {Schweitzer-Stenner}},\ }\href {\doibase 10.1021/jp052986a} {\bibfield
  {journal} {\bibinfo  {journal} {J. Phys. Chem. A}\ }\textbf {\bibinfo
  {volume} {109}},\ \bibinfo {pages} {10493} (\bibinfo {year}
  {2005})}\BibitemShut {NoStop}%
\bibitem [{\citenamefont {S\'{a}nchez-Carrera}\ \emph
  {et~al.}(2010)\citenamefont {S\'{a}nchez-Carrera}, \citenamefont {Delgado},
  \citenamefont {Ferrón}, \citenamefont {Osuna}, \citenamefont {Hernández},
  \citenamefont {Navarrete},\ and\ \citenamefont
  {Aspuru-Guzik}}]{SANCHEZCARRERA20101701}%
  \BibitemOpen
  \bibfield  {author} {\bibinfo {author} {\bibfnamefont {R.~S.}\ \bibnamefont
  {S\'{a}nchez-Carrera}}, \bibinfo {author} {\bibfnamefont {M.~C.~R.}\
  \bibnamefont {Delgado}}, \bibinfo {author} {\bibfnamefont {C.~C.}\
  \bibnamefont {Ferrón}}, \bibinfo {author} {\bibfnamefont {R.~M.}\
  \bibnamefont {Osuna}}, \bibinfo {author} {\bibfnamefont {V.}~\bibnamefont
  {Hernández}}, \bibinfo {author} {\bibfnamefont {J.~T.~L.}\ \bibnamefont
  {Navarrete}}, \ and\ \bibinfo {author} {\bibfnamefont {A.}~\bibnamefont
  {Aspuru-Guzik}},\ }\href {\doibase
  https://doi.org/10.1016/j.orgel.2010.07.001} {\bibfield  {journal} {\bibinfo
  {journal} {Org. Electron.}\ }\textbf {\bibinfo {volume} {11}},\ \bibinfo
  {pages} {1701 } (\bibinfo {year} {2010})}\BibitemShut {NoStop}%
\bibitem [{\citenamefont {Mollow}(1969)}]{mollow1969power}%
  \BibitemOpen
  \bibfield  {author} {\bibinfo {author} {\bibfnamefont {B.}~\bibnamefont
  {Mollow}},\ }\href@noop {} {\bibfield  {journal} {\bibinfo  {journal} {Phys.
  Rev.}\ }\textbf {\bibinfo {volume} {188}} (\bibinfo {year}
  {1969})}\BibitemShut {NoStop}%
\bibitem [{\citenamefont {Ge}\ \emph {et~al.}(2013)\citenamefont {Ge},
  \citenamefont {Van~Vlack}, \citenamefont {Yao}, \citenamefont {Young},\ and\
  \citenamefont {Hughes}}]{Ge2013mollow}%
  \BibitemOpen
  \bibfield  {author} {\bibinfo {author} {\bibfnamefont {R.-C.}\ \bibnamefont
  {Ge}}, \bibinfo {author} {\bibfnamefont {C.}~\bibnamefont {Van~Vlack}},
  \bibinfo {author} {\bibfnamefont {P.}~\bibnamefont {Yao}}, \bibinfo {author}
  {\bibfnamefont {J.~F.}\ \bibnamefont {Young}}, \ and\ \bibinfo {author}
  {\bibfnamefont {S.}~\bibnamefont {Hughes}},\ }\href {\doibase
  10.1103/PhysRevB.87.205425} {\bibfield  {journal} {\bibinfo  {journal} {Phys.
  Rev. B}\ }\textbf {\bibinfo {volume} {87}},\ \bibinfo {pages} {205425}
  (\bibinfo {year} {2013})}\BibitemShut {NoStop}%
\bibitem [{\citenamefont {Wrigge}\ \emph {et~al.}(2008)\citenamefont {Wrigge},
  \citenamefont {Gerhardt}, \citenamefont {Hwang}, \citenamefont {Zumofen},\
  and\ \citenamefont {Sandoghdar}}]{wrigge2008efficient}%
  \BibitemOpen
  \bibfield  {author} {\bibinfo {author} {\bibfnamefont {G.}~\bibnamefont
  {Wrigge}}, \bibinfo {author} {\bibfnamefont {I.}~\bibnamefont {Gerhardt}},
  \bibinfo {author} {\bibfnamefont {J.}~\bibnamefont {Hwang}}, \bibinfo
  {author} {\bibfnamefont {G.}~\bibnamefont {Zumofen}}, \ and\ \bibinfo
  {author} {\bibfnamefont {V.}~\bibnamefont {Sandoghdar}},\ }\href@noop {}
  {\bibfield  {journal} {\bibinfo  {journal} {Nat. Phys.}\ }\textbf {\bibinfo
  {volume} {4}},\ \bibinfo {pages} {60} (\bibinfo {year} {2008})}\BibitemShut
  {NoStop}%
\bibitem [{\citenamefont {Mu{\~{n}}oz}\ \emph {et~al.}(2018)\citenamefont
  {Mu{\~{n}}oz}, \citenamefont {Laussy}, \citenamefont {del Valle},
  \citenamefont {Tejedor},\ and\ \citenamefont
  {Gonz\'{a}lez-Tudela}}]{SanchezMunoz2018}%
  \BibitemOpen
  \bibfield  {author} {\bibinfo {author} {\bibfnamefont {C.~S.}\ \bibnamefont
  {Mu{\~{n}}oz}}, \bibinfo {author} {\bibfnamefont {F.~P.}\ \bibnamefont
  {Laussy}}, \bibinfo {author} {\bibfnamefont {E.}~\bibnamefont {del Valle}},
  \bibinfo {author} {\bibfnamefont {C.}~\bibnamefont {Tejedor}}, \ and\
  \bibinfo {author} {\bibfnamefont {A.}~\bibnamefont {Gonz\'{a}lez-Tudela}},\
  }\href {\doibase 10.1364/OPTICA.5.000014} {\bibfield  {journal} {\bibinfo
  {journal} {Optica}\ }\textbf {\bibinfo {volume} {5}},\ \bibinfo {pages} {14}
  (\bibinfo {year} {2018})}\BibitemShut {NoStop}%
\bibitem [{\citenamefont {Miroshnichenko}\ \emph {et~al.}(2010)\citenamefont
  {Miroshnichenko}, \citenamefont {Flach},\ and\ \citenamefont
  {Kivshar}}]{Miroshnichenko2010}%
  \BibitemOpen
  \bibfield  {author} {\bibinfo {author} {\bibfnamefont {A.~E.}\ \bibnamefont
  {Miroshnichenko}}, \bibinfo {author} {\bibfnamefont {S.}~\bibnamefont
  {Flach}}, \ and\ \bibinfo {author} {\bibfnamefont {Y.~S.}\ \bibnamefont
  {Kivshar}},\ }\href {\doibase 10.1103/RevModPhys.82.2257} {\bibfield
  {journal} {\bibinfo  {journal} {Rev. Mod. Phys.}\ }\textbf {\bibinfo {volume}
  {82}},\ \bibinfo {pages} {2257} (\bibinfo {year} {2010})}\BibitemShut
  {NoStop}%
\bibitem [{\citenamefont {Kabuss}\ \emph {et~al.}(2011)\citenamefont {Kabuss},
  \citenamefont {Carmele}, \citenamefont {Richter},\ and\ \citenamefont
  {Knorr}}]{Kabuss2011}%
  \BibitemOpen
  \bibfield  {author} {\bibinfo {author} {\bibfnamefont {J.}~\bibnamefont
  {Kabuss}}, \bibinfo {author} {\bibfnamefont {A.}~\bibnamefont {Carmele}},
  \bibinfo {author} {\bibfnamefont {M.}~\bibnamefont {Richter}}, \ and\
  \bibinfo {author} {\bibfnamefont {A.}~\bibnamefont {Knorr}},\ }\href
  {\doibase 10.1103/PhysRevB.84.125324} {\bibfield  {journal} {\bibinfo
  {journal} {Phys. Rev. B}\ }\textbf {\bibinfo {volume} {84}},\ \bibinfo
  {pages} {125324} (\bibinfo {year} {2011})}\BibitemShut {NoStop}%
\bibitem [{\citenamefont {Cohen-Tannoudji}\ and\ \citenamefont
  {Reynaud}(1977{\natexlab{a}})}]{cohen1977modification}%
  \BibitemOpen
  \bibfield  {author} {\bibinfo {author} {\bibfnamefont {C.}~\bibnamefont
  {Cohen-Tannoudji}}\ and\ \bibinfo {author} {\bibfnamefont {S.}~\bibnamefont
  {Reynaud}},\ }\href@noop {} {\bibfield  {journal} {\bibinfo  {journal} {J.
  Phys. B: At. Mol. Phys.}\ }\textbf {\bibinfo {volume} {10}},\ \bibinfo
  {pages} {365} (\bibinfo {year} {1977}{\natexlab{a}})}\BibitemShut {NoStop}%
\bibitem [{\citenamefont {Agarwal}\ and\ \citenamefont
  {Jha}(1979)}]{agarwal1979theory}%
  \BibitemOpen
  \bibfield  {author} {\bibinfo {author} {\bibfnamefont {G.~S.}\ \bibnamefont
  {Agarwal}}\ and\ \bibinfo {author} {\bibfnamefont {S.~S.}\ \bibnamefont
  {Jha}},\ }\href@noop {} {\bibfield  {journal} {\bibinfo  {journal} {J. Phys.
  B: At. Mol. Phys.}\ }\textbf {\bibinfo {volume} {12}},\ \bibinfo {pages}
  {2655} (\bibinfo {year} {1979})}\BibitemShut {NoStop}%
\bibitem [{\citenamefont {Am-Shallem}\ \emph {et~al.}(2015)\citenamefont
  {Am-Shallem}, \citenamefont {Levy}, \citenamefont {Schaefer},\ and\
  \citenamefont {Kosloff}}]{am2015three}%
  \BibitemOpen
  \bibfield  {author} {\bibinfo {author} {\bibfnamefont {M.}~\bibnamefont
  {Am-Shallem}}, \bibinfo {author} {\bibfnamefont {A.}~\bibnamefont {Levy}},
  \bibinfo {author} {\bibfnamefont {I.}~\bibnamefont {Schaefer}}, \ and\
  \bibinfo {author} {\bibfnamefont {R.}~\bibnamefont {Kosloff}},\ }\href@noop
  {} {\bibfield  {journal} {\bibinfo  {journal} {arXiv preprint
  arXiv:1510.08634}\ } (\bibinfo {year} {2015})}\BibitemShut {NoStop}%
\bibitem [{\citenamefont {Tan}(1999)}]{tan1999quantum}%
  \BibitemOpen
  \bibfield  {author} {\bibinfo {author} {\bibfnamefont {S.~M.}\ \bibnamefont
  {Tan}},\ }\href@noop {} {\emph {\bibinfo {title} {A quantum optics toolbox
  for {M}atlab 5}}} (\bibinfo {year} {1999}),\ \bibinfo {note}
  {https://copilot.caltech.edu/documents/230-qousersguide.pdf}\BibitemShut
  {NoStop}%
\bibitem [{\citenamefont {Rabl}(2010)}]{rabl2010cooling}%
  \BibitemOpen
  \bibfield  {author} {\bibinfo {author} {\bibfnamefont {P.}~\bibnamefont
  {Rabl}},\ }\href@noop {} {\bibfield  {journal} {\bibinfo  {journal} {Phys.
  Rev. B}\ }\textbf {\bibinfo {volume} {82}},\ \bibinfo {pages} {165320}
  (\bibinfo {year} {2010})}\BibitemShut {NoStop}%
\bibitem [{\citenamefont {Cohen-Tannoudji}\ and\ \citenamefont
  {Reynaud}(1994)}]{cohen1994dressed}%
  \BibitemOpen
  \bibfield  {author} {\bibinfo {author} {\bibfnamefont {C.}~\bibnamefont
  {Cohen-Tannoudji}}\ and\ \bibinfo {author} {\bibfnamefont {S.}~\bibnamefont
  {Reynaud}},\ }\href@noop {} {\bibfield  {journal} {\bibinfo  {journal} {Atoms
  in Electromagnetic Fields}\ }\textbf {\bibinfo {volume} {1}},\ \bibinfo
  {pages} {310} (\bibinfo {year} {1994})}\BibitemShut {NoStop}%
\bibitem [{\citenamefont {Cohen-Tannoudji}\ and\ \citenamefont
  {Reynaud}(1977{\natexlab{b}})}]{Tanoudji77dressedmulti}%
  \BibitemOpen
  \bibfield  {author} {\bibinfo {author} {\bibfnamefont {C.}~\bibnamefont
  {Cohen-Tannoudji}}\ and\ \bibinfo {author} {\bibfnamefont {S.}~\bibnamefont
  {Reynaud}},\ }\href {http://stacks.iop.org/0022-3700/10/i=3/a=005} {\bibfield
   {journal} {\bibinfo  {journal} {J. Phys. B: At. Mol. Phys.}\ }\textbf
  {\bibinfo {volume} {10}},\ \bibinfo {pages} {345} (\bibinfo {year}
  {1977}{\natexlab{b}})}\BibitemShut {NoStop}%
\bibitem [{\citenamefont {Scully}\ and\ \citenamefont
  {Zubairy}(1997)}]{Scully1997}%
  \BibitemOpen
  \bibfield  {author} {\bibinfo {author} {\bibfnamefont {M.~O.}\ \bibnamefont
  {Scully}}\ and\ \bibinfo {author} {\bibfnamefont {M.~S.}\ \bibnamefont
  {Zubairy}},\ }\href@noop {} {\emph {\bibinfo {title} {Quantum optics}}}\
  (\bibinfo  {publisher} {Cambridge {U}niversity {P}ress},\ \bibinfo {year}
  {1997})\BibitemShut {NoStop}%
\end{thebibliography}%

\end{document}